\documentclass{article}

\usepackage{microtype}
\usepackage{graphicx}
\usepackage{subfigure}
\usepackage{booktabs}

\usepackage{hyperref}

\usepackage{xspace}

\usepackage{float}
\usepackage{subcaption}
\usepackage{subfigure}
\usepackage{tikz}
\usepackage{amsmath}
\usepackage{subcaption}
\usepackage{graphicx}
\usepackage{caption}
\usepackage{xurl}
\usepackage{enumitem}
\usepackage{float}
\usepackage{listings}
\usepackage[normalem]{ulem}
\usepackage[dvipsnames]{xcolor}

\newcommand{\sref}[1]{Section \ref{#1}}
\newcommand{\fref}[1]{Figure \ref{#1}}
\newcommand{\aref}[1]{Appendix \ref{#1}}
\newcommand{\algoref}[1]{Algorithm \ref{#1}}

\newcommand{\cam}[1]{{\color{black}#1}}
\newcommand{\fin}[1]{{\color{black}#1}}

\newcommand{\proj}{CRAFT\xspace}
\newcommand{\proje}{CRA}

\newcommand{\del}[1]{}

\newcommand*\circled[1]{\tikz[baseline=(char.base)]{
            \node[shape=circle,fill,inner sep=1pt] (char) {\textcolor{white}{#1}};}}
\newcommand*\bcircled[1]{\textbf{\tikz[baseline=(char.base)]{
            \node[shape=circle,fill,inner sep=1pt] (char) {\textcolor{white}{#1}};}}}

\usepackage[accepted]{mlsys2026}

\AddToShipoutPictureBG*{
  \AtPageUpperLeft{
    \hspace*{\paperwidth}
    \raisebox{-68pt}{
      \llap{
        \href{https://www.acm.org/publications/policies/artifact-review-and-badging-current}{
          \includegraphics[height=65pt]{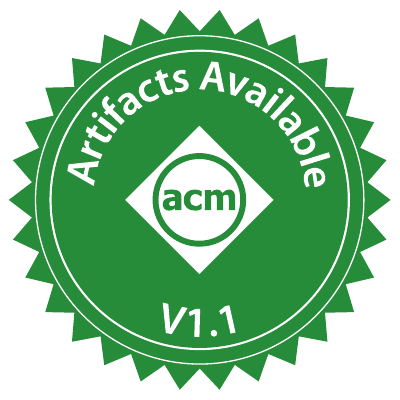}}
        \hspace{1pt}
        \href{https://www.acm.org/publications/policies/artifact-review-and-badging-current}{
          \includegraphics[height=65pt]{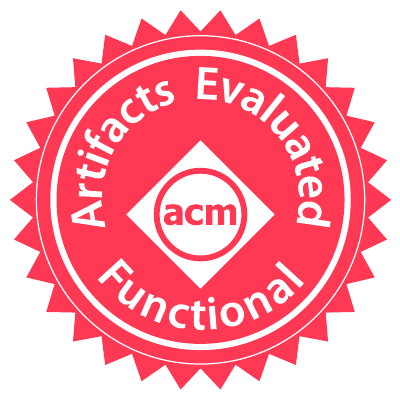}}
        \hspace{80pt}
      }
    }
  }
}

\mlsystitlerunning{CRAFT: Fine-Grained Cost-Aware Expert Replication For Efficient Mixture-of-Experts Serving}

\begin{document}

\twocolumn[
\mlsystitle{CRAFT: Fine-Grained Cost-Aware Expert Replication For Efficient Mixture-of-Experts Serving}



\mlsyssetsymbol{equal}{*}

\begin{mlsysauthorlist}
\mlsysauthor{Adrian Zhao}{equal,am,to}
\mlsysauthor{Zhenkun Cai}{am}
\mlsysauthor{Zhenyu Song}{am}
\mlsysauthor{Lingfan Yu}{am}
\mlsysauthor{Haozheng Fan}{am}
\mlsysauthor{Jun Wu}{am}
\mlsysauthor{Yida Wang}{am}
\mlsysauthor{Nandita Vijaykumar}{am,to}
\end{mlsysauthorlist}

\mlsysaffiliation{to}{Department of Computer Science, University of Toronto, Toronto, Ontario, Canada}
\mlsysaffiliation{am}{Amazon, Santa Clara, California, USA}

\mlsyscorrespondingauthor{Adrian Zhao}{adrianz@cs.toronto.edu}

\mlsyskeywords{Mixture of Experts, MoE, ML Inference, Machine Learning, MLSys}

\vskip 0.3in

\begin{abstract}
Mixture-of-Experts (MoE) has recently emerged as the mainstream architecture for efficiently scaling large language models while maintaining near-constant computational cost.
Expert parallelism distributes parameters by partitioning experts across devices, but this introduces token-level load imbalance during inference. 
Expert replication is a widely adopted load-balancing technique in serving frameworks that alleviates load imbalance in large-scale deployments by replicating experts with high loads. 
In this work, we demonstrate that existing replication schemes often \emph{over-replicate}, with many replicas providing marginal improvement. Replicas consume substantial GPU memory, which may lead to resource contention and throughput degradation.
We present \proj, an efficient expert replication framework that maximizes load balance under a given memory budget by performing fine-grained, per-layer replication based on the estimated replication benefit. 
\proj can be seamlessly integrated into existing serving frameworks without any additional training or model changes.
Our evaluation shows that \proj increases end-to-end serving throughput by $1.14\times$ on average (up to $1.2\times$) over existing replication techniques in large-scale deployments with models ranging from hundreds of billions to a trillion parameters.
\end{abstract}
]



\printAffiliationsAndNotice{}  

\section{Introduction}
Mixture-of-Experts (MoE) alleviates the steep compute and memory scaling of dense transformers by replacing the dense feed-forward blocks with a router and a pool of experts \cite{scalelaw,trainllm,SMoE}.
By activating only a subset of experts per token, MoE models decouple total parameters from per‑token FLOPs and achieve near‑constant inference cost as capacity scales \cite{glam}. It has become the foundation for recent large models across various domains, such as LLMs \cite{dpsr1,qwen3,llama4,kimik2}, image and video diffusion models \cite{wan22,remixdit}, and vision transformers \cite{adamv}, delivering competitive performance at substantially lower compute than dense counterparts. 
However, deploying MoE models incurs substantial operational costs, primarily due to the massive GPU memory footprint required to host all expert parameters \cite{deepspeed,switch}. Therefore, system-level optimizations that improve inference throughput yield significant cost-performance benefits by maximizing resource utilization and reducing GPU memory pressure \cite{nvmixtral,nvllama4,amdllama4}.

To effectively distribute large expert parameters across devices, Expert Parallelism (EP) partitions experts across devices, with each device holding weights of a subset of experts. EP uses two all‑to‑all collectives to dispatch tokens to the devices that host the assigned experts and then recombines them \cite{switch,gshard}. Despite its promising scalability, EP introduces a new system bottleneck during deployment \textemdash~ \textit{expert load imbalance}\cam{, where \textit{expert load} is the number of tokens routed to an expert}. The MoE router routes tokens to experts based on input distributions and often sends a disproportionate number of tokens to a few \textit{hot} experts while other experts remain \textit{cold} \cite{towardeffinf,megascale,sida}. Thus, the cumulative token loads across GPUs become imbalanced, leading to: (i) idleness on GPUs with low load and (ii) network congestion during dispatch and combine due to imbalanced traffic \cite{lina,netmoe,fastermoe,occult}. This degrades communication and computation efficiency during MoE inference, necessitating load-balancing strategies for EP.

\textit{Expert placement} and \textit{expert replication} are two load-balancing techniques applicable to any MoE model and are widely adopted in modern LLM serving frameworks \cite{sglang,vllm,tensorrt-llm,deepspeed}. Expert placement leverages expert load distribution to assign experts to devices such that the expected tokens per device and traffic patterns are balanced \cite{exflow,moetuner,lina}. However, expert placement fails in layers where a few experts dominate most of the token load, making a balanced packing impossible. Expert replication addresses this by replicating hot experts \cite{dpsv3}. This mitigates extreme load imbalance by distributing token loads from a single hot expert to multiple devices. \cam{We refer to the number of additional experts allocated during replication as \textit{replicas}.}

Although expert replication is an effective load-balancing strategy that mitigates extreme expert load imbalance, it requires substantial GPU memory to store the replica weights. The most widely adopted replication technique EPLB \cite{dpsv3} allocates one replica per layer per GPU, incurring significant GPU memory overhead on large MoE models with many MoE layers. Large-scale LLM deployments are already memory-intensive due to high parameter counts and the large KV cache necessary to efficiently process long-context requests \cite{vllm,flashattn2}. Therefore, the memory overhead limits the practicality of replication for large-scale MoE deployments. Expert placement does not incur additional memory overhead, but it cannot handle extreme load skew during inference. Thus, efficiently achieving balanced expert load in large-scale deployment remains an important unsolved problem for large MoE models.

\begin{figure}[!t]
    \centering
    \includegraphics[width=1\linewidth]{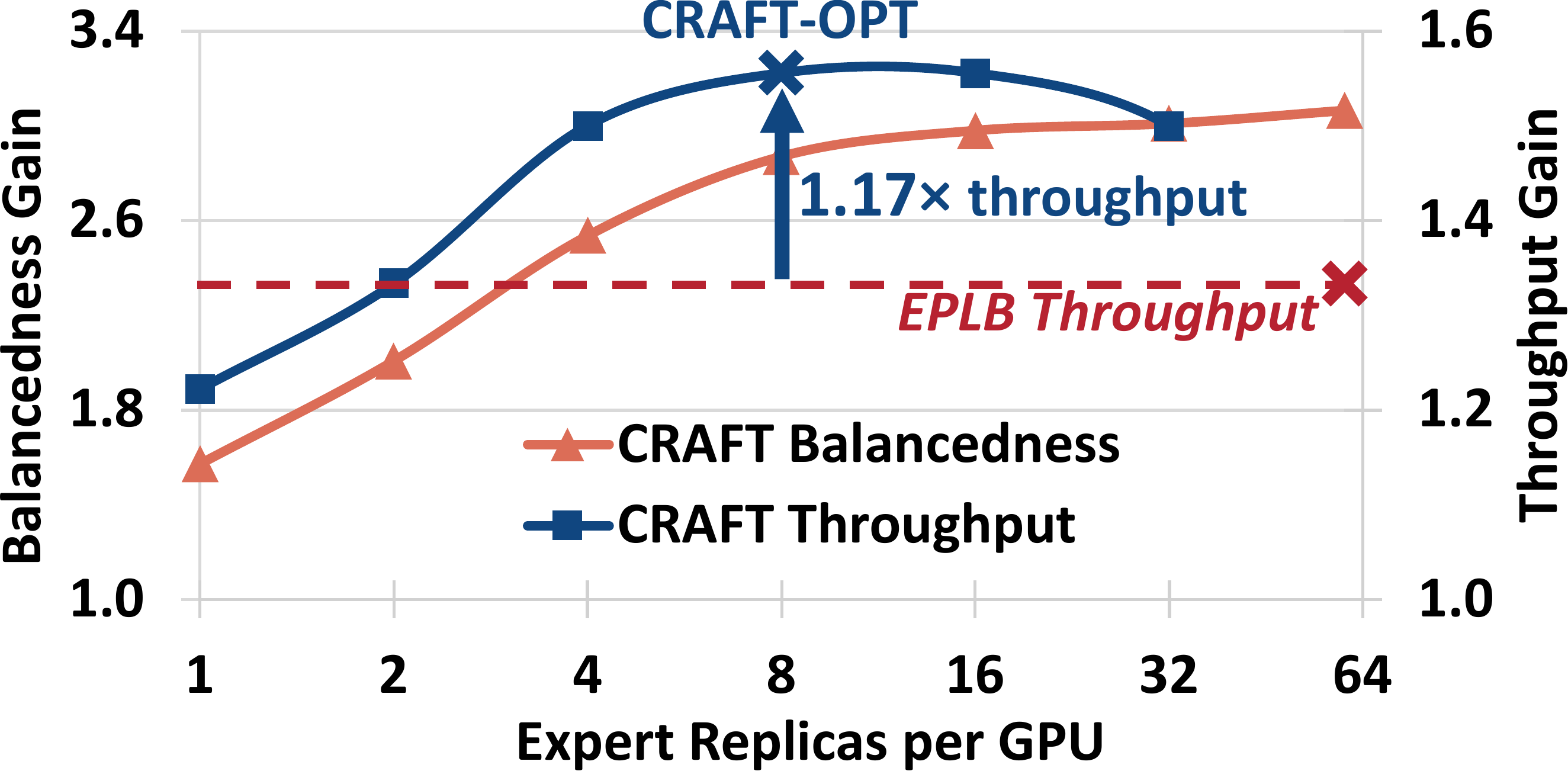}
    \caption{Balancedness gain (orange) and throughput gain (blue) of \proj on Kimi-K2-1000B deployed on 64 GPUs, normalized to the baseline with only expert placement (refer to \texttt{KE8} configuration in \sref{sec:setup}). EPLB is configured with 60 replicas per GPU on Kimi-K2 with 60 MoE layers (at minimum one replica per layer per GPU). \cam{The X-axis represents the number of replicas on a base-2 logarithmic scale, where more replicas imply higher memory consumption (to the right).}}
    \label{fig:tptbal1}
\end{figure}

In this work, we perform a detailed characterization of the benefits and overheads of replication as a means to achieve load balance.  We make two key observations: 
(1) As the number of replicas increases, balancedness gains diminish rapidly. (2) MoE layers benefit differently from replication and can be roughly categorized into two types: layers with few experts dominating token load (high-skew) benefit substantially from replication, while layers with balanced token load (low-skew) benefit little.
Leveraging these observations, we introduce \proj, an end-to-end cost-aware replica allocation framework for efficient expert replication. The key idea of \proj is to selectively allocate replicas at fine, per-layer granularity to layers that benefit more from replication. We develop a cost model that estimates the per-layer balancedness gain with replication (i.e., \textit{replication benefit}). Based on this cost model, \proj computes a per-layer replication plan with optimal balancedness by assigning more replicas to layers with higher estimated gains. 

\fref{fig:tptbal1} demonstrates the effectiveness of \proj: (1) \proj allocates replicas for each layer based on estimated performance benefits. This enables replication under flexible memory budgets that are much lower than EPLB (depicted by the blue curve), reducing memory usage while preserving balancedness. (2) \proj with the optimal number of replicas (\proj-OPT) achieves $1.17\times$ higher throughput compared to EPLB. The number of replicas can either be manually set by the user to meet specific memory constraints, or \proj can automatically set a value that optimizes replication efficiency. We implement \proj in three steps: First, we analyze expert load distributions and estimate per-layer benefits across different replica counts. Then, we use dynamic programming to select the optimal number of replicas per layer under the memory budget. Finally, we reserve memory and assign experts evenly across devices to minimize capacity imbalance, producing a complete replication plan for deployment.

Our contributions are summarized as follows:

\begin{itemize}
    \item To our knowledge, this is the first detailed characterization of how throughput and load balancedness scale with replication. We demonstrate that replication quickly yields diminishing returns due to the trade-off between memory overhead and balanced load, and identify layer-specific behavior as critical to achieving efficiency with replication.
    \item We introduce \proj, an end-to-end expert replication framework designed for efficient expert replication in large-scale MoE deployments. \proj integrates directly into existing LLM serving frameworks \cite{dpsv3,tensorrt-llm,sglang,vllm}, offering significant throughput gains and reduced operational costs \cite{infcost}. \proj applies to any MoE architecture and requires no additional training or model changes.
    \item We evaluate \proj on a state-of-the-art LLM serving framework SGLang \cite{sglang} for a wide range of datasets and MoE models. We demonstrate that \proj achieves consistent throughput gains, outperforming EPLB by $1.14\times$ on average (up to $1.2\times$) on an 8-node cluster.
\end{itemize}

\section{Background}
\subsection{Mixture of Experts}
\label{sec:moe}

Mixture-of-Experts (MoE) replaces a dense feed-forward block with a fixed number of \textit{experts} and a learned \textit{router} that assigns each input token to a sparse subset of experts. Each \textit{expert} is an independently parameterized feed-forward block consisting of an up-projection linear layer, a non-linear activation, and a down-projection linear layer. The \textit{router} typically implements a weighted top-$K$ routing mechanism with a lightweight gating network \cite{SMoE}. For each token, the gating network computes a score for each expert and selects the experts with the top-$K$ scores for activation. In practice, $K$ is often set to 8 in large-scale MoE models \cite{qwen3,dpsr1,kimik2}. After passing through the feed-forward blocks of the activated experts, their outputs are gathered and aggregated into a weighted sum. Because each token only activates a few experts, MoE enables the scaling of model parameters by adding more experts without substantially increasing computation costs \cite{glam}.

\subsection{Expert Load Imbalance}
\label{sec:loadimbal}

Unlike traditional dense models, where the compute load across devices is uniform, MoE with EP introduces device-level token load variance due to its dynamic routing. During MoE inference, the router activates experts based on the input tokens \cite{SMoE}. Real-world natural language tokens follow a Zipfian distribution, in which a few tokens are extremely frequent while most others are rare \cite{zipfian,zipf}. This leads to expert load imbalance, producing high-load \textit{hot} experts and low-load \textit{cold} experts at each MoE layer \cite{fastermoe,expchoice,towardeffinf,flexmoe}.

When deploying MoE models with EP, the expert load imbalance becomes device‑level load imbalance: GPUs that host hot experts become performance bottlenecks, causing stalls on GPUs with cold experts and network contention during the all‑to‑all. \fref{fig:placerep} illustrates an example of the expert load distribution of an MoE layer with 8 experts. The baseline placement strategy simply assigns experts to devices based on their ID. This leads to a significant load skew, where GPU 1 carries a much higher load than other GPUs, resulting in suboptimal performance.

\subsection{Inference Load-Balancing Techniques}
\label{sec:placerep}

\begin{figure}[!htb]
    \centering
    \includegraphics[width=1\linewidth]{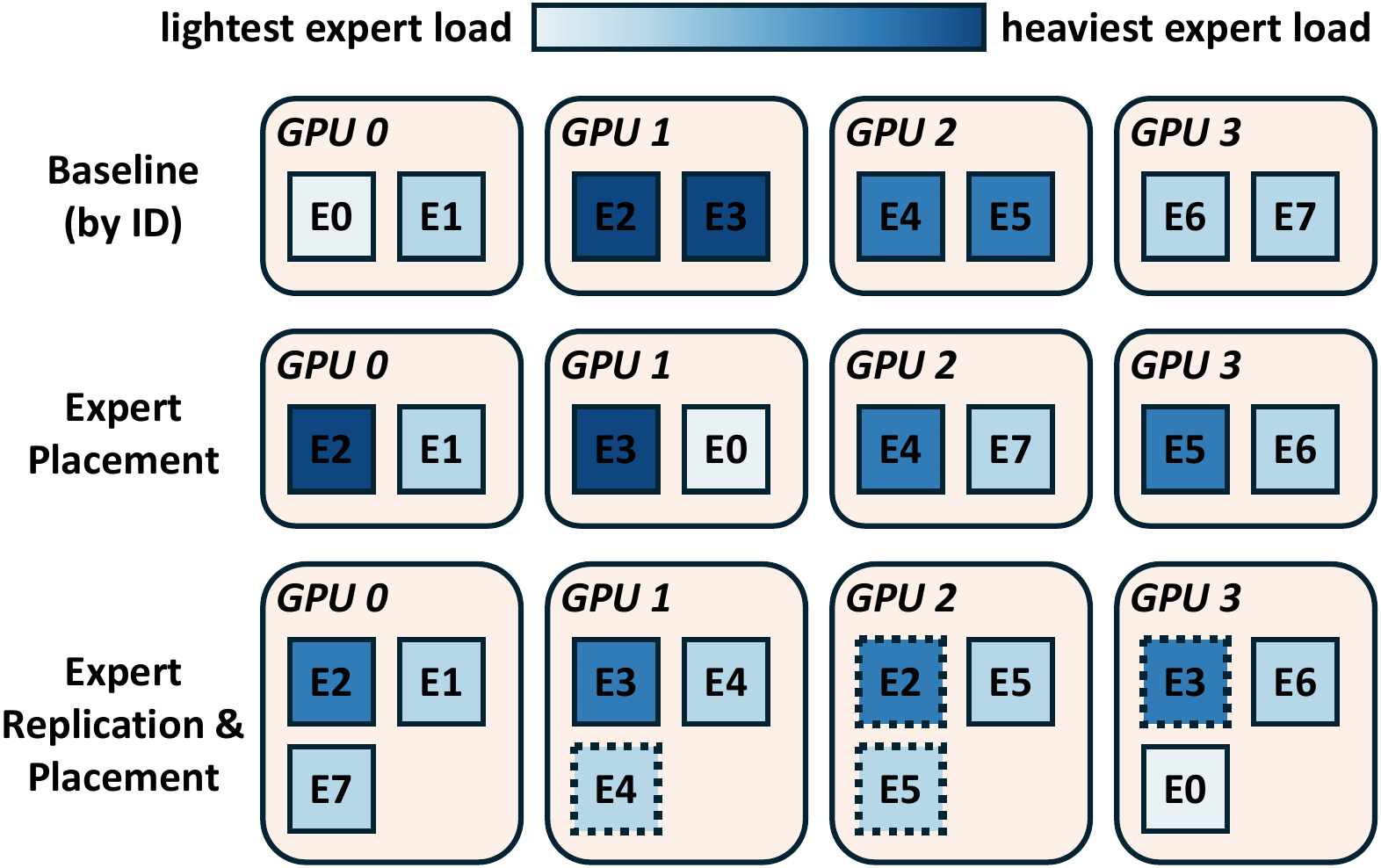}
    \caption{Expert load distribution under different EP optimizations on an MoE layer with 8 experts. Color density represents the number of tokens assigned to the expert (expert load) during inference, and dotted experts represent replicas allocated during replication.}
    \label{fig:placerep}
\end{figure}

Despite the dynamic nature of MoE routing, the expert load at each layer typically follows a distinct distribution. Existing load-balancing techniques profile expert load distributions by recording token loads across three dimensions \textemdash~ batch, layer, and experts. Analysis of the load distribution identifies hot and cold experts in each layer, providing opportunities for load-balancing optimizations.

\textbf{Expert placement} is a widely adopted EP load-balancing technique that alleviates load imbalance by co-locating hot experts with cold experts, thereby minimizing load variance across devices \cite{dpsv3,moetuner}. \fref{fig:placerep} demonstrates the effectiveness of expert placement, as the GPU load is more balanced compared to the baseline placement. However, expert placement falls short when the load distribution is significantly skewed, in which a few experts dominate the token load. For example, in \fref{fig:placerep}, if experts 2 and 3 constitute the majority of the total token load, the GPU load post-placement would remain imbalanced.

\textbf{Expert replication} further improves load balance by replicating hot experts and spreading their load across replicas. Expert Parallelism Load Balancer (EPLB) \cite{dpsv3} is a widely adopted load balancer in modern LLM serving frameworks \cite{sglang,vllm,tensorrt-llm,deepspeed} that supports both expert placement and expert replication. It employs \textit{uniform replication}, allocating the same number of replicas per layer to each GPU. \cam{The minimum number of replicas per layer with \textit{uniform replication} is equal to the number of GPUs (1 extra replica per GPU).} In \fref{fig:placerep}, four replicas are created for experts 2, 3, 4, and 5, and each GPU reserves memory for one additional expert slot to host these replicas. Then, expert placement maps the experts (including replicas) across GPUs. As shown in the figure, expert replication can achieve near-perfect GPU load balance.

\section{Motivation}
\label{sec:motivation}

\begin{figure*}[!htb]
    \centering
    \subfigure[\texttt{DE8}]{
        \includegraphics[width=0.329\textwidth]{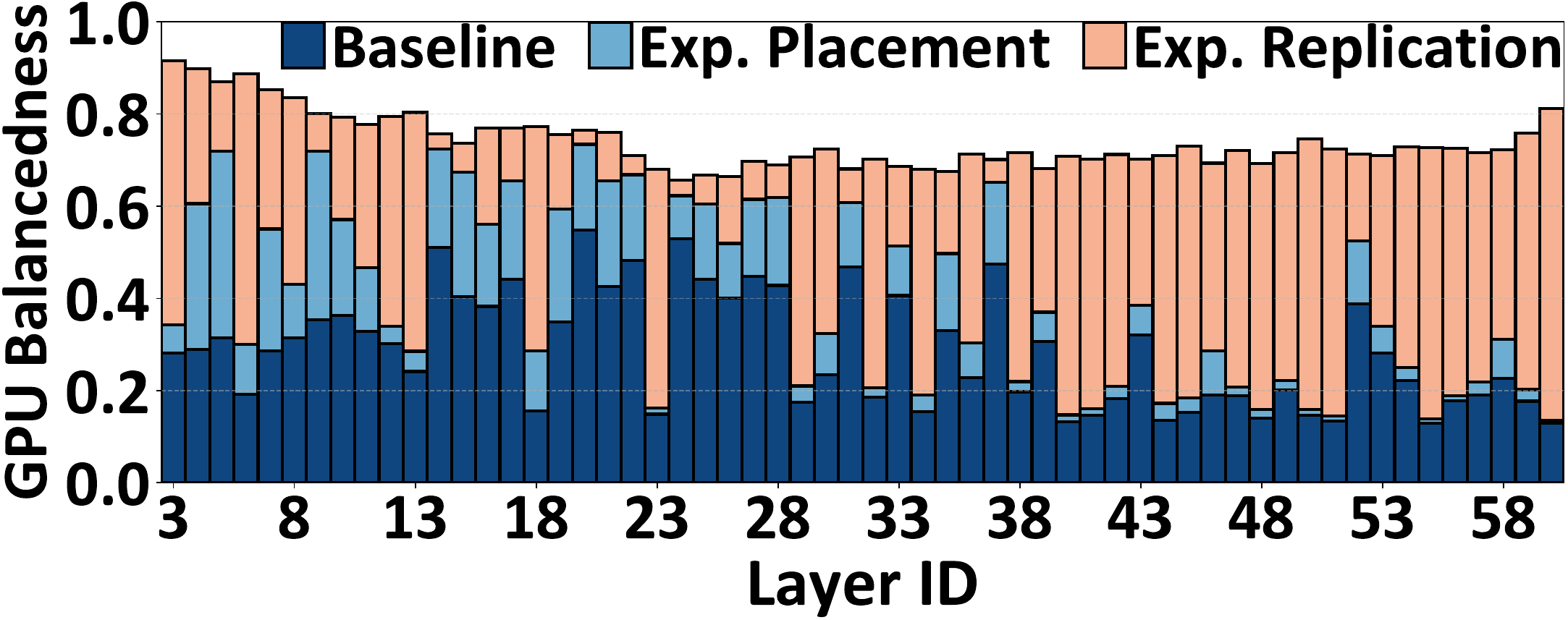}
        \label{fig:bdde8}
     } 
    \subfigure[\texttt{KJ6}]{
        \includegraphics[width=0.314\textwidth]{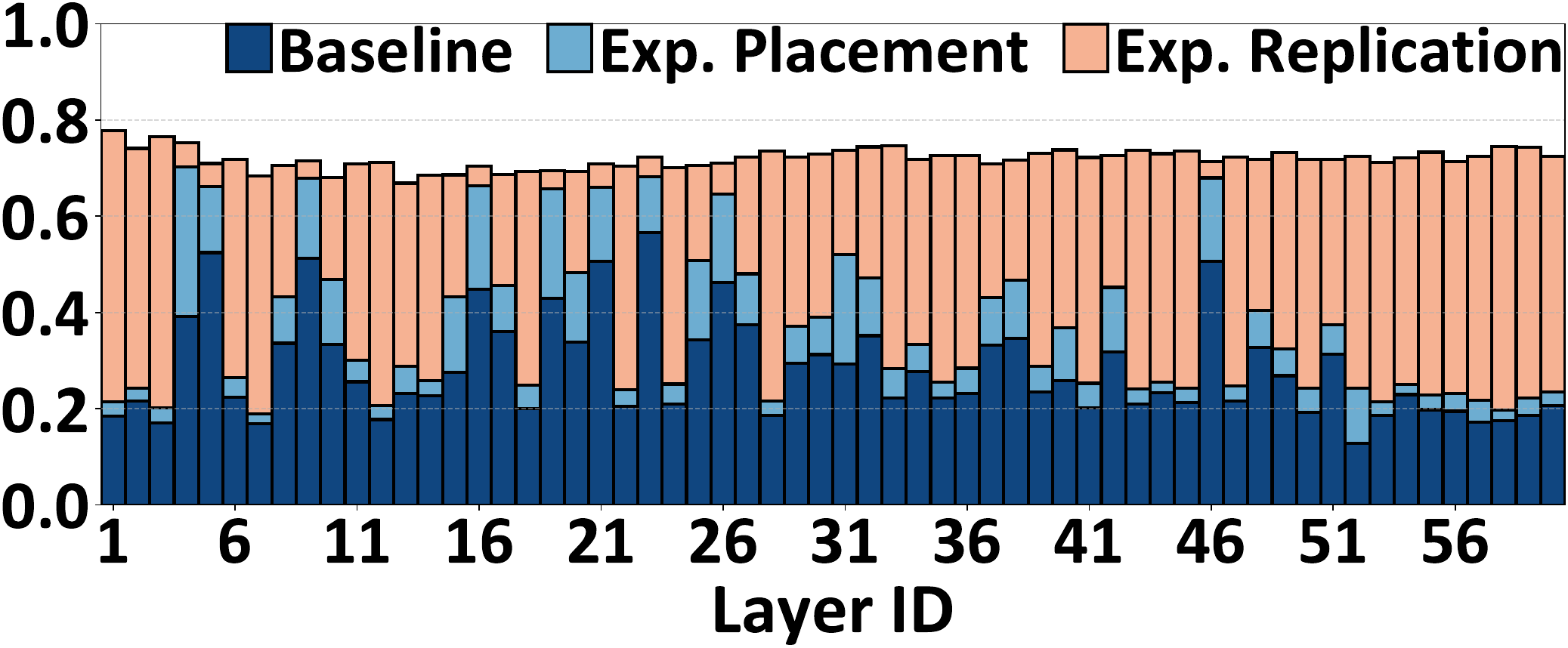}
        \label{fig:bdkj6}
    } 
    \subfigure[\texttt{KA6}]{
        \includegraphics[width=0.314\textwidth]{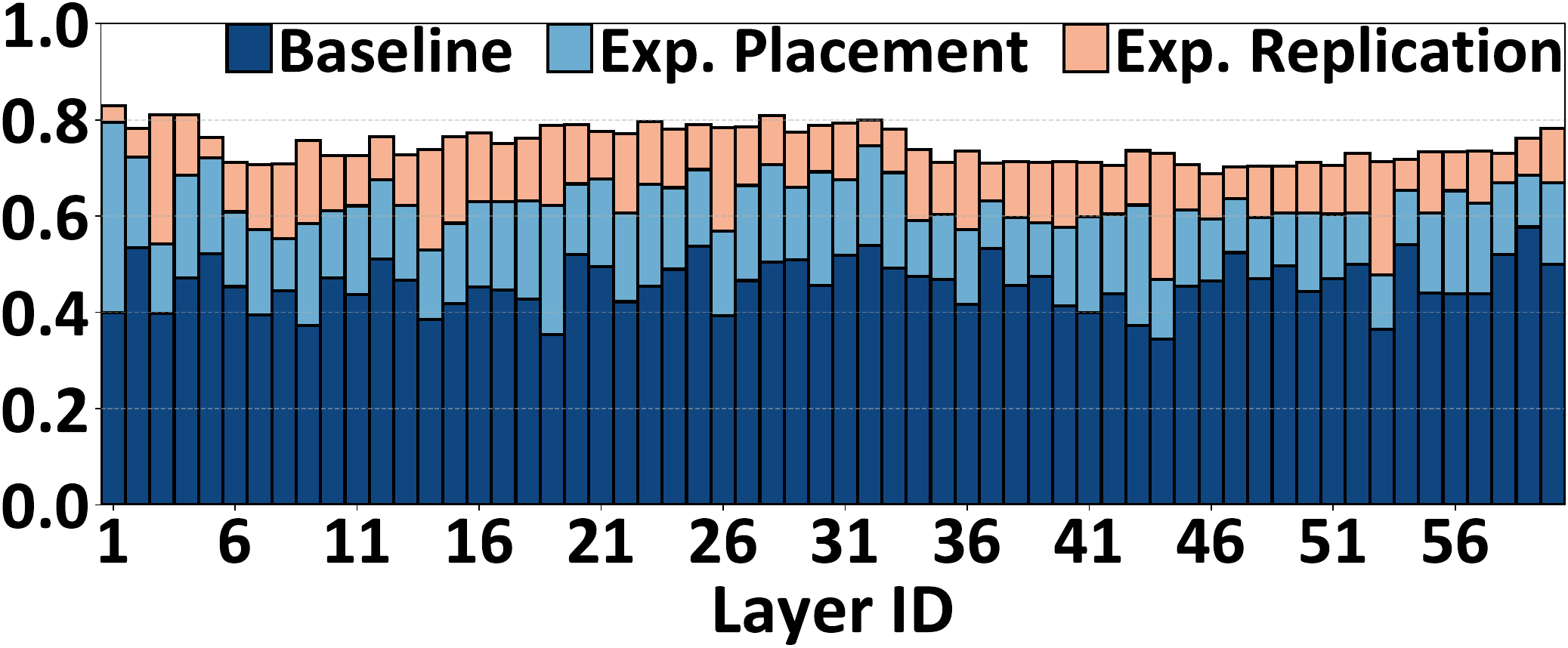}
        \label{fig:bdka6}
    } 
    \subfigure[\texttt{DL8}]{
        \includegraphics[width=0.329\textwidth]{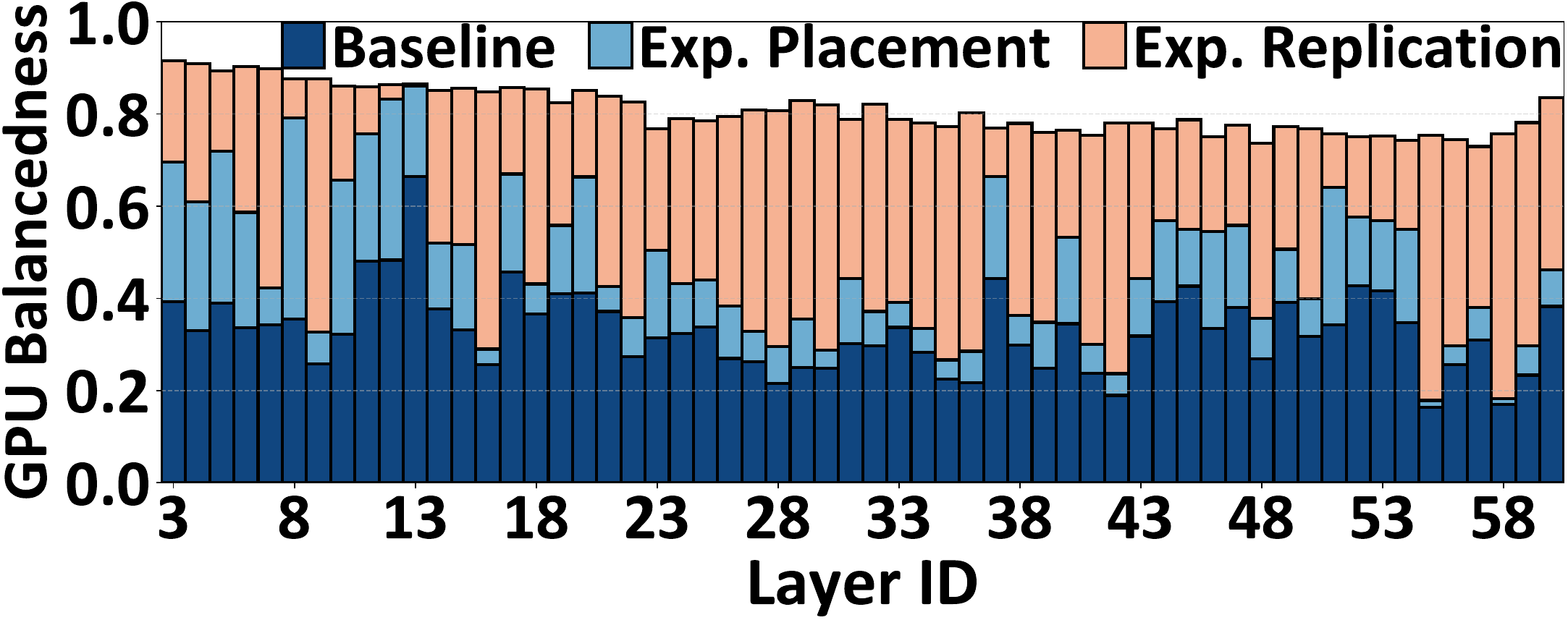}
        \label{fig:bddl8}
    } 
    \subfigure[\texttt{KJ12}]{
        \includegraphics[width=0.314\textwidth]{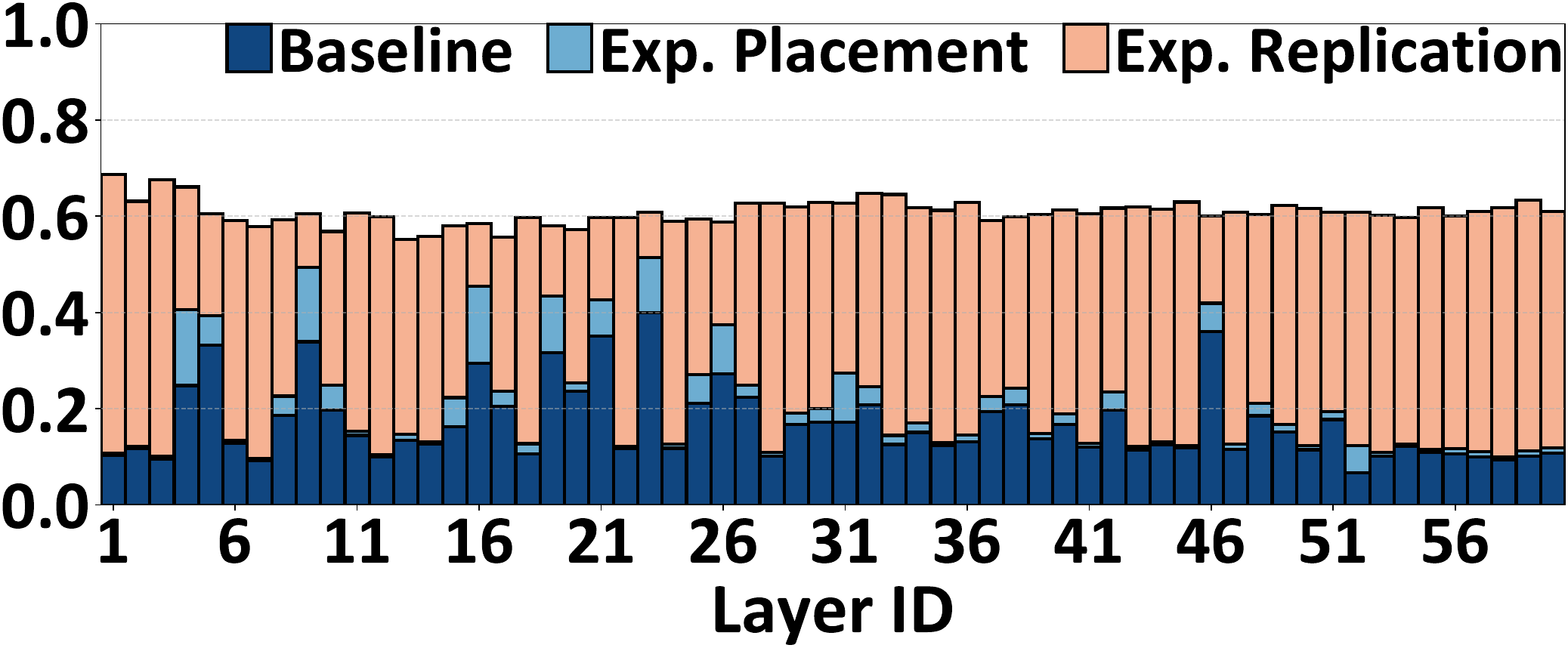}
        \label{fig:bdkj12}
    } 
    \subfigure[\texttt{KA12}]{
        \includegraphics[width=0.314\textwidth]{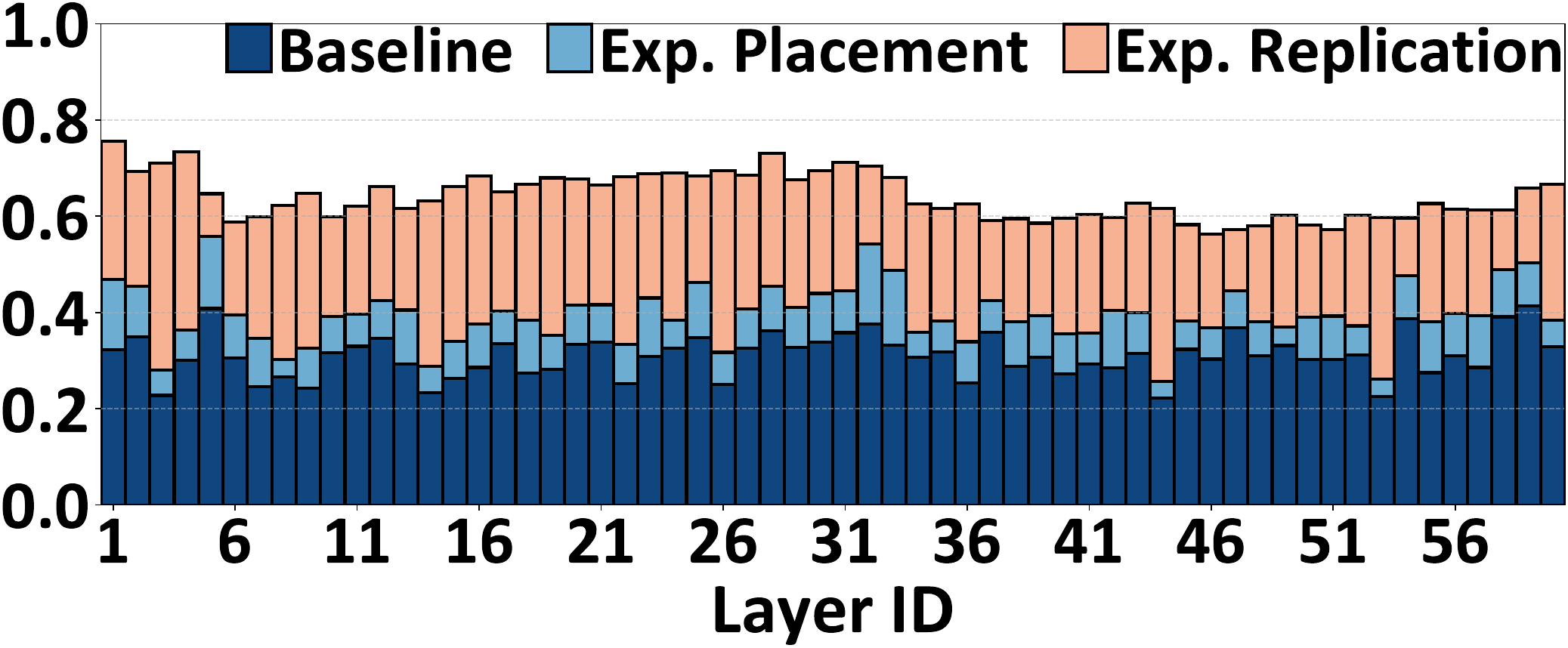}
        \label{fig:bdka12}
    } 
    \caption{Breakdown of the final post-replication balancedness by each technique's contribution across various configurations. The total height of a bar represents the GPU balancedness in an MoE layer with expert replication. A technique constituting a higher portion of the bar implies higher contribution to the overall balancedness. Layers excluded from the figures are non-MoE dense layers.}
    \label{fig:breakdown}
\end{figure*}

Although expert replication with placement mitigates most expert-level load imbalance, it introduces an important trade-off between GPU memory usage and load balance.
In this section, we present a detailed analysis of the effectiveness of expert replication.

\subsection{Experimental Setup}
\label{sec:setup}

\textbf{System. }Our experiments are conducted on a cluster of AWS EC2 p4de.24xlarge instances, each equipped with 8 NVIDIA A100 GPUs (80 GB each) interconnected via NVLink. Inter-node communication uses the Elastic Fabric Adapter (EFA) \cite{efa} with direct peer-to-peer (P2P) GPU communication support. All evaluations use CUDA 12.8 and NCCL 2.26.2 \cite{nccl}.

\textbf{Metrics.} \textit{Expert load} is measured as the total number of tokens dispatched to an expert after routing. We evaluate load balance using \textit{balancedness}, computed as the average load divided by the max load (higher means more balanced).

\textbf{Models and Workloads. }We use two large-scale MoE models in bfloat16 format: \texttt{D} \textemdash~ DeepSeek-R1-671B with 58 MoE layers (layers 3 - 60) and 256 experts \cite{dpsr1}, and \texttt{K} \textemdash~ Kimi-K2-1000B with 60 MoE layers (layers 1 - 60) and 384 experts \cite{kimik2}. Both models employ top-8 expert routing. \cam{For evaluation, we use diverse long-sequence workloads from large-scale datasets ranging from hundreds of millions to trillions of tokens. The workloads consist of diverse text spanning a broad range of domains, including law, science, entertainment, literature, news articles and technical documents in various formats such as PDFs, passages, and research papers: (1) two splits from the FinePDFs dataset \cite{finepdfs}: \texttt{E} \textemdash~ deu\_Latn split with German PDF files, and \texttt{J} \textemdash~ jpn\_Jpan split with Japanese and Chinese PDF files, (2) \texttt{L} \textemdash~ Lambada dataset \cite{lambada} with long narrative passages, (3) \texttt{A} \textemdash~RedPajama-Data-1T dataset \cite{redpajama} arXiv split with academic research papers.} All inputs are chunked to 4096 tokens with a fixed-size output length of 256 tokens. \cam{For each workload, we randomly sample a pool of sequences and run 3000 inference batches to collect the expert load distribution (\sref{sec:placerep}).}

\textbf{Setup Configuration Syntax. }We use a two-letter combination to represent different model-workload setups, followed by a number indicating the cluster size. For example, \texttt{DE8} represents the DeepSeek-R1 model and the FinePDF deu\_Latn workload on a cluster of 8 nodes.

\subsection{Replication Effectiveness on Load-Balancing}
\label{sec:repeffect}

We analyze the effectiveness of EPLB placement and replication by measuring the achieved GPU load balancedness for each layer.
\fref{fig:breakdown} illustrates the breakdown of the balancedness gains when applying expert replication. We make two observations.

\begin{figure}[!htb]
    \centering
    \subfigure[Layer 20 - low-skew layer (expert load range 0 - 6K)]{
        \includegraphics[width=0.48\textwidth]{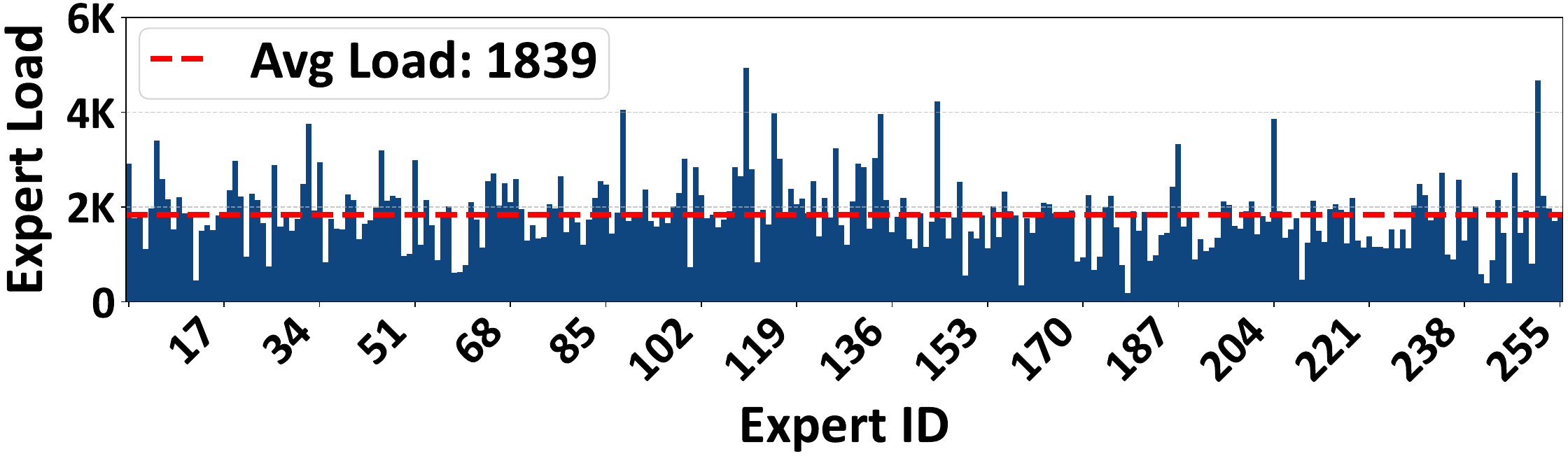}
        \label{fig:ballaydist}
     } 
    \subfigure[Layer 51 - high-skew layer (expert load range 0 - 60K)]{
        \includegraphics[width=0.48\textwidth]{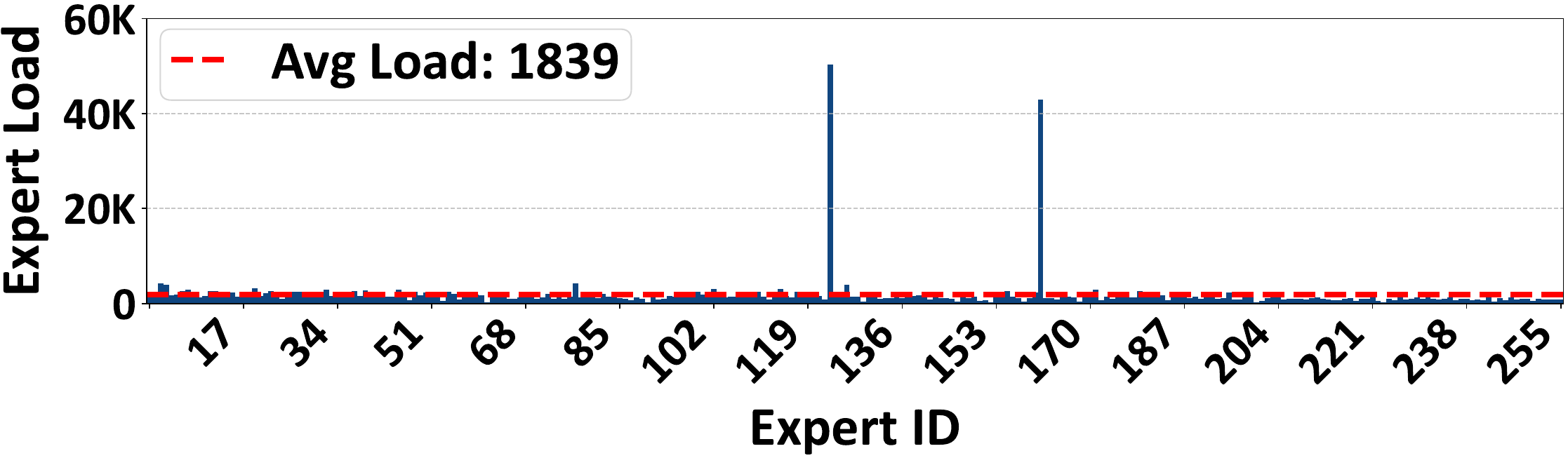}
        \label{fig:skewlaydist}
    } 
    \caption{Average expert load distribution of \texttt{DE8}. Total load is identical on both layers; the red line marks the average load.}
    \label{fig:explddist}
\end{figure}

\textbf{Observation 1: The effectiveness of replication varies across layers. }Although the effectiveness of replication matches or exceeds expert placement in many layers, some layers benefit little. To characterize this variation, we examine two representative layers in \fref{fig:bdde8} \textemdash~layer 20 (least effective) and layer 51 (most effective). \fref{fig:explddist} illustrates their average expert load distributions. Layer 20 exhibits a balanced, dispersed distribution. This explains its high baseline balancedness with placement, as the peak-to-mean load ratio is only approximately $2.5\times$, allowing for effective hot–cold expert co-location. In contrast, layer 51 exhibits strong skew: the hottest expert receives $>27\times$ the average load, and the top two experts account for $\approx 20\%$ of the total load. Placement alone cannot achieve load balance because GPUs hosting these hot experts are inevitably overloaded. Replication mitigates this skew by distributing hot-expert traffic across replicas, substantially reducing the peak load and improving balancedness. We can thus roughly categorize all layers into these two types: high-skew layers (e.g., layer 51) \textemdash~ replication-effective layers, where the maximum expert load significantly exceeds the average expert load ($>10\times$); and low-skew layers (e.g., layer 20) \textemdash~ replication-ineffective layers, where the maximum load is similar to the average load.

\textbf{Observation 2: As cluster size increases, overall load balancedness decreases and replication becomes more effective. }In Figures \ref{fig:bdkj6}, \ref{fig:bdka6}, \ref{fig:bdkj12}, and \ref{fig:bdka12}, increasing cluster size from 6 to 12 nodes decreases baseline balancedness and placement effectiveness across layers. This is because increasing the number of GPUs reduces the number of experts per GPU. This reduces opportunities for hot-cold expert co-location and weakens the averaging effect on each device, making the system more susceptible to skewed load distributions. Replication remains effective because the overall load distribution remains unchanged, and replicating hot experts still reduces peak load and improves balancedness. Moreover, Figures \ref{fig:bdka6}, \ref{fig:bdka12} show that setup \texttt{KA} is relatively balanced with strong baseline balancedness and placement effectiveness at 6 nodes. However, as cluster size increases to 12, baseline balancedness substantially degrades, where replication becomes more effective than placement. This underscores the importance of expert replication at scale.

\subsection{Replication Scaling}
\label{sec:repscale}

To mitigate performance degradation from replication memory overhead and identify the optimal balancedness-memory trade-off, we analyze how balancedness scales with replica count at fine-grained per-layer granularity. We study both aggregated balancedness across layers and per-layer balancedness scaling, and we make two observations.

\begin{figure}[!htb]
    \centering
    \includegraphics[width=1.0\linewidth]{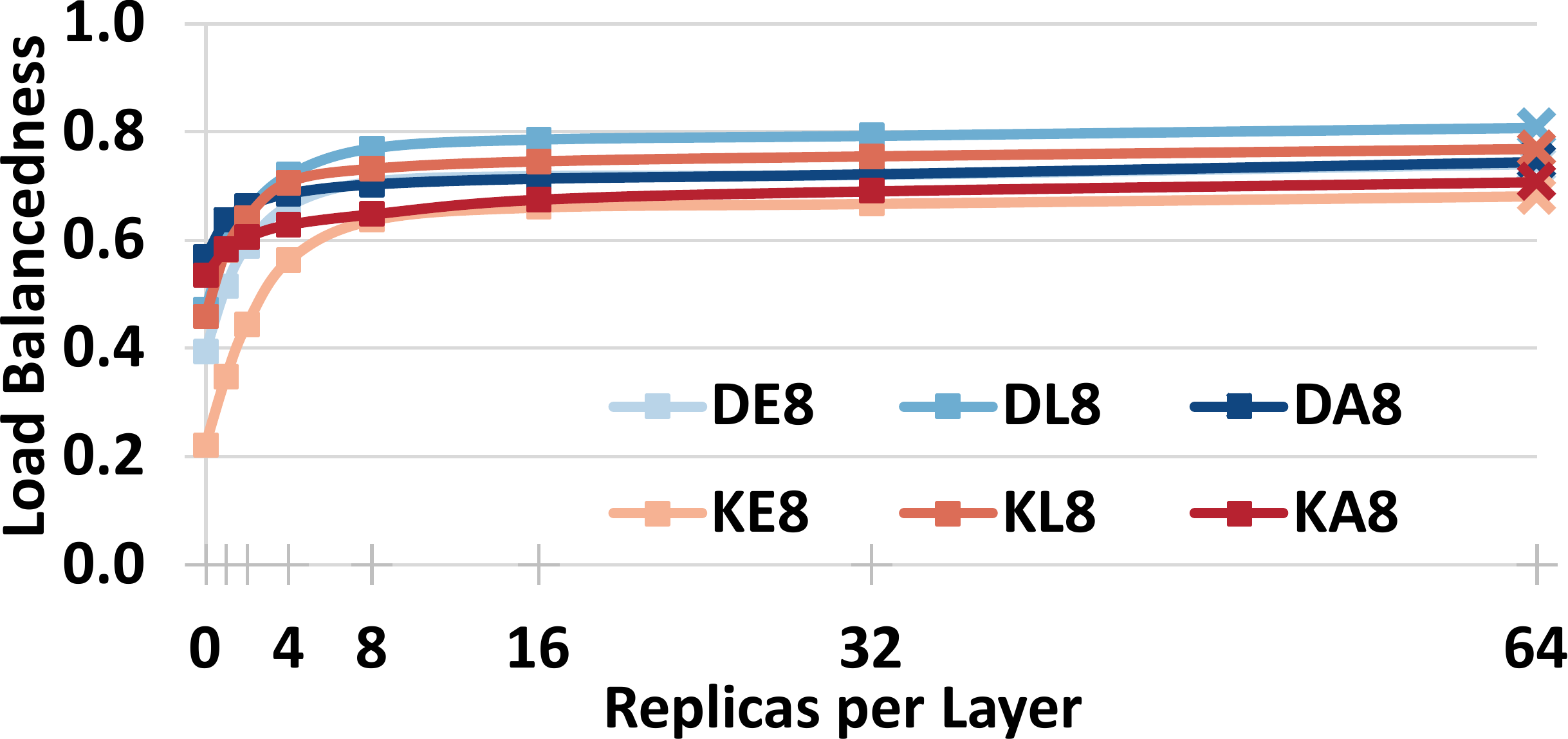}
    \caption{Load balancedness under varying per-layer replica counts across configurations. Balancedness is aggregated across all MoE layers. Each layer is allocated the same number of replicas; zero denotes the placement-only baseline. $\times$ indicates the minimum uniform replication (one replica per layer per GPU, \sref{sec:placerep}).}
    \label{fig:repbalscale}
\end{figure}

\textbf{Observation 3: Load balancedness scales sublinearly with the number of replicas. }\fref{fig:repbalscale} depicts the aggregated GPU balancedness. Across configurations, initial replication reduces peak expert load and improves load balance. After replicating hot experts sufficiently, placement alone balances the remaining load, and additional replicas yield little or no benefit. Doubling replicas beyond 16 offers negligible balancedness gains but doubles memory use. \cam{Existing \textit{uniform replication} limits the minimum number of replicas (\sref{sec:placerep}), where it allocates at minimum 63 additional replicas per layer in a 64-GPU system (one GPU holds the original expert copy). This indicates significant memory inefficiency, where 63 replicas are excessive as most replicas do not contribute to load balance.}

\begin{figure}[!htb]
    \centering
    \includegraphics[width=\linewidth]{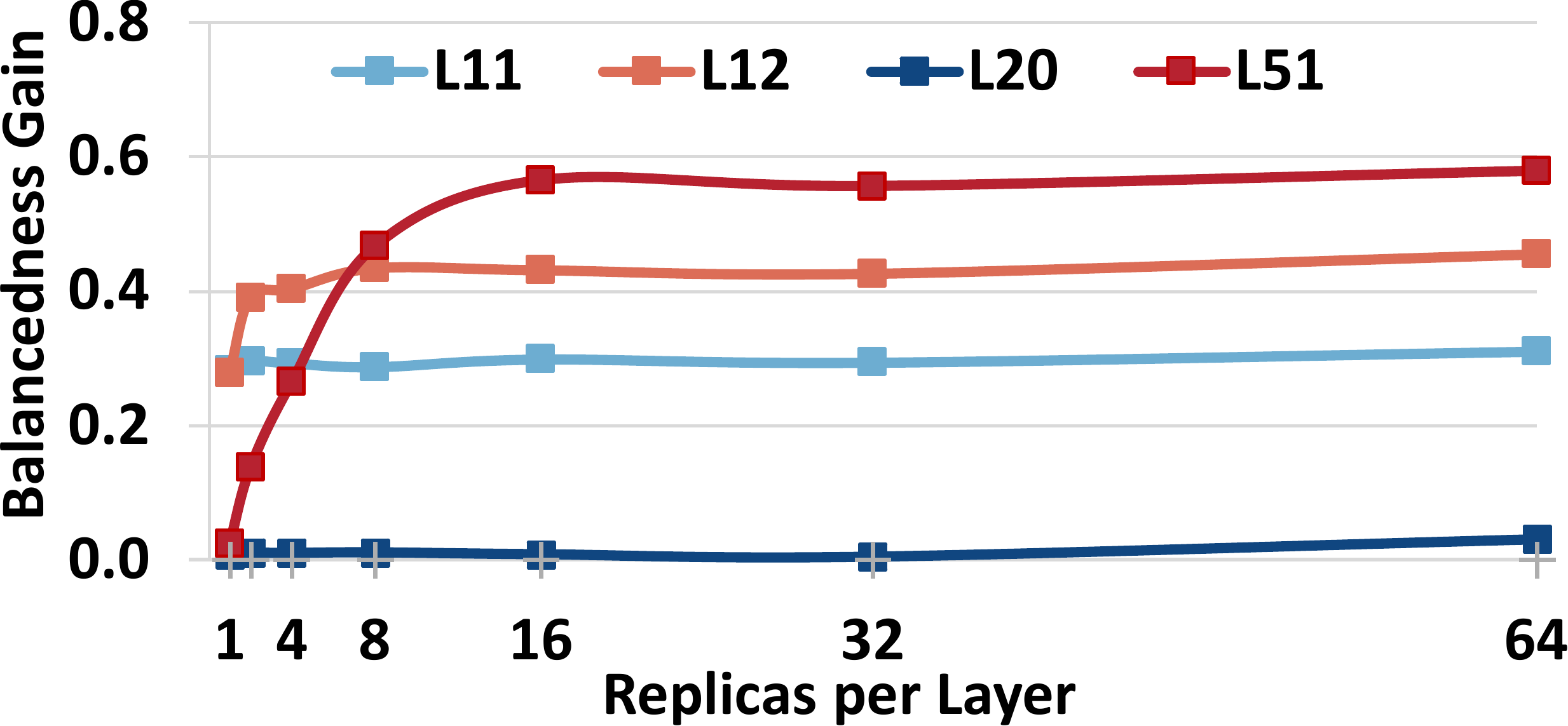}
    \caption{Per-layer load balancedness gain in 4 sample layers under varying replica counts with configuration \texttt{DE8}.} 
    \label{fig:laybal}
\end{figure}

\textbf{Observation 4: The effective replica count varies across layers. }\fref{fig:laybal} depicts balancedness gains at different replica counts for four layers: aforementioned balanced layer 20 and skewed layer 51 (\fref{fig:explddist}), and two layers 11 and 12 with moderate skewness. We observe that (1) the replica count at which gains plateau varies by layer, and (2) benefits diminish beyond 16 replicas across layers, confirming Observation 3. Layer 20 gains little from replication, consistent with its already-balanced distribution (Observation 1). For layers 11, 12, and 51, the replica counts at which balancedness gains diminish are 1, 8, and 16, respectively. Hence, replication should be allocated at fine, per-layer granularity to match layer-specific needs; a fixed replica count over-allocates balanced layers and under-allocates skewed layers, leading to wasted memory.

\section{Design}
\label{sec:design}
\begin{figure*}[!htb]
    \centering
    \includegraphics[width=1\linewidth]{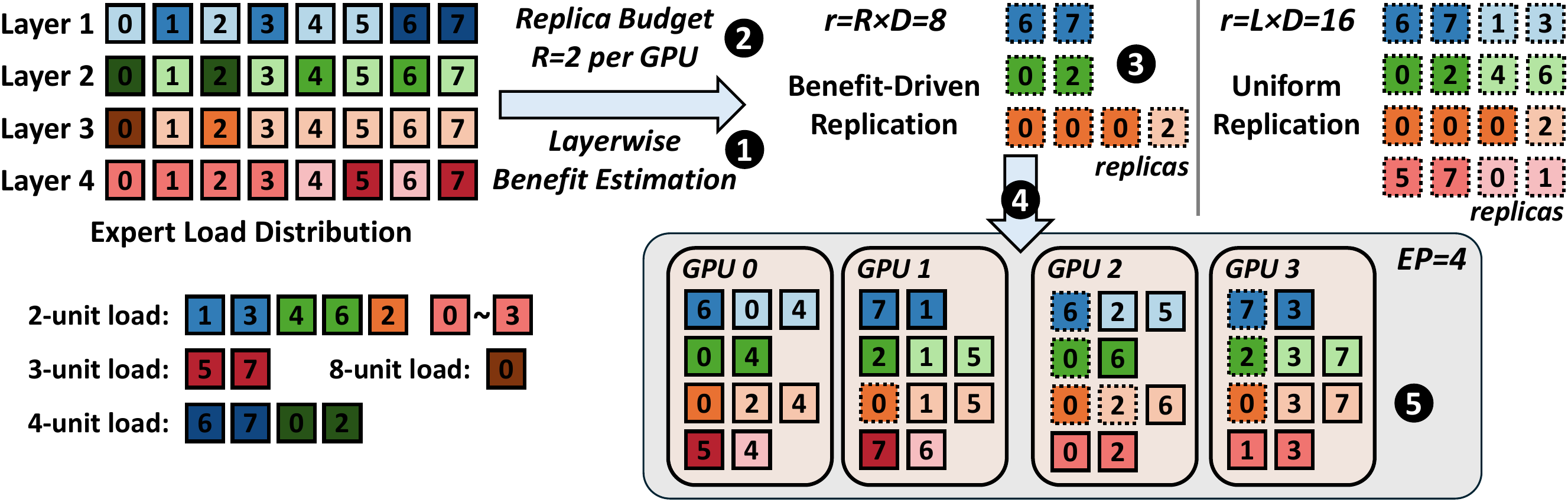}
    \caption{\proj workflow. We assume a system with 4 devices, 4 layers, and 8 experts per layer ($L=D=4$). Each layer has total load of 16 units. Darker color represents heavier load, and experts without load indication have 1 unit load. Dotted boxes present replicas. \proj achieves perfect load balance while using only half the number of replicas compared to uniform replication (top-right). }
    \label{fig:ufoflow}
\end{figure*}

We introduce \proj, an end-to-end replica allocation framework that enables replication under low memory budgets while preserving near-optimal load balance.
We define the balancedness gain from replication as \textit{replication benefit}, and the number of experts stored on each GPU as \textit{expert capacity}. 
The core idea of \proj is to independently allocate replicas at the granularity of a layer, in proportion to the layer-specific replication benefit.
This benefit-driven policy maximizes per-replica effectiveness and sustains high balancedness gains.

\subsection{Design Challenges of \proj}
\label{sec:designchallenge}

Implementing per-layer fine-grained replica allocation requires addressing the following challenges:

\textbf{Challenge 1: How do we determine the optimal number of replicas to allocate to each layer to maximize the overall balancedness gain? }As discussed in \sref{sec:repscale}, balancedness scales differently with replica count across layers. Selecting each layer's locally optimal replica count is insufficient, because layers benefit from replication differently, and the total number of replicas is constrained by the replication memory budget.

\textbf{Challenge 2: With varying replica counts per layer, how do we avoid memory imbalance when assigning replicas to devices? }The uniform replication baseline allocates one replica per layer per device, so assignment is simple and memory usage is trivially balanced across devices. Under benefit-driven replication, layers have fewer replicas than devices, which challenges replica-to-device assignment for two reasons: (1) The expert capacity across GPUs for each layer may differ. This complicates expert placement within each layer and introduces additional imbalance, because each GPU may hold a different number of experts. (2) Replica memory allocation must remain consistent across devices to maintain a uniform KV cache size. Inconsistent KV cache sizes lead to memory fragmentation, where the maximum concurrency is determined by the device with the lowest KV cache size within a data-parallel (DP) rank.

\subsection{\proj Workflow}

\fref{fig:ufoflow} illustrates the high-level \proj workflow on an example with four MoE layers ($L=4$) and four GPUs ($D=4$). The replication factor $R$ denotes the number of replicas per GPU, quantifying the replication memory usage. The total number of replicas $r$ is calculated as $r=R\times D$. \proj allocates replicas in three steps. First, we collect the balancedness scaling and estimate per-layer replication benefit by analyzing the expert load distribution \circled{1}. Second, the value of $R$ is determined either manually by the user or automatically according to the balancedness scaling that optimizes for high replication efficiency \circled{2}. The example in \fref{fig:ufoflow} uses $R=2$ ($r=8$). Lastly, the replica allocation algorithm computes an allocation plan of $r$ replicas that maximizes balancedness \circled{3}. As replica counts vary across layers, replica-to-device mapping becomes challenging (Challenge 2). \proj employs an interleaved expert assignment to minimize the intra-layer expert capacity imbalance across devices \circled{4}. Then, a capacity-aware greedy placement algorithm assigns experts to devices \circled{5}. Compared to uniform replication, which allocates $r=L\times D=16$ replicas uniformly across layers and devices, \proj achieves load balance with half the memory budget. We present the design details of \proj in the following sections.

\subsubsection{Benefit-Driven Replica Allocation}
\label{sec:repalloc}

In this section, we formulate an optimization problem for benefit-driven replica allocation to address Challenge 1 (\sref{sec:designchallenge}). The goal is to maximize the balancedness gain (i.e., replication benefit) under the $r$-replica budget. The allocation has three steps:

\textbf{Step 1: Estimate per-layer replication benefit }\circled{1}\textbf{.} For each layer, we evaluate the replication benefit of $K=\operatorname{log}_2D+1$ different replica counts chosen from a base-2 geometric progression over $[1,D]$. We replay each inference batch from the expert load distribution and measure the balancedness gain at each chosen replica count (details in \aref{sec:odalgo1}). This replay-based analysis yields an $L\times K$ balancedness gain matrix $T$ that quantifies per-layer replication benefits across replica counts.

\textbf{Step 2: Determine replication factor $R$ }\circled{2}\textbf{.} \proj allows the user to manually set $R$ that meets specific memory constraints. Alternatively, \proj supports the automatic selection of an $R$ that optimizes for replication memory efficiency by analyzing the balancedness gains obtained in Step 1, as illustrated in \fref{fig:repbalscale}. Specifically, \proj selects the $R$ that yields the highest per-replica balancedness gain, which avoids allocating replicas with diminishing benefits (\sref{sec:repscale}). We demonstrate the effectiveness of this method by showing a strong correlation between diminished balancedness gains and end-to-end throughput degradation in \sref{sec:excrep}.

\textbf{Step 3: Compute an $r$-replica allocation plan with maximum benefit }\circled{3}\textbf{.} We transform the allocation problem into a Multiple-Choice Knapsack Problem (MCKP) \textemdash~ selecting one replica count per layer under the $r$-replica constraint that maximizes total replication benefit in $T$. Although MCKP is NP-hard, the parameters $D$, $K$, and $L$ remain small even at large-scale deployment, enabling an efficient dynamic programming solution in pseudo-polynomial time \cite{mckp}. We provide a detailed implementation in \aref{sec:odalgo2}.

\fref{fig:ufoflow} shows an example optimal replication plan from benefit-driven replica allocation. Layer 3 is the most skewed and receives 4 replicas (highest benefit). Layers 1 and 2 are moderately skewed and receive 2 replicas each (moderate benefit). Layer 4 is already balanced under expert placement and receives no replicas (zero benefit). This allocation achieves optimal balancedness using far fewer replicas than uniform replication.

\subsubsection{Capacity-Aware Expert Assignment and Placement}
\label{sec:interleaveassign}

\proj addresses Challenge 2 with capacity-aware expert assignment, which uses a greedy algorithm that iteratively assigns experts for each layer under two objectives:

\textbf{Primary Objective: Each additional expert (replica) is always assigned to one of the GPUs with the fewest experts.} Since $r$ is a multiple of $D$ ($r=R\times D$, $R$ is the number of replicas per GPU, and $D$ is the number of GPUs), replication memory usage remains consistent across devices. This constraint does not cause excessive replication, because $D$ and $L$ are of the same order in practice; the minimum setting $R=1,\,r=D$ incurs marginal memory overhead under benefit-driven replication (\sref{sec:repalloc}). This constraint also bounds per-layer, device-level capacity differences to at most one, reducing expert capacity skew.

\textbf{Secondary Objective: Within each layer, additional experts are assigned in an interleaved manner across devices and nodes. }While satisfying the primary objective, many GPUs may tie for the fewest experts. As a tie-breaking measure, we evenly distribute experts among nodes. For example, in \fref{fig:ufoflow}, the two additional experts in layer 1 are assigned to GPU 0 (node 0) and GPU 2 (node 1); those in layer 2 are assigned to GPU 1 (node 0) and GPU 3 (node 1). This interleaved assignment balances per-layer node-level expert capacity and further improves overall load balance.

We provide the detailed implementation of the greedy assignment algorithm in \aref{sec:odalgo3}. With these two objectives, capacity-aware expert assignment yields an optimized GPU memory allocation plan that minimizes expert capacity imbalance \circled{4}. We then apply a greedy placement algorithm that iteratively assigns the most-loaded expert to the least-loaded device, following the standard strategy \cite{dpsv3} while respecting the varying expert capacity across devices \circled{5}. After placement, \proj produces a complete end-to-end replication plan.

\subsection{\cam{Adapting to Shifting Workload Distributions}}
\label{sec:workloadshift}

\cam{Changes in the workload distribution during runtime may necessitate online adaptation in expert replication plans. Thus, mainstream LLM serving frameworks \cite{sglang,vllm,tensorrt-llm} and large-scale MoE deployment practices \cite{nvl72,largevllm,sglangdps,moreh} adopt \textit{periodic expert rebalancing}, initially proposed by EPLB, to adapt to shifting workloads. Current online periodic rebalancing mechanisms comprise three steps: (1) the framework samples and records the expert load distribution post-routing; (2) at each rebalancing window (configurable, with a 10-minute default in EPLB), the framework triggers rebalancing and asynchronously sends the recorded distribution to EPLB to compute a new expert-device assignment map; (3) the framework relocates expert weights across devices based on the generated assignment.
These steps can be performed asynchronously by pinning expert weights in CPU memory and streaming the weight updates to GPU across multiple iterations, overlapped with ongoing inference batches. Thus, such weight updates can be performed efficiently with negligible impact on user experience and service-level objectives \cite{rtreloc1,rtreloc2,sglangdps}.

\proj can serve as a drop-in replacement for EPLB in existing serving frameworks that support online periodic expert rebalancing, substituting EPLB in step 2. When workload shifts occur, \proj can adopt periodic rebalancing to improve load balancedness over time. In \sref{sec:shifteval}, we analyze various workload shift and load distribution patterns, identifying scenarios where rebalancing is beneficial and where it is unnecessary.

}

\section{Evaluation}
\label{sec:evaluation}

In this section, we address the following questions:

\textbf{Question \circled{1}: }How does the balancedness-memory trade-off from replication affect end-to-end performance?

\textbf{Question \circled{2}: }What end-to-end performance improvement does \proj deliver compared to existing approaches? 

\textbf{Question \circled{3}: }How does inference performance scale with the cluster size under \proj?

\textbf{Question \circled{4}: }Does \proj introduce any additional overhead in the serving framework?

\begin{figure*}[!htb]
    \centering

    \subfigure[\texttt{KE6}]{
        \includegraphics[width=0.329\textwidth]{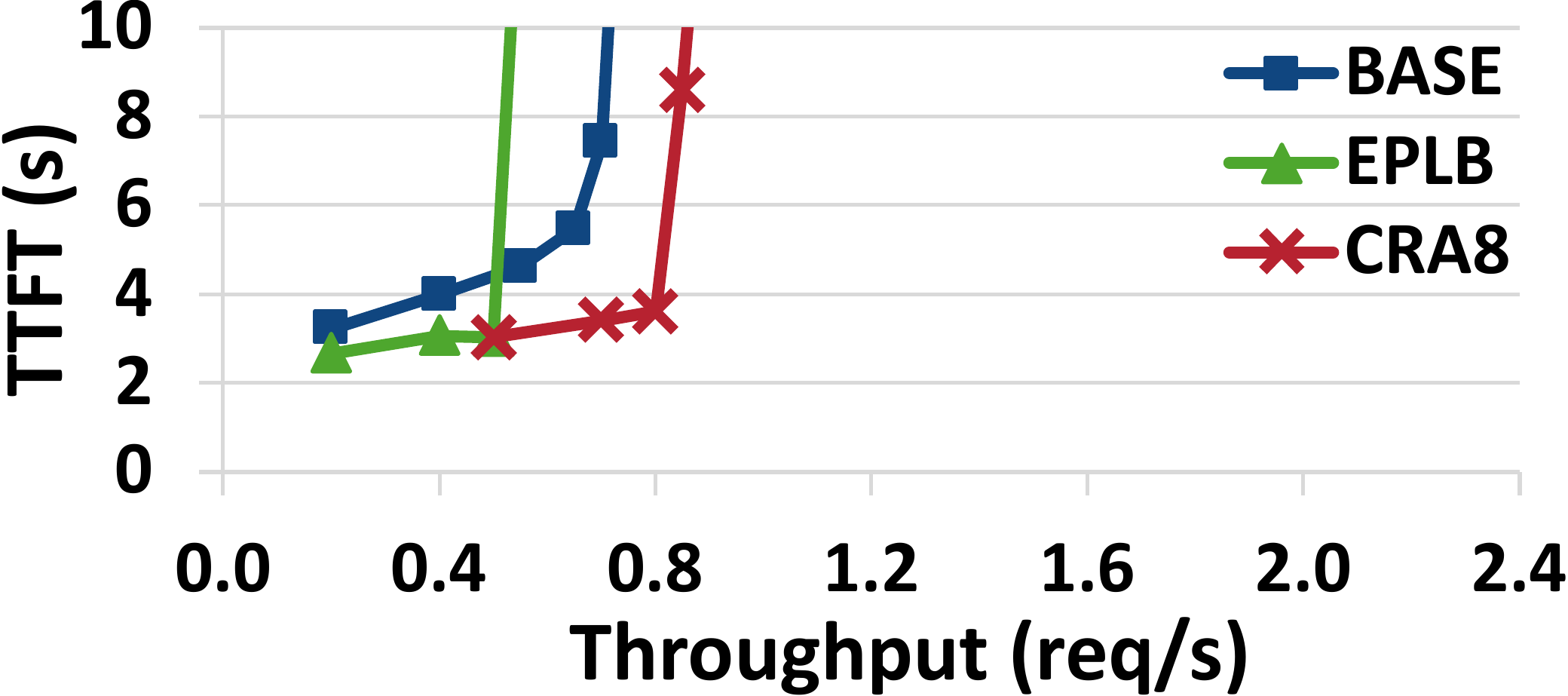}
        \label{fig:ke6}
    } 
    \subfigure[\texttt{KE8}]{
        \includegraphics[width=0.314\textwidth]{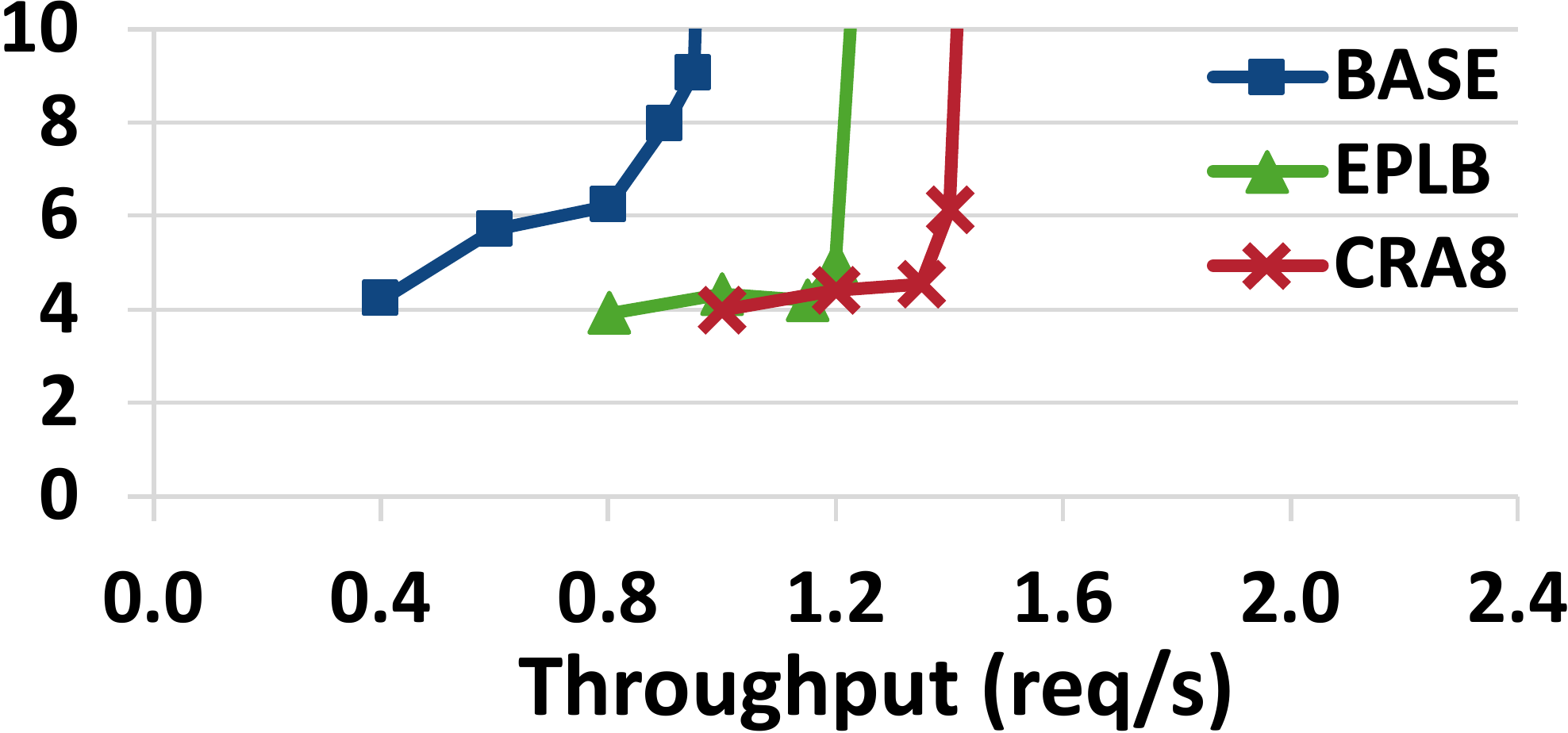}
        \label{fig:ke8}
    } 
    \subfigure[\texttt{KE12}]{
        \includegraphics[width=0.314\textwidth]{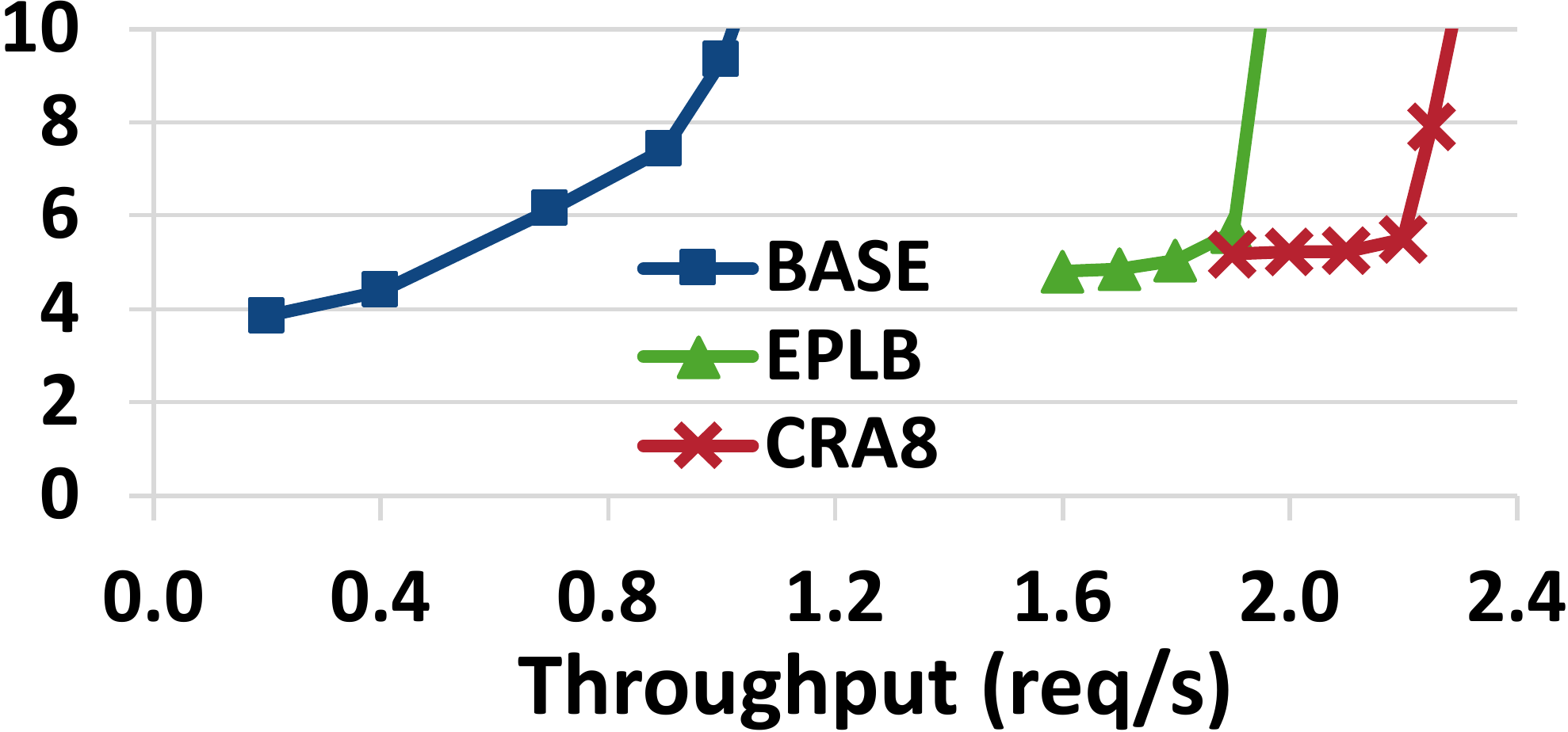}
        \label{fig:ke12}
    } 
    \subfigure[\texttt{KJ6}]{
        \includegraphics[width=0.329\textwidth]{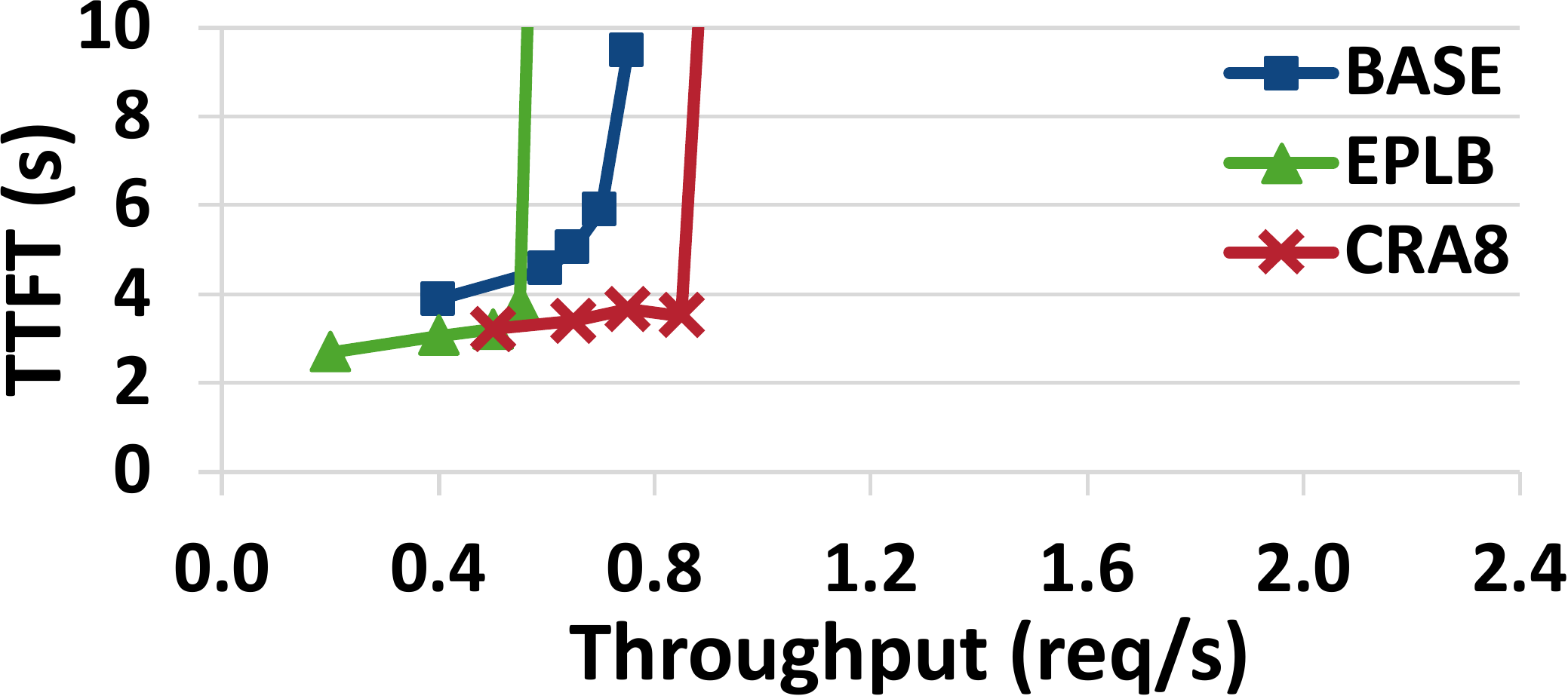}
        \label{fig:kj6}
    } 
    \subfigure[\texttt{KJ8}]{
        \includegraphics[width=0.314\textwidth]{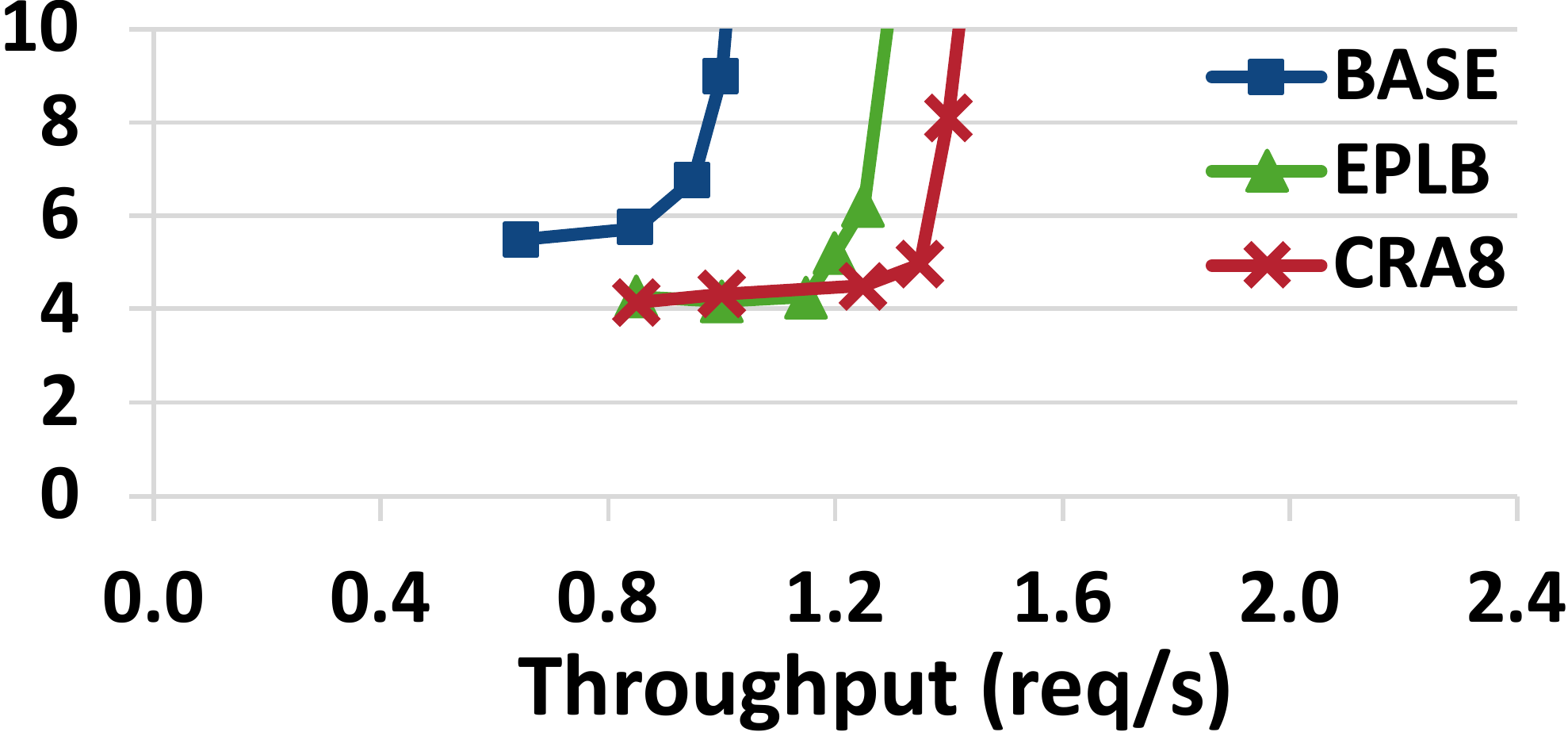}
        \label{fig:kj8}
    } 
    \subfigure[\texttt{KJ12}]{
        \includegraphics[width=0.314\textwidth]{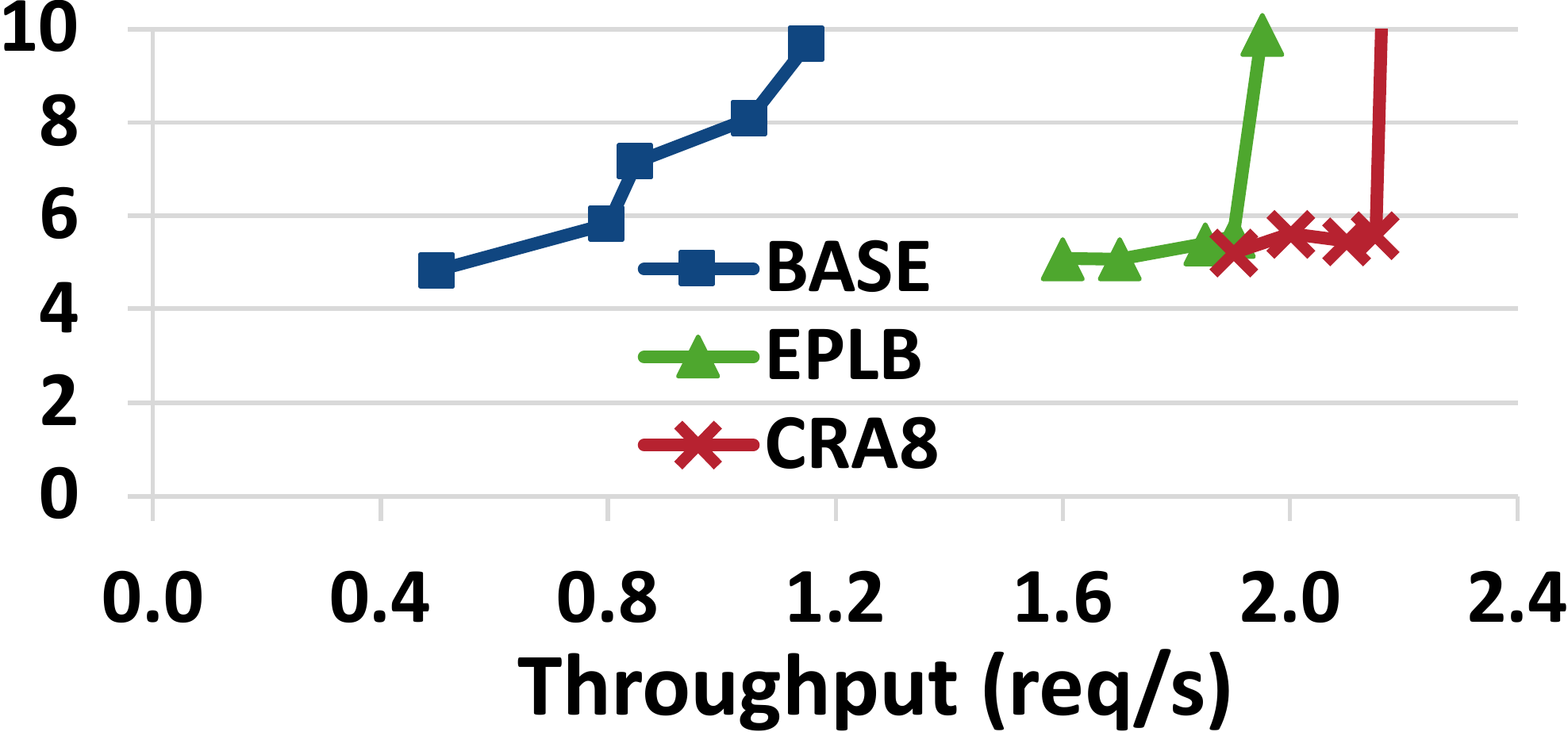}
        \label{fig:kj12}
    } 
    \subfigure[\texttt{KL6}]{
        \includegraphics[width=0.329\textwidth]{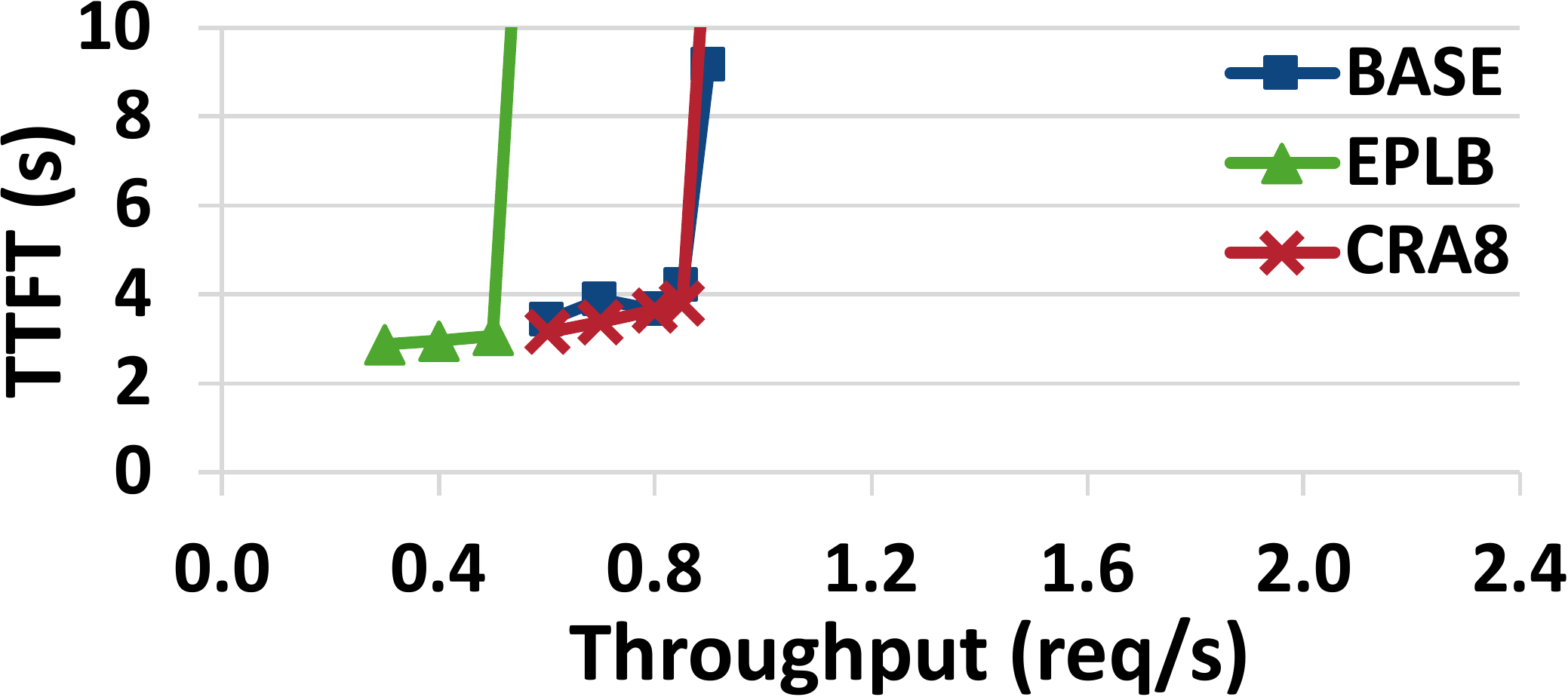}
        \label{fig:kl6}
    } 
    \subfigure[\texttt{KL8}]{
        \includegraphics[width=0.314\textwidth]{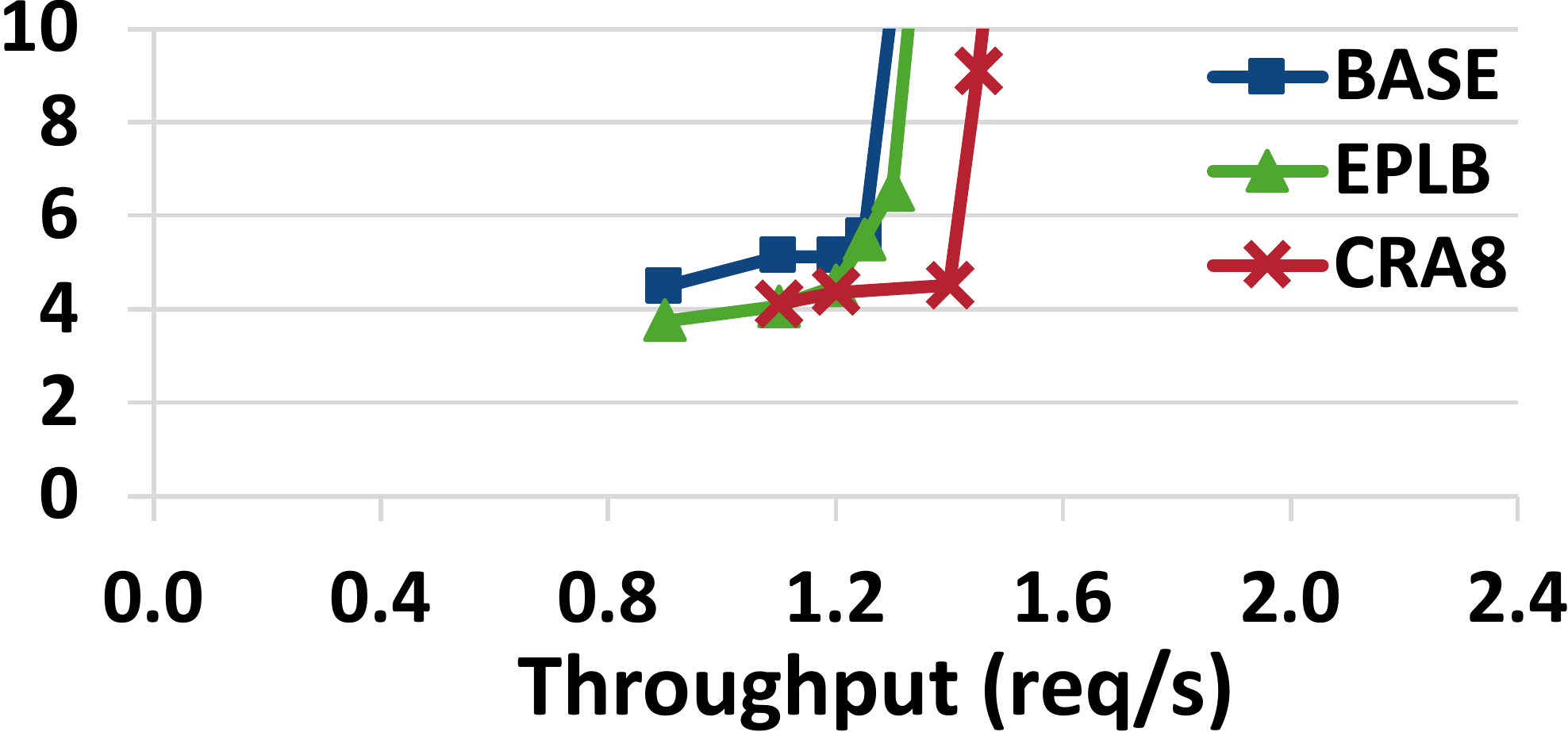}
        \label{fig:kl8}
    } 
    \subfigure[\texttt{KL12}]{
        \includegraphics[width=0.314\textwidth]{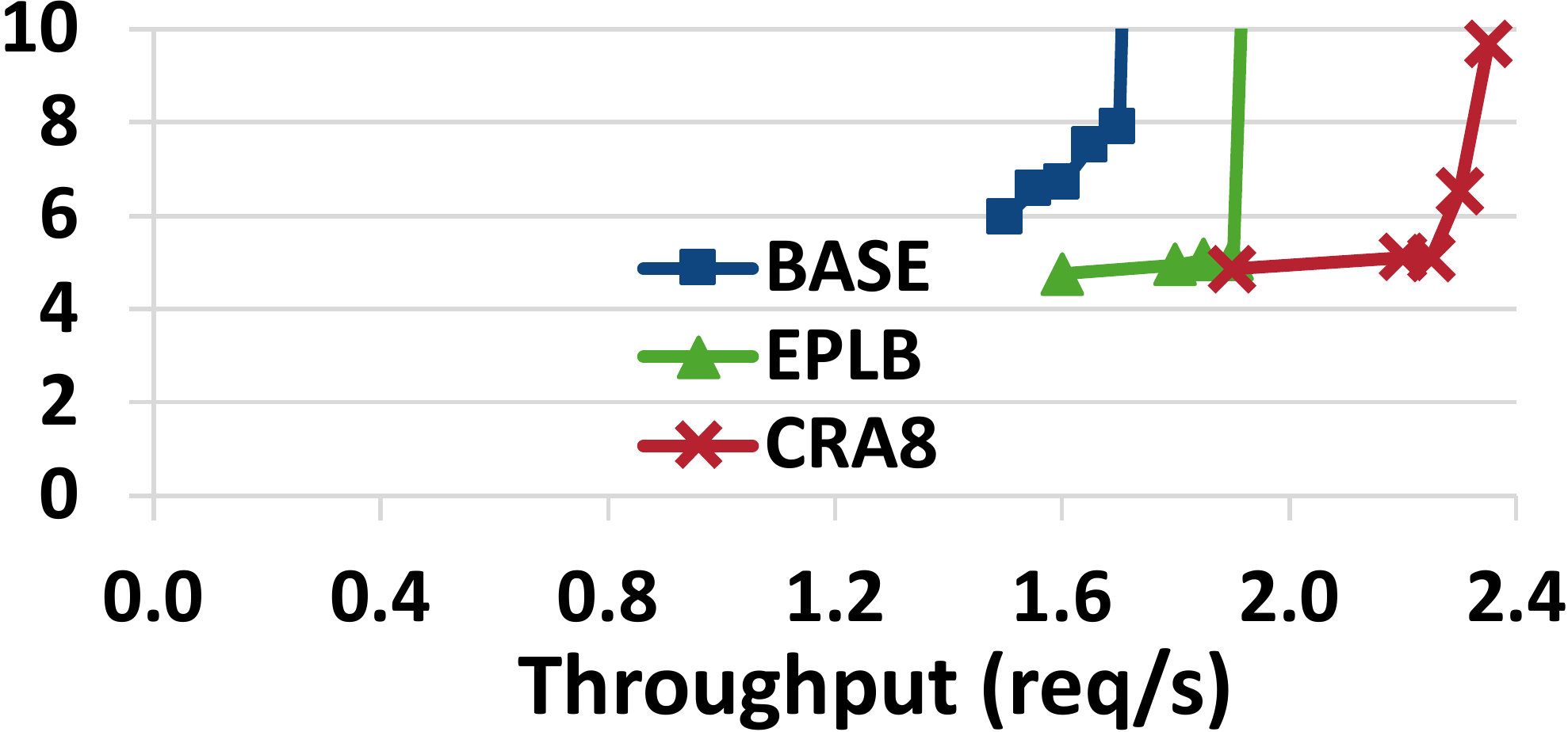}
        \label{fig:kl12}
    } 
    
    \subfigure[\texttt{KA8}]{
        \includegraphics[width=0.329\textwidth]{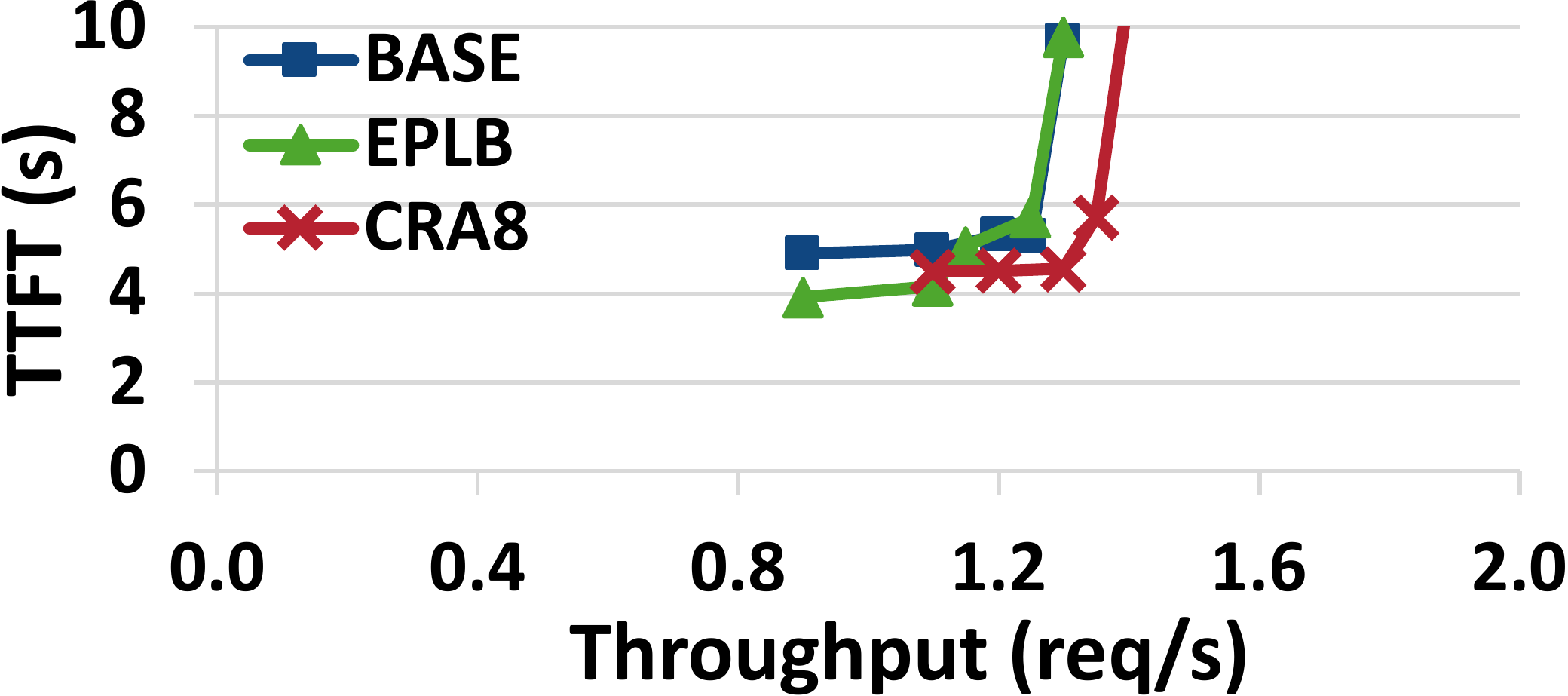}
        \label{fig:ka}
    } 
    \subfigure[\texttt{DE8}]{
        \includegraphics[width=0.314\textwidth]{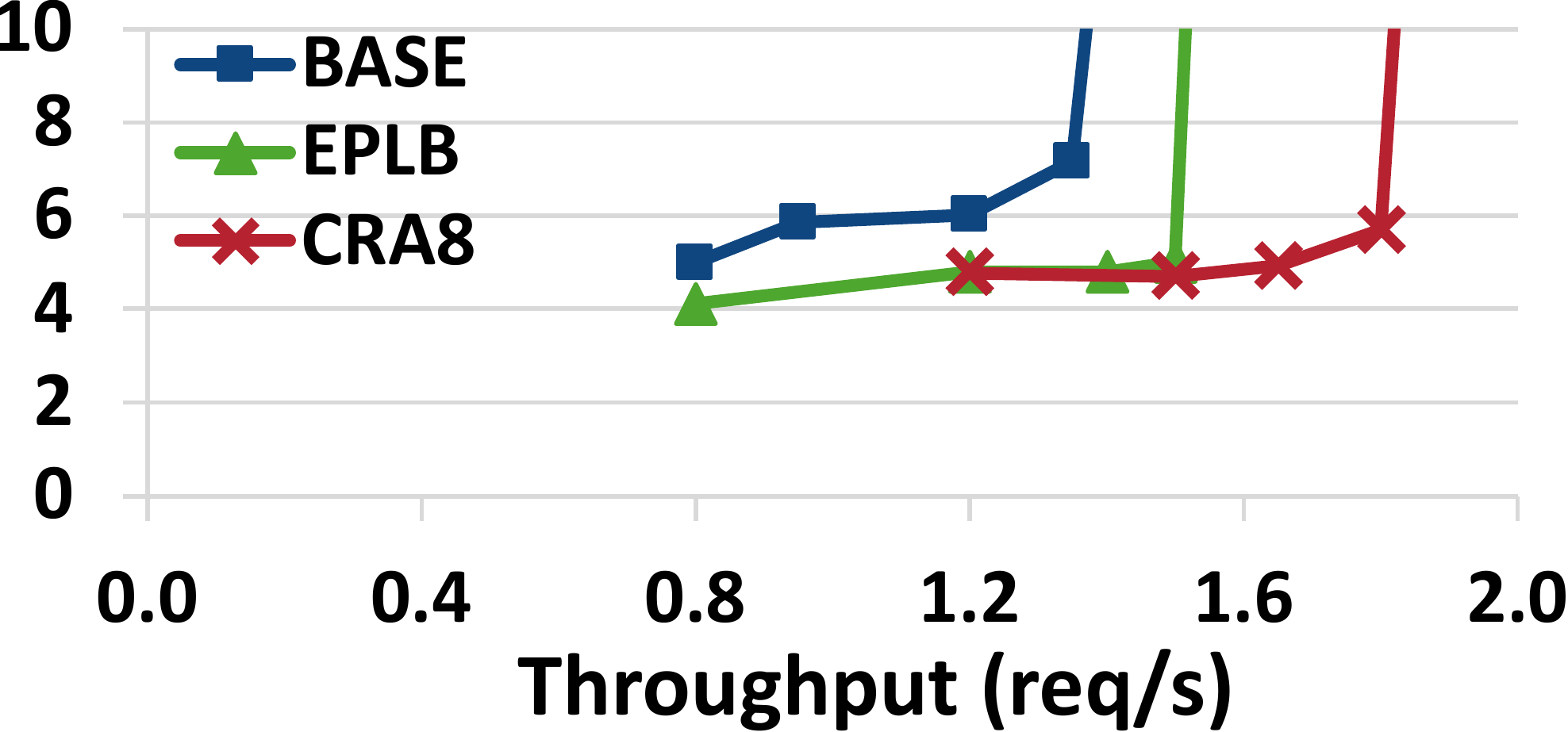}
        \label{fig:de}
    } 
    \subfigure[\texttt{DJ8}]{
        \includegraphics[width=0.314\textwidth]{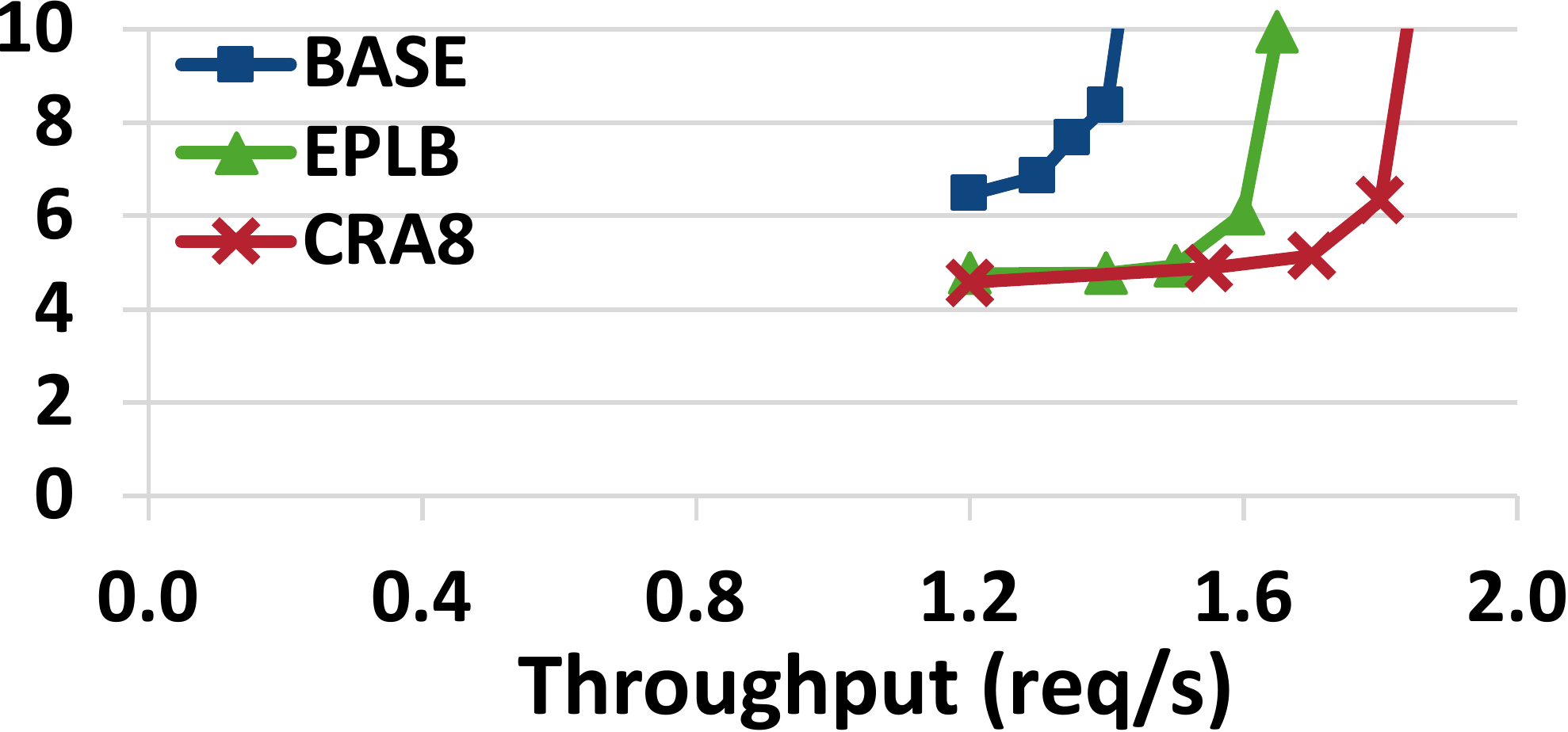}
        \label{fig:dj}
    } 
    \subfigure[\texttt{DL8}]{
        \includegraphics[width=0.329\textwidth]{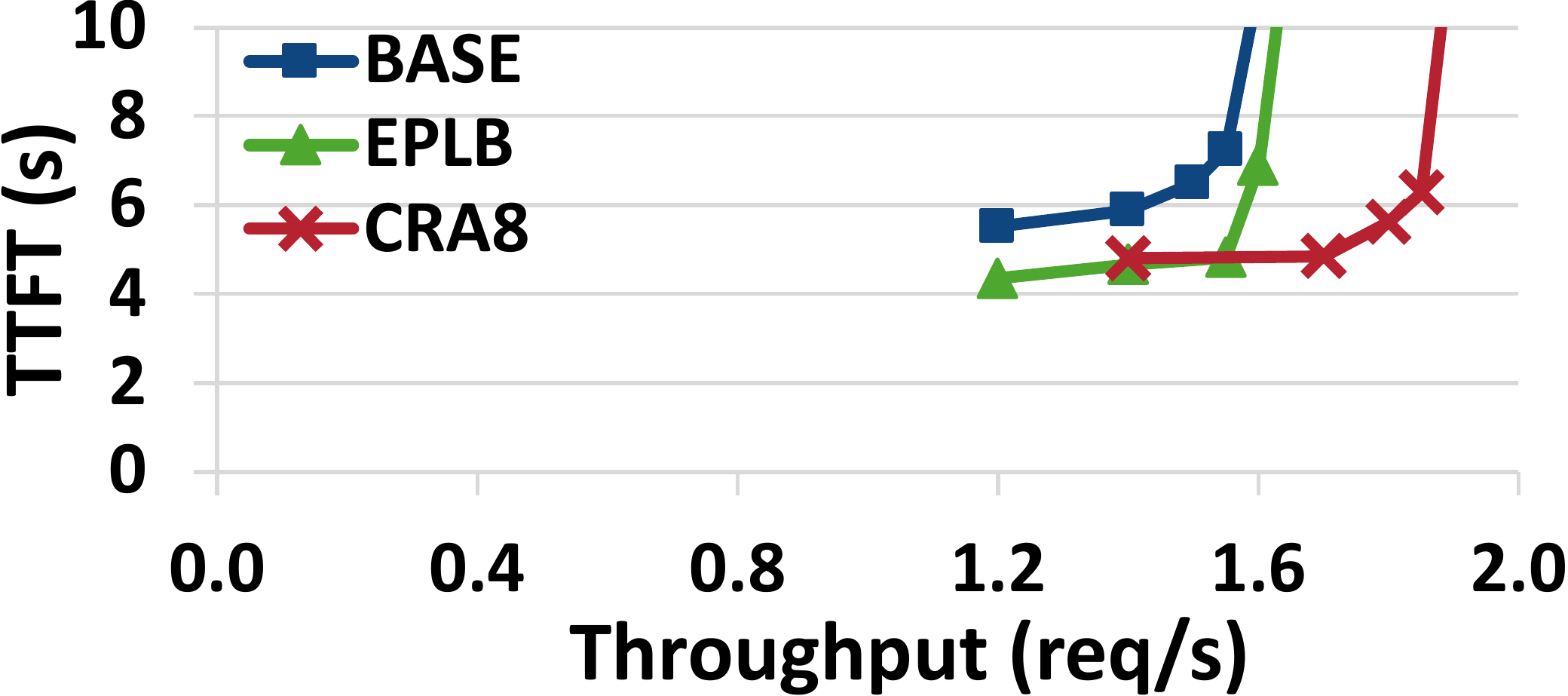}
        \label{fig:dl}
    } 
    \subfigure[\texttt{DA8}]{
        \includegraphics[width=0.314\textwidth]{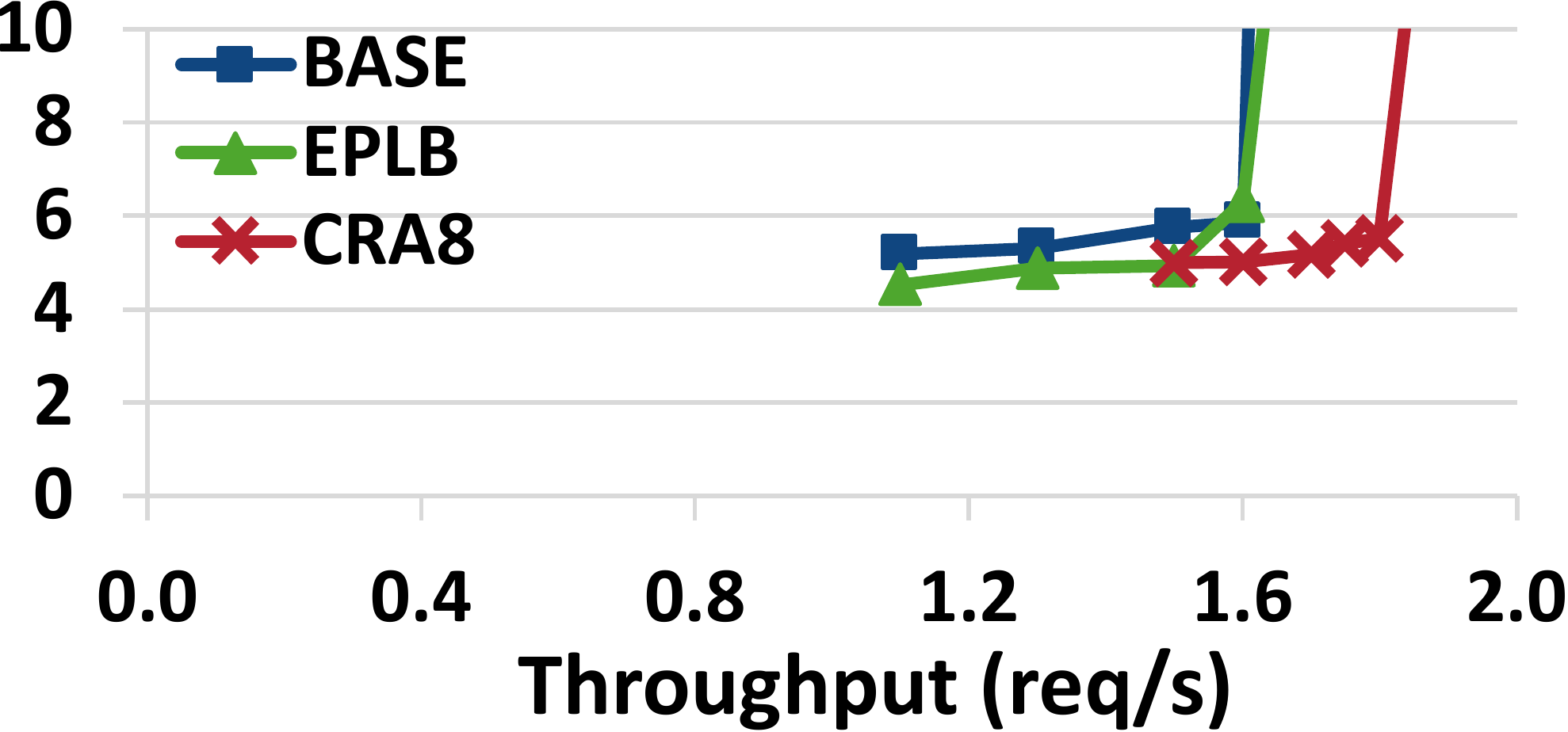}
        \label{fig:da}
    } 
    \caption{End-to-end throughput to TTFT curves across different system setups for \texttt{BASE}, \texttt{EPLB} and \texttt{\proje8}.}
    \label{fig:e2e}
\end{figure*}

\begin{figure*}[!htb]
    \centering
    \subfigure[\texttt{DE8}]{
        \includegraphics[width=0.343\textwidth]{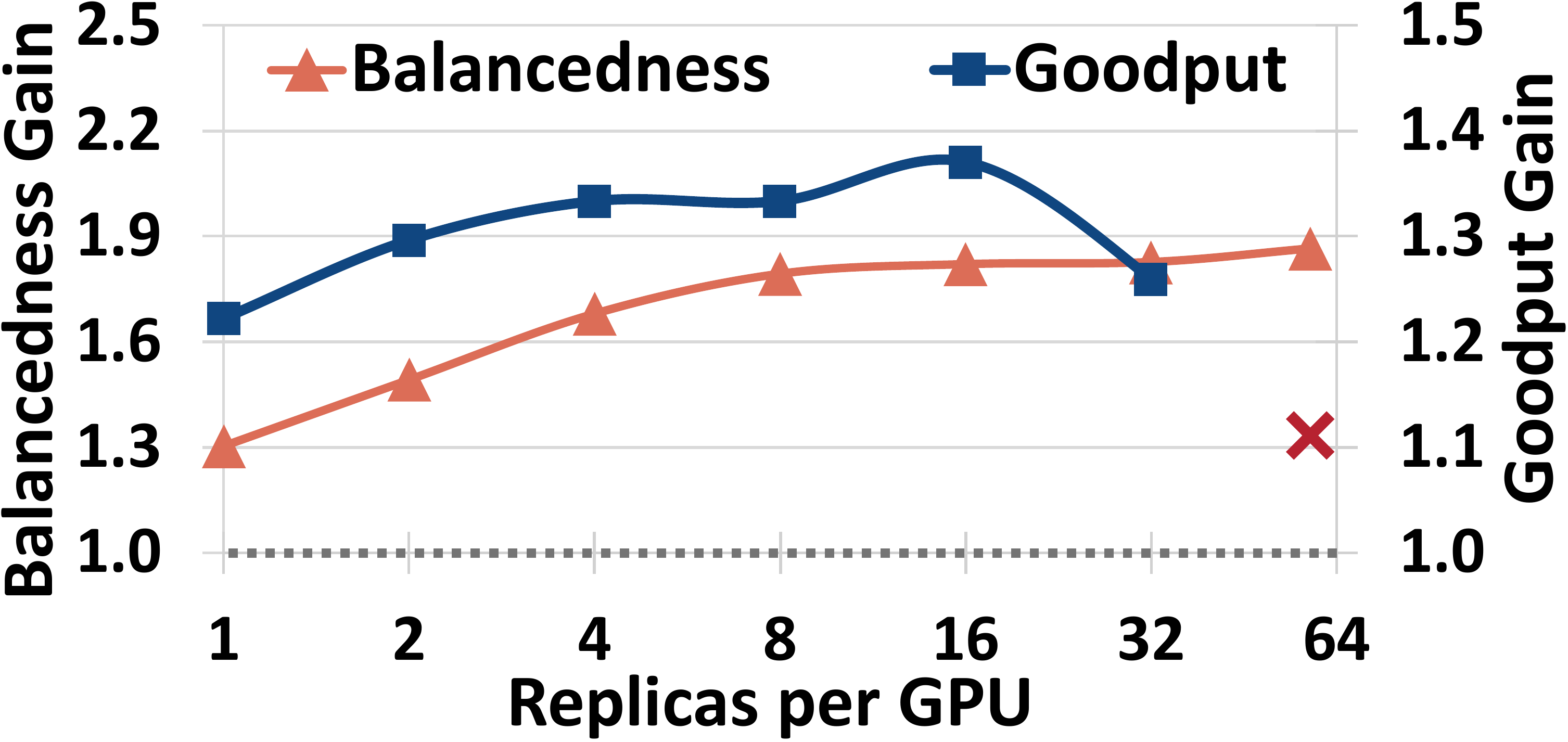}
        \label{fig:btde8}
    } 
    \subfigure[\texttt{KE6}]{
        \includegraphics[width=0.307\textwidth]{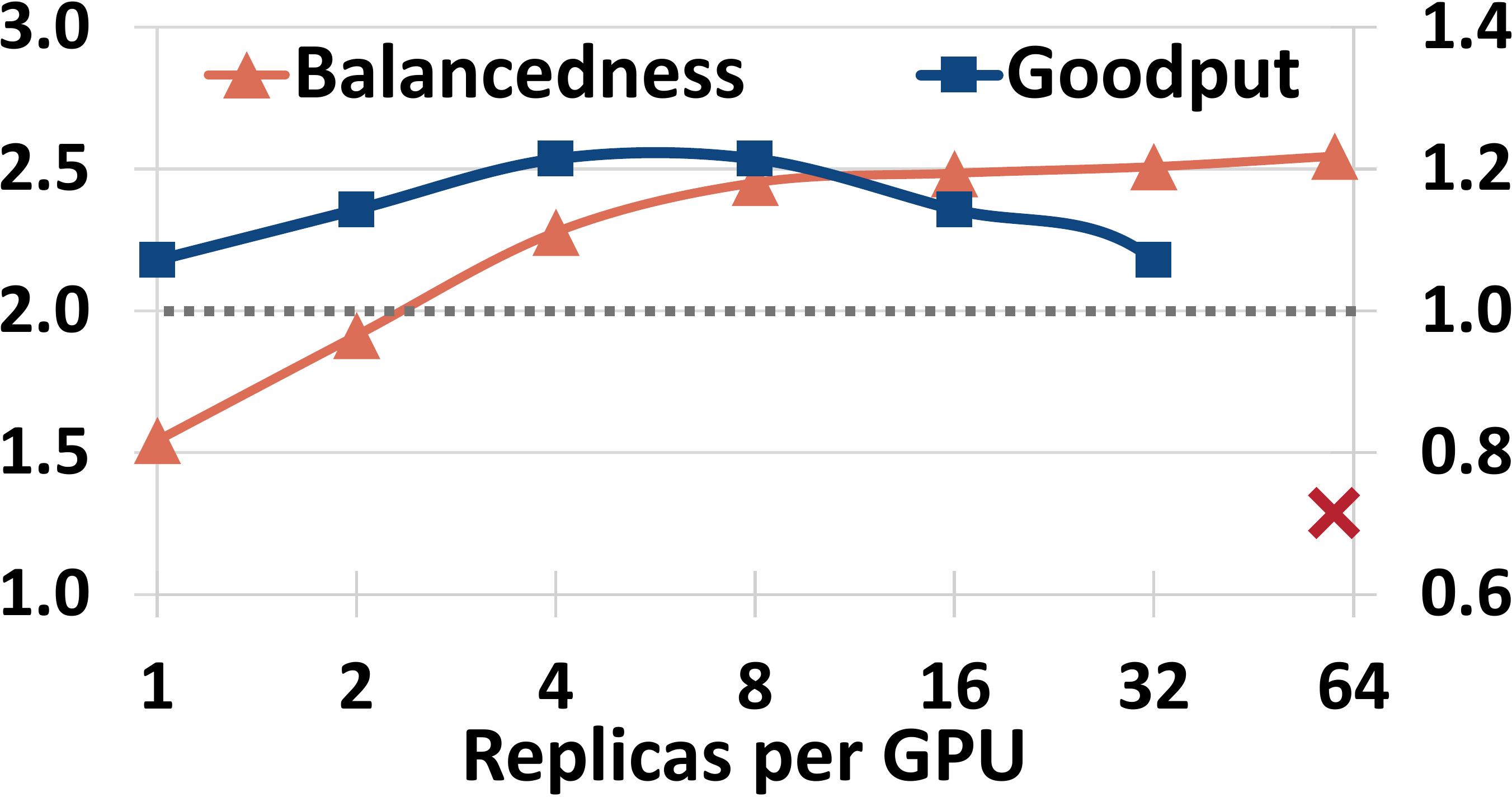}
        \label{fig:btke6}
    } 
    \subfigure[\texttt{KJ6}]{
        \includegraphics[width=0.307\textwidth]{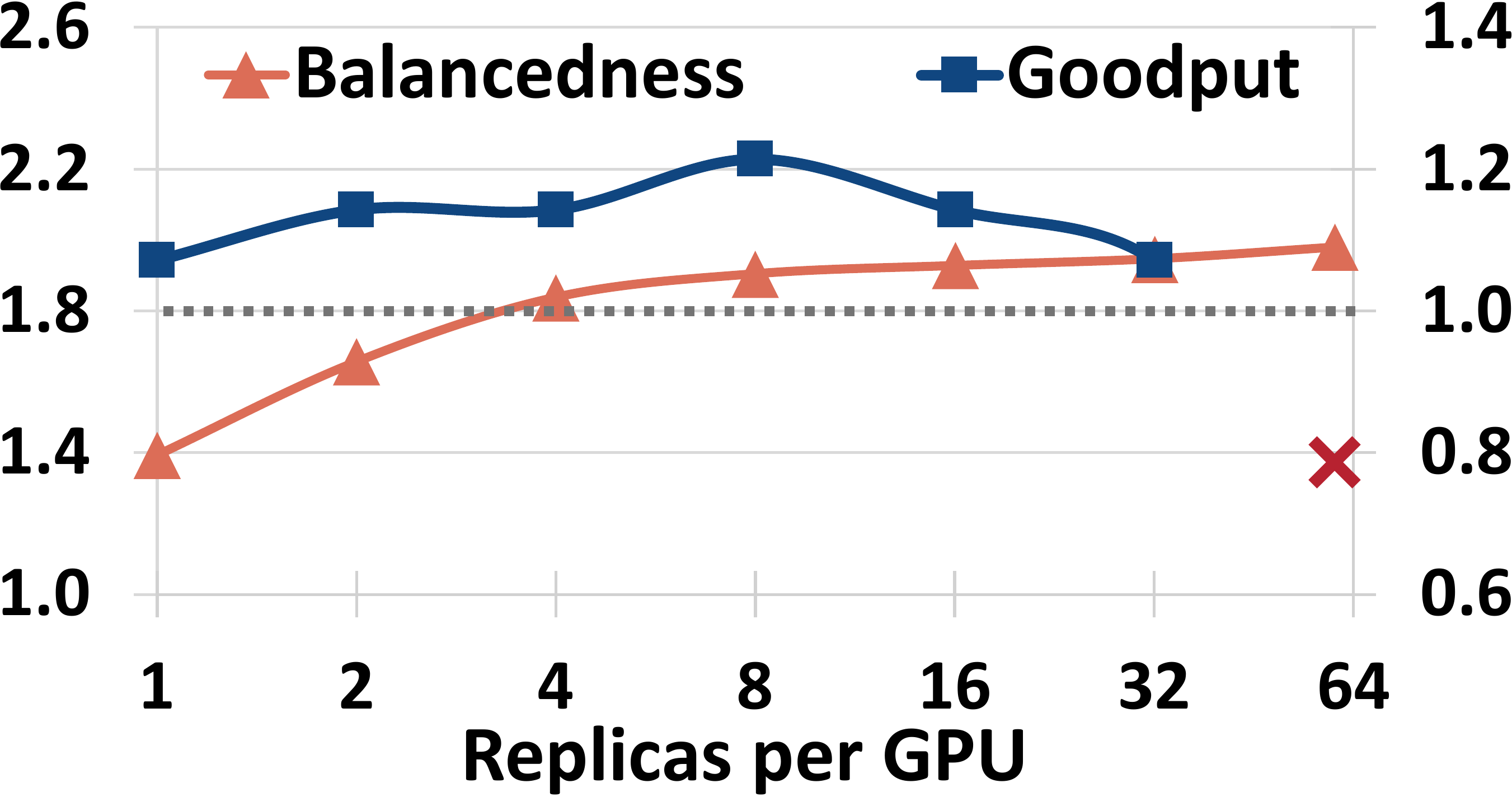}
        \label{fig:btkj6}
    } 
    \subfigure[\texttt{DL8}]{
        \includegraphics[width=0.343\textwidth]{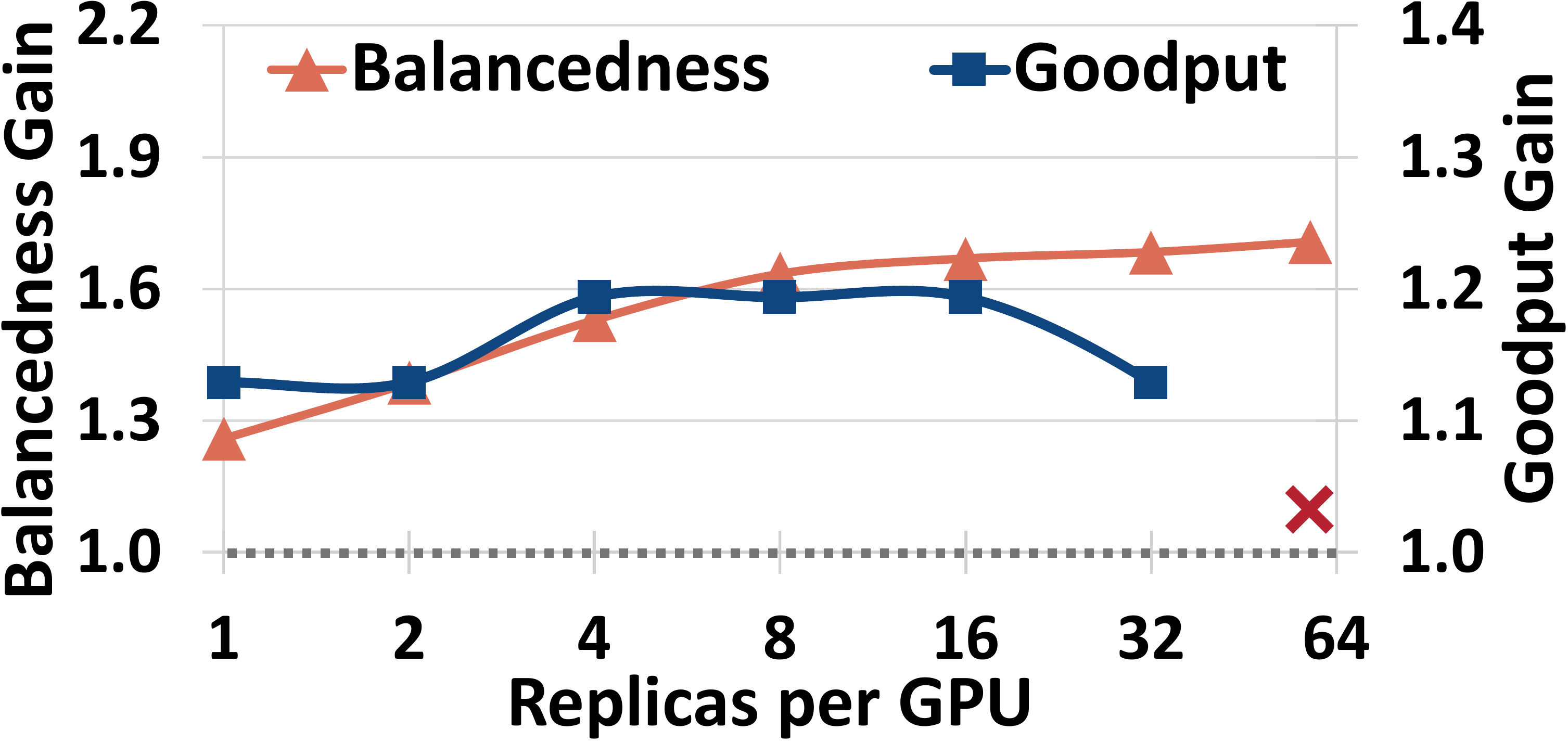}
        \label{fig:btdl8}
    } 
    \subfigure[\texttt{KE12}]{
        \includegraphics[width=0.307\textwidth]{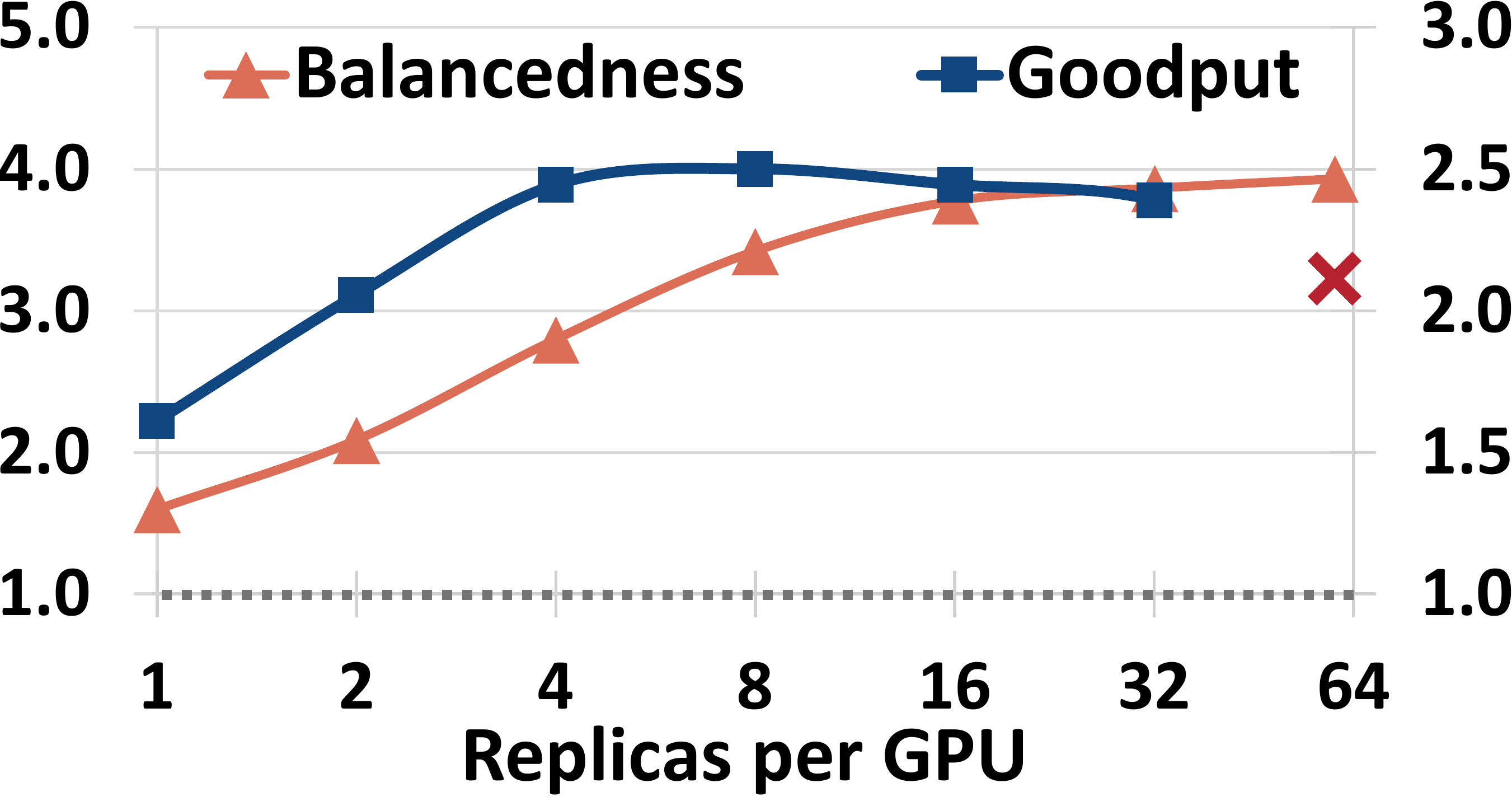}
        \label{fig:btke12}
    } 
    \subfigure[\texttt{KJ12}]{
        \includegraphics[width=0.307\textwidth]{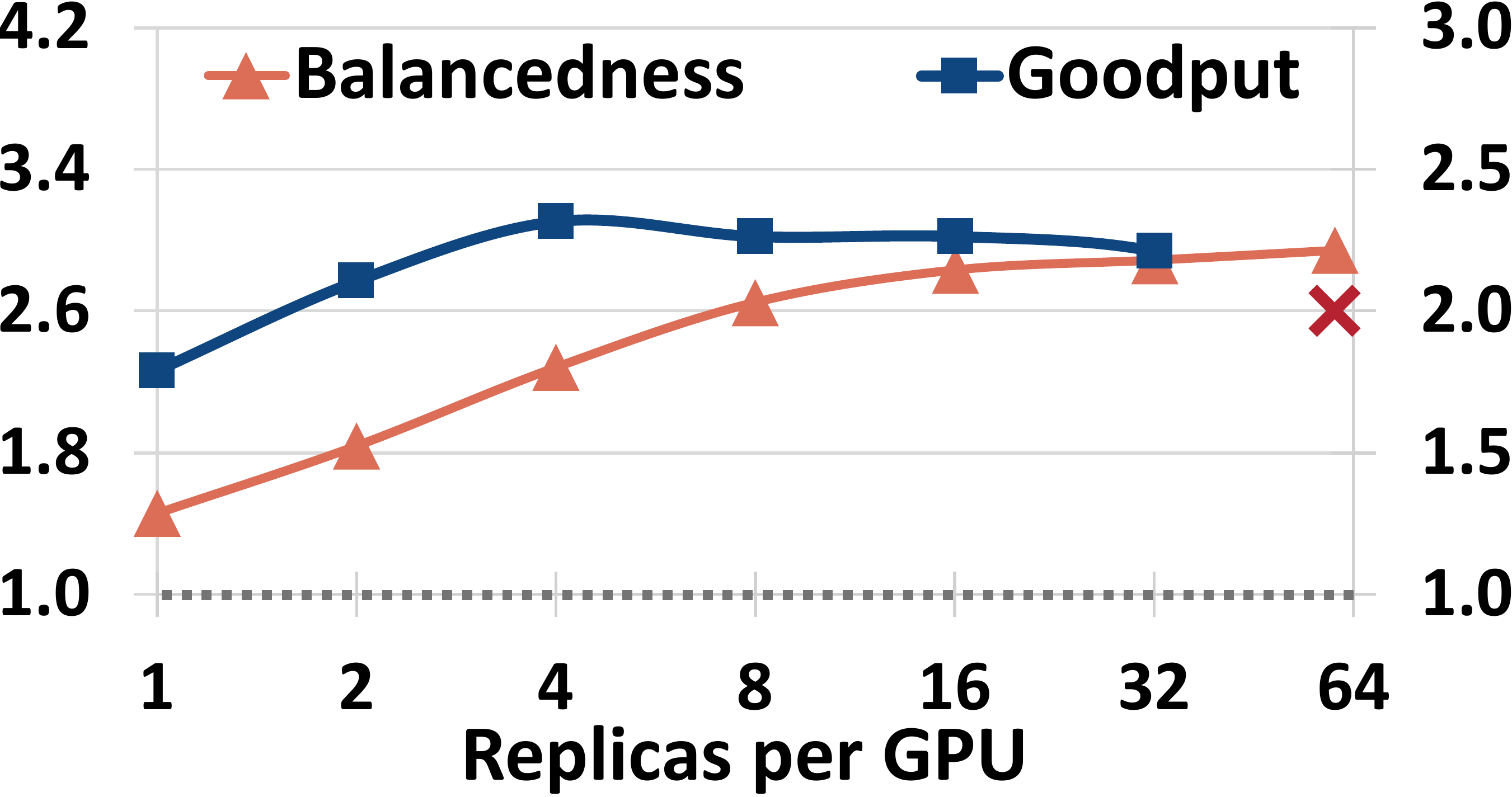}
        \label{fig:btkj12}
    } 
    \caption{\texttt{\proje} end-to-end goodput gain (right, blue/red) and balancedness gain (left, orange) with different numbers of replicas allocated per GPU (replication factor $R$), normalized to \texttt{BASE}. Red $\times$ represents \texttt{EPLB} goodput (one replica per MoE layer per GPU). Goodput gains below the dotted line represent slowdowns compared to \texttt{BASE}. Number of replicas per GPU is presented on a log scale.}
    \label{fig:tptbal2}
\end{figure*}

\subsection{Experimental Baseline}
\label{sec:baseline}

\textbf{System, Models and Workloads. }Refer to \sref{sec:setup}.

\textbf{Baseline. }We use SGLang v0.4.8 \cite{sglang} with Expert Parallelism Load Balancer (EPLB) \cite{dpsv3} as the baseline. We adopt the standard large-scale deployment parallelism configuration: data parallelism (DP), head-parallel tensor-parallel (TP) attention, and EP. We implement dynamic all-to-all EP dispatch and combine with PyTorch and NCCL P2P collectives with EFA support. We enable mixed chunked prefill with a maximum prefill chunk size of 4096. \cam{The batch sizes are dynamic depending on request arrivals and KV cache availability at runtime.} EPLB is the most widely adopted EP load-balancing algorithm supporting expert placement and replication in mainstream LLM serving frameworks \cite{sglang,vllm,deepspeed,tensorrt-llm}. We implement and integrate \proj into SGLang as a direct replacement for EPLB and evaluate its performance. \cam{Profiling sequences used to obtain the expert load distributions for EPLB and \proj are randomly sampled from the diverse datasets (\sref{sec:setup}). The samples used for profiling are excluded from the evaluation inputs.} We evaluate three setups \textemdash~ \texttt{BASE}: EPLB placement with no replication, \texttt{EPLB}: minimum EPLB replication with $L$ replicas per GPU ($L$ is the number of MoE layers), and \texttt{\proje}: our approach, suffixed with the replication factor $R$. In our deployment setup, $R=8$ is generally the best-performing setting selected by \proj across configurations (\sref{sec:repalloc}), and we use it for the subsequent end-to-end evaluations. We include an evaluation of the effect of $R$ in \aref{sec:ratioscale}.

\textbf{Metrics. }We report \textit{goodput} as the primary metric in our experiments, defined as the maximum sustained throughput the system delivers before requests incur prolonged queuing delays. For prefill and decode performance, we report \textit{time-to-first-token} (TTFT) and \textit{inter-token-latency} (ITL).

\subsection{End-to-end Performance}
\label{sec:e2eeval}

We address Question \bcircled{2} by measuring end-to-end serving throughput across configurations. \fref{fig:e2e} plots request rate (throughput) versus time-to-first-token (TTFT) latency for each configuration. \cam{The request arrival times are modeled as a Poisson process.} The knee point marks saturation, where further increasing request rates overloads the system and TTFT becomes dominated by queuing time rather than prefill time. This knee point defines the system goodput. \fin{A higher goodput directly reduces operational costs and is crucial in real deployments, as the number of GPUs required to meet a target peak request rate under a fixed latency budget scales inversely with the per-instance goodput, dominating the total cost of ownership~\cite{infcost}.} We make three observations:

\textbf{\proj consistently achieves higher goodput than EPLB. } \fref{fig:e2e} demonstrates that compared to \texttt{EPLB}, \texttt{\proje8} achieves on average $1.15\times$ higher goodput (up to $1.2\times$) on DeepSeek-R1 and $1.12\times$ (up to $1.17\times$) on Kimi-K2 in an 8-node cluster (setups suffixed with 8). This is because \texttt{\proje8} allocates $7.25\times$ and $7.5\times$ fewer replicas than \texttt{EPLB} while achieving most of the balancedness gains on DeepSeek-R1 and Kimi-K2, respectively. The saved memory enables a larger KV cache and larger sequence batches without significant slowdown from load imbalance.

\textbf{\proj improves TTFT. }\cam{Expert load imbalance is more significant during prefill due to high token counts from long-sequence requests (\sref{sec:setup}), which leads to long TTFTs \cite{moegps, megascaleinfer}.} At a fixed request rate below saturation, both \texttt{EPLB} and \texttt{\proje8} achieve lower TTFT than \texttt{BASE}. \fref{fig:e2e} illustrates that \texttt{\proje8} reduces TTFT by $29\%$ on average (up to $58\%$) compared to \texttt{BASE}, closely matching \texttt{EPLB} ($30\%$ on average, up to $59\%$). This demonstrates that \proj's benefit-driven replication reduces load imbalance and improves computational efficiency to a similar extent as \texttt{EPLB} during prefill.

\textbf{\proj is robust across datasets with varying load imbalance. }We analyze dataset load distribution skewness as in \fref{fig:breakdown}. We observe that datasets \texttt{E} and \texttt{J} are highly skewed (low \texttt{BASE} balancedness), whereas datasets \texttt{L} and \texttt{A} are less skewed (high \texttt{BASE} balancedness). Compared to \texttt{BASE}, \texttt{\proje8} improves goodput by $1.42\times$ on highly skewed datasets and by $1.14\times$ on less skewed datasets on average, demonstrating broad applicability. In comparison, \texttt{EPLB} yields a $1.24\times$ goodput gain on highly skewed datasets but only $1.02\times$ on less skewed datasets on average. This is because less skewed datasets benefit less from replication, where excessive replication from \texttt{EPLB} yields little benefit.

\subsection{\proj Balancedness-Memory Trade-Off}
\label{sec:excrep}

Expert replication mitigates load imbalance but consumes additional GPU memory. In this section, we address Question \circled{1} by empirically quantifying this trade-off in \proj and its impact on performance. \fref{fig:tptbal2} depicts balancedness and end-to-end goodput gain of \texttt{CRA} at different values of the replication factor $R$, normalized to \texttt{BASE}. \texttt{EPLB} goodput is depicted by the rightmost red $\times$ in each figure. We make the following observations.

\textbf{Excess replication leads to suboptimal goodput. }The balancedness gain curves illustrate that \texttt{\proje} achieves most of the balancedness gains with far fewer replicas compared to \texttt{EPLB}. Allocating replicas excessively yields diminished benefits and impedes goodput because it reduces KV cache size. On \texttt{D*8}, \texttt{K*6}, and \texttt{K*12} (\texttt{*} denotes any workload), \texttt{EPLB} reduces KV cache size by $19\%$, $75\%$, and $24\%$ respectively as memory usage depends on model and cluster size. Goodput drops when the concurrency loss from a smaller KV cache outweighs the benefit of higher balancedness. This explains why \texttt{EPLB} achieves lower throughput than \texttt{BASE} in Figures \ref{fig:btke6} and \ref{fig:btkj6} with \texttt{K*6} \textemdash~ the $75\%$ KV cache reduction outweighs the balancedness gains.

\begin{figure*}[!htb]
    \centering
    \subfigure[\texttt{SYN1} workload mix ($\sigma^2=0.001$)]{
        \includegraphics[width=0.487\textwidth]{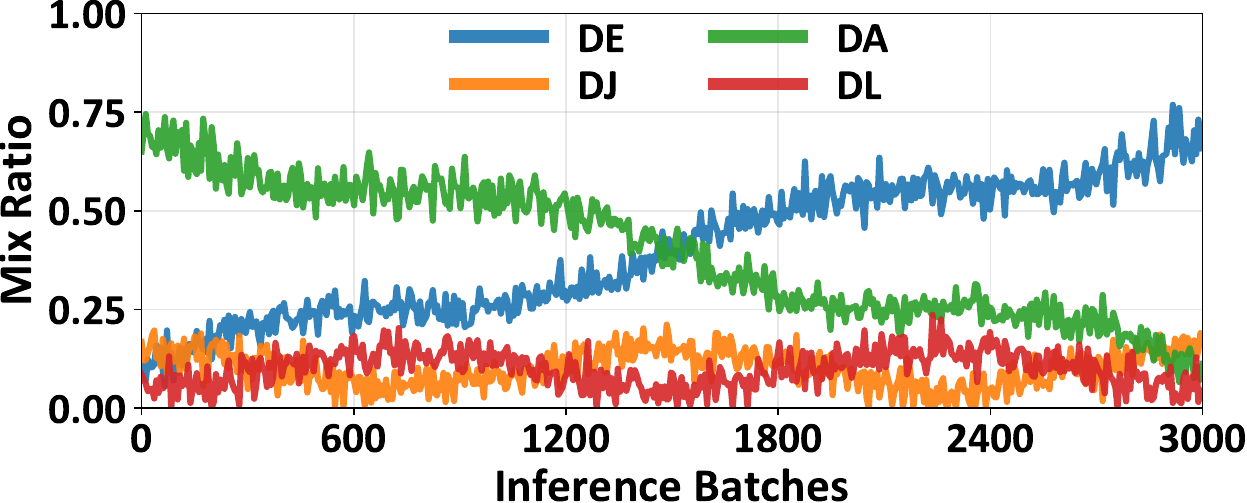}
        \label{fig:shift1mix}
    }
    \subfigure[\texttt{SYN1} balancedness ($\sigma^2=0.001$, no rebalancing)]{
        \includegraphics[width=0.487\textwidth]{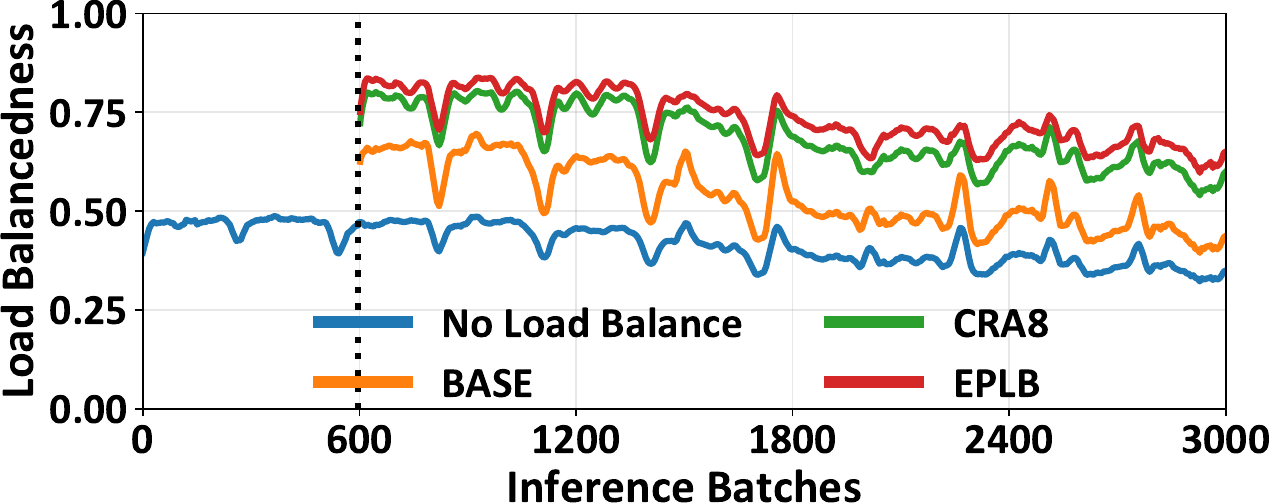}
        \label{fig:shift1bal1}
    }
    \subfigure[\texttt{SYN1} balancedness ($\sigma^2=0.001$, with rebalancing)]{
        \includegraphics[width=0.487\textwidth]{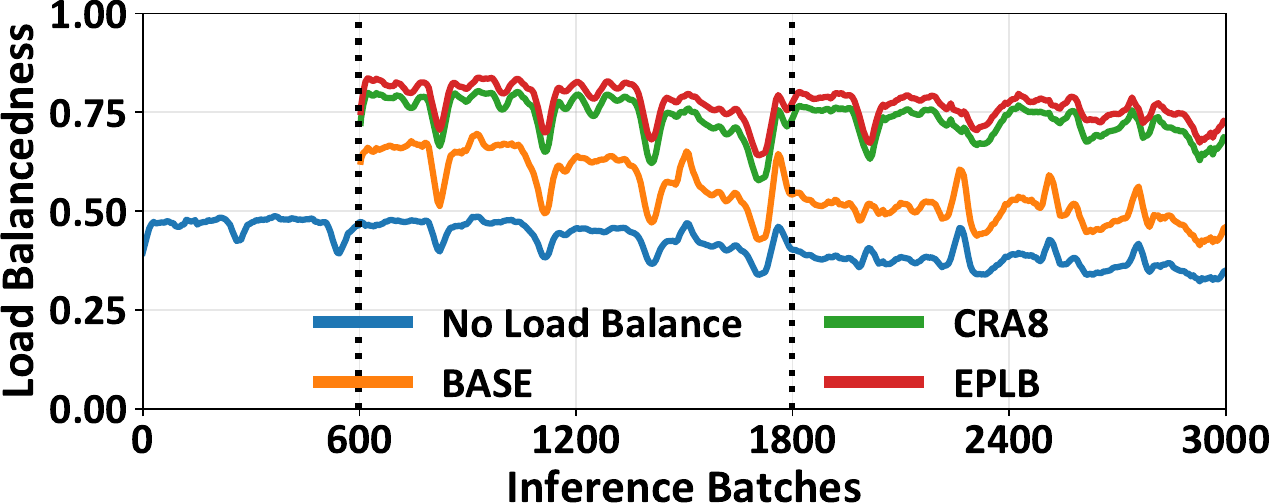}
        \label{fig:shift1bal2}
    } 
    \subfigure[\texttt{SYN1} balancedness ($\sigma^2=0.5$, no rebalancing)]{
        \includegraphics[width=0.487\textwidth]{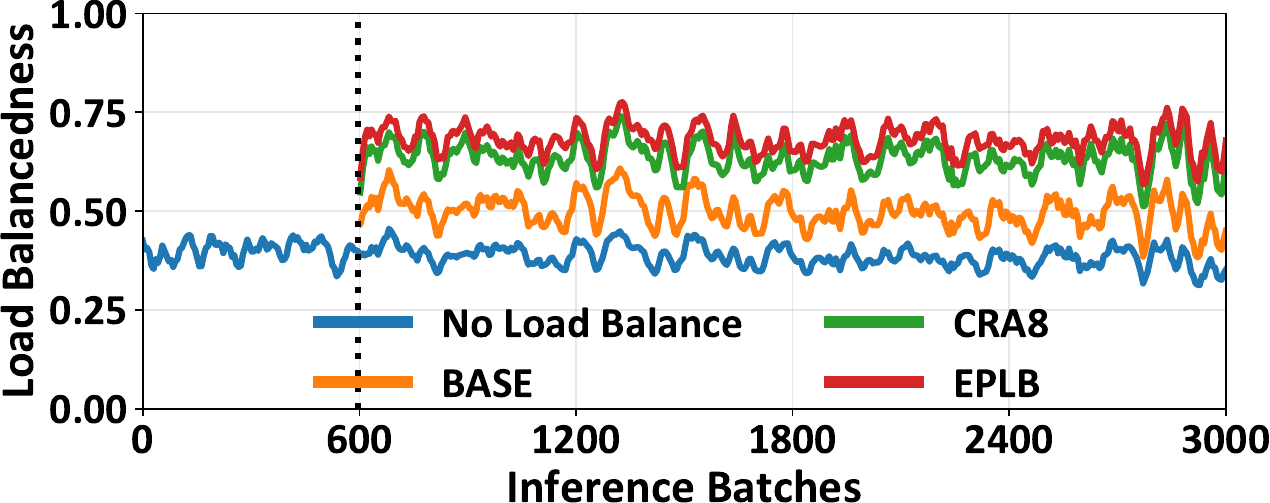}
        \label{fig:shift1bal3}
    } 
    
    \caption{\cam{\texttt{SYN1} shifting workload mix ratios and achieved load balancedness over time on DeepSeek-R1-671B (\texttt{D}) across 8 nodes. \fref{fig:shift1mix} depicts the ratio of each dataset over time with burstiness variance $\sigma^2=0.001$. Figures \ref{fig:shift1bal1} - \ref{fig:shift1bal3} depict the achieved load balancedness under various configurations. The dotted lines represent triggered online periodic rebalancing, and data lines are smoothed for clearer presentation. Load-balancing techniques (\texttt{BASE}, \texttt{EPLB}, \texttt{CRA8}) are activated after the initial warmup of 600 iterations. Workloads with higher $\sigma^2$ only exhibit higher degree of burstiness; the shift and diurnal patterns remain identical.}}
    \label{fig:shift}
\end{figure*}

\subsection{Scalability with Cluster Size}
\label{sec:clusterscale}

To answer Question \bcircled{3}, we study how goodput scales with different cluster sizes and setups. Figures \ref{fig:ke6}-\ref{fig:kl12} depict the throughput–TTFT curves for cluster sizes of 6, 8, and 12 in \texttt{KE}, \texttt{KJ} and \texttt{KL}. Since the model size is fixed, smaller clusters imply more experts and less KV cache per device. Kimi-K2 (\texttt{K}) has 384 experts without replication, so each GPU holds 8, 6, and 4 experts for cluster sizes of 6, 8, and 12, respectively. We also scale the sequence batch size by setting each node as a DP rank. We make two observations:

\textbf{\proj scales goodput efficiently with large cluster sizes and outperforms EPLB. } 
The goodput scaling of \texttt{BASE} is poor, with unstable and elevated TTFT. This confirms Observation 2 in \sref{sec:repeffect}: placement-only baseline suffers from high load imbalance at larger clusters. \texttt{\proje8} scales by $1.65\times$ and $1.6\times$ on average when increasing the cluster size from 6 to 8 and from 8 to 12, respectively, outperforming \texttt{EPLB} and demonstrating good scalability.

\textbf{The memory overhead of EPLB replication impedes goodput more significantly at smaller clusters. }
In Figures \ref{fig:ke6}, \ref{fig:kj6}, and \ref{fig:kl6}, \texttt{EPLB} goodput is on average $46\%$ lower compared to \texttt{BASE}. The KV cache in a 6-node cluster is very limited; \texttt{EPLB} further shrinks it by $75\%$ through excessive replication, significantly reducing goodput (\sref{sec:excrep}). In comparison, \texttt{\proje8} allocates far fewer replicas, reducing KV cache size by only $6\%$. Thus, \texttt{\proje8} still achieves $1.14\times$ higher goodput than \texttt{BASE} on average.

\subsection{Overhead Analysis}

We answer Question \circled{4} and analyze both online and offline overhead of \proj. At inference time, \proj incurs no additional overhead. We present an inter-token-latency (ITL) evaluation in \aref{sec:itl}. During initialization, the overhead of \proj is dominated by replication benefit estimation, as we sweep replica counts and replay the load distribution across all layers (\sref{sec:repalloc}). In our experiments, this process takes approximately 10 seconds, which is negligible since framework initialization typically takes minutes. With online periodic rebalancing, the estimation overhead can be overlapped with running batches on CPU.

\subsection{\cam{Impact of Workload Shifts}}
\label{sec:shifteval}

\cam{In this section, we perform an ablation of the impact of shifts in workload distributions on distribution-based load-balancing techniques, \proj and EPLB. In online deployments, workloads may change over time, leading to drift in load distributions. For example, prior datacenter studies \cite{mlaas,servegen,characterizing,reqsto} identify two common characteristics of datacenter workloads\textemdash long-term diurnal shifts and short-term stochastic bursty arrivals. LLM-serving requests specifically may exhibit diurnal fluctuations and notably bursty request arrivals \cite{servegen}. At the same time, consecutive requests are typically highly correlated due to shared context, common vocabulary, and request continuity \cite{emoe}. To analyze these scenarios where (1) workload shifts exhibit stochastic burstiness, and (2) diurnal workload shifts that occur gradually over time, we mix the four evaluation datasets and create four synthetic shifting workloads: \texttt{SYN1}, \texttt{SYN2}, \texttt{SYN3} and \texttt{SYN4}, each defined by the start and end ratios of datasets \texttt{E}, \texttt{J}, \texttt{A}, and \texttt{L}. We synthesize the workloads in a four-step process: (1) the dataset ratios are initially modeled as linear trends from their start ratios to their end ratios. (2) we assign sinusoidal waves with distinct phase shifts to model workload-specific diurnal patterns. (3) we inject zero-mean Gaussian noise $\mathbf{\epsilon}_t \sim \mathcal{N}(\mathbf{0}, \sigma^2 \mathbf{I})$ into the mix ratio of each dataset to model short-term bursts, where $\sigma^2$ is the variance (degree of burstiness) and $I$ is the identity matrix ensuring independent noise across datasets. (4) we normalize the ratios of all datasets within each inference batch to sum to 1.0. For each synthetic workload, we evaluate two different burstiness levels: $\sigma^2=0.001$ and $0.5$. \fref{fig:shift} depicts the workload mix ratios and the achieved load balancedness on \texttt{SYN1}. We use a rebalancing window of 1200 iterations, which is approximately 20 minutes on our setup \textemdash~2× longer than EPLB's 10-minute default (\sref{sec:workloadshift}). \fin{Despite this less frequent rebalancing, we find that periodic rebalancing at this interval is sufficient to sustain high overall balancedness.} The results of \texttt{SYN2}, \texttt{SYN3} and \texttt{SYN4} are included in \aref{sec:shifteval2}. We make three observations. 
First, for shifting workloads that have lower burstiness and more gradual distribution changes ($\sigma^2=0.001$), the balancedness degrades over time for both EPLB and \proj as expected (\fref{fig:shift1bal1}). With periodic online rebalancing, the balancedness improves to reflect the change in distribution (\fref{fig:shift1bal2}). 
Second, for shifting workloads that have higher burstiness ($\sigma^2=0.5$), the balancedness exhibits higher fluctuations where the short-term bursts become more significant, leading to drastic changes in the load distributions (\fref{fig:shift1bal3}). Thus, the contribution of each dataset to the load distribution becomes more even, and the initial plan learned by EPLB and \proj remains effective despite the shifting workload. 
Third, \proj achieves comparable load balancedness to EPLB, and better than the baseline with no load-balancing, across all evaluated configurations.
\texttt{CRA8} achieves this while consuming $7.25\times$ less memory than \texttt{EPLB}, enabling the benefit of a larger KV cache and improved throughput, as demonstrated in Sections \ref{sec:e2eeval} and \ref{sec:clusterscale}.

}

\subsection{\cam{\proj Speedup Breakdown}}
\label{sec:timebreak}

\begin{figure}[!htb]
    \centering
    \includegraphics[width=1\linewidth]{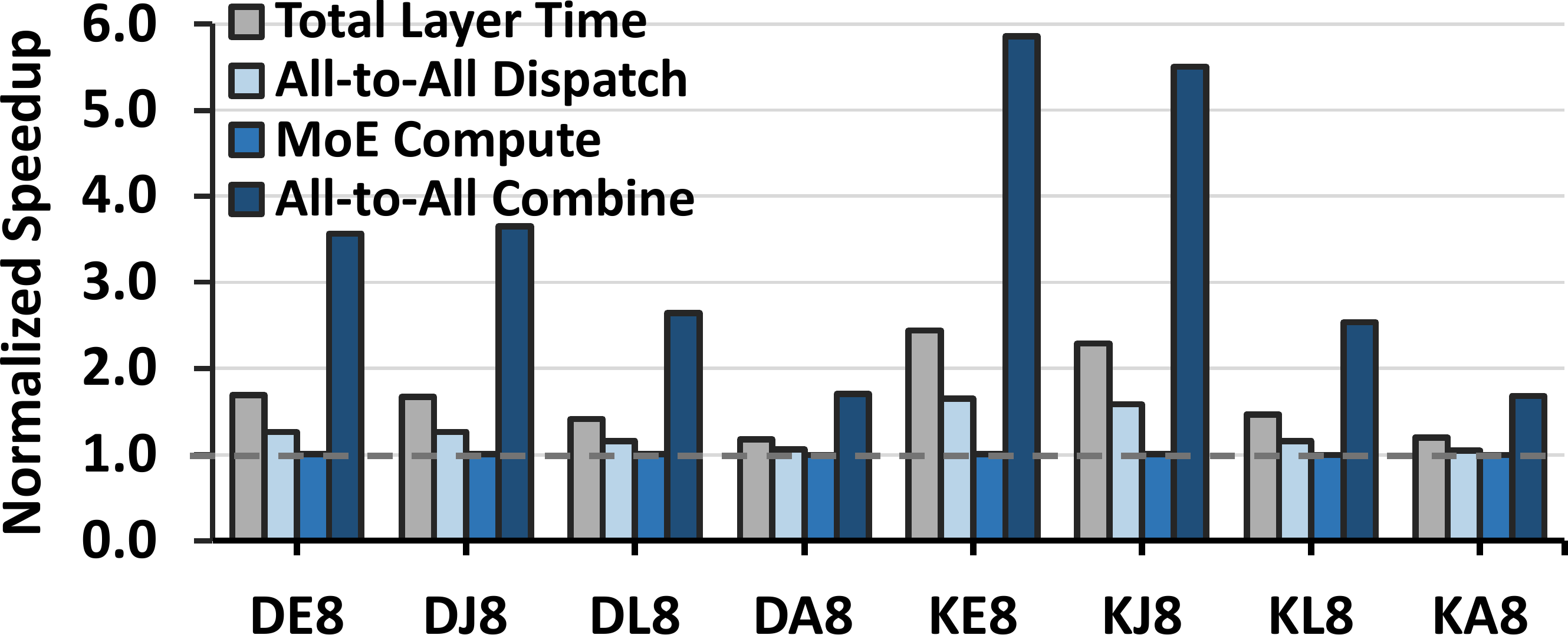}
    \caption{\cam{\texttt{CRA8} speedup of total layer time (including non-MoE blocks) and time of each MoE block step, normalized to \texttt{BASE}.}}
    \label{fig:laybreak}
\end{figure}

\cam{In this ablation, we break down the end-to-end speedup and analyze how different layer components benefit from \proj. \fref{fig:laybreak} depicts the speedups of total layer time and main MoE components with \texttt{CRA8}, normalized to \texttt{BASE}. The speedups are collected with NVIDIA Nsight Systems \cite{nsys} and aggregated over 10 iterations, all layers, and all GPUs. We make three observations: (1) MoE runtime only improves marginally because the token count is constant, and runtime is aggregated across GPUs and is approximately proportional to token count and compute. (2) All-to-All Dispatch exhibits moderate speedup, because replication eliminates network link congestion by evenly distributing load. (3) \fin{All-to-All Combine speeds up significantly, but its contribution to end-to-end speedup (\sref{sec:e2eeval}) is limited as it constitutes only a fraction of total layer runtime in the baseline.} Under high load imbalance, cold devices stall on All-to-All Combine, waiting for hot devices. When load is evenly distributed, MoE runtime aligns across devices, eliminating stalls.}

\section{Related Work}
\label{sec:relatedwork}

\textbf{Expert Placement. } Many existing works aim to achieve optimal expert placement during inference to reduce all-to-all traffic and mitigate load imbalance. A recent line of work reduces all-to-all traffic via topology-aware placement that co-locates experts with high activation affinity \cite{smartmoe,flexmoe,tutel,gracemoe}. ExFlow \cite{exflow} and MoETuner \cite{moetuner} formulate placement as integer programs with topology and load constraints. Occult \cite{occult} constructs an expert-collaboration graph and derives placements by clustering. As discussed in \sref{sec:placerep}, placement-only strategies cannot address load imbalance in extremely skewed workloads. However, we note that these approaches are orthogonal and can be integrated into \proj as a replacement for the current greedy placement (\sref{sec:interleaveassign}), further improving  communication efficiency.

\textbf{Expert Replication. } Replication is widely used to mitigate load imbalance during distributed MoE training. FasterMoE \cite{fastermoe} introduces shadow experts, selectively replicating hot experts to improve throughput. SwiftMoE \cite{swiftmoe} adopts adaptive, non-uniform replication during training that tracks and replicates hot experts on the fly. However, efficient replication during training relies on overlapping replica weight transfers with gradient synchronization, and applying it during inference introduces an additional communication phase with prohibitive latency overheads. Concurrent work GraceMoE \cite{gracemoe} groups experts to reduce communication overhead. It allocates replicas dynamically for each layer, but restricts replication within a single group, since excessive replication impairs the effectiveness of grouping. \proj's fine-grained replication is orthogonal to the grouping strategy and can potentially improve load balancedness by replicating across multiple groups. 

\textbf{Routing Prediction and Alteration. }Many works use load distributions and routing patterns during inference to predict future expert activations \cite{swapmoe,moelightning,klotski,kxformer,promoe,duoserve,prescope,expflow}. They typically combine host offloading with prefetching or caching. However, mispredictions can introduce high latency, and many target resource-constrained environments. Other works modify the routing mechanism to aid prediction and load balance \cite{pregated,adapmoe,expchoice}. These post-training techniques add deployment overhead and may compromise model accuracy.

\textbf{Expert Sharding. } Expert sharding is a variant of tensor parallelism that distributes expert weights across devices. Early MoE systems combine sharding with EP to scale the expert parameters \cite{gshard,deepspeed}. Recent studies show that expert sharding can achieve near-perfect load balance during inference \cite{moeshard}. However, cross-node global token reduction incurs high inter-node communication overhead, significantly limiting its scalability \cite{deepspeedted,megascale}.

\section{Conclusion}
In this work, we demonstrate the importance of fine-grained replication for mitigating expert load imbalance at low memory costs. We propose \proj, an intuitive and practical framework that allocates replicas based on estimated benefits to improve inference throughput. \proj can be flexibly integrated into mainstream serving frameworks, delivering significantly higher throughput and efficiency with no runtime overhead.

\pagebreak


\bibliography{refs}
\bibliographystyle{mlsys2026}

\appendix
\newpage

\section{\proj Algorithms}
\label{sec:odalgo}

In this section, we present the pseudocode for the three main algorithms in \proj, detailing the step-by-step logic of each algorithm. When integrating \proj, we leverage vectorized operations to achieve maximum efficiency.

\begin{minipage}{\columnwidth}

\subsection{Estimate Per-Layer Replication Benefit}
\label{sec:odalgo1}

In \proj, the input positive replica counts are chosen from a base-2 geometric progression over $[1,D]$ for computational efficiency (\sref{sec:repalloc}).

\begin{algorithm}[H]
   \caption{Per-Layer Replication Benefit Estimation}
   \label{alg:odalgo1}
\begin{algorithmic}
   \STATE {\bfseries Input 1:} list of positive replica counts $R$
   \STATE {\bfseries Input 2:} number of GPUs $D$ and nodes $N$
   \STATE {\bfseries Input 3:} offline load distribution $W$ of shape $B\times L\times E$: $B$ - \# inference batches, $L$ - \# MoE layers, $E$ - \# experts per MoE layer, $W[b][\ell][e]$ - token load of expert $e$ at layer $\ell$ in batch $b$
   \STATE {\bfseries Output:} map $T$: keys - replica count $r\in R$ ; values - length $L$ lists with per-layer GPU balancedness gain compared to placement-only.
    \STATE {\bfseries Auxiliary Function:}
    \STATE \hspace{1em} $\textsc{Placement}(W,E,N)$: returns an expert placement plan $P$ given $L\times E$ load distribution $W$, expert count $E$ including replicas, and number of nodes $N$. We use a greedy placement algorithm, similar to EPLB. Placement plan $P$ describes expert-to-device mapping and the number of replicas assigned for each expert.
   \STATE $W_s=W.sum(dim=0)\;\;\;$ // $W_s$ shape ($L\times E$)
   \STATE $base=0.0\;\;\;$ // placement-only baseline balancedness
   \FOR{$r$ {\bfseries in} $[0] + R$}
   \STATE placement plan $P=$ \textsc{Placement}($W_s,\;E+r,\;N$)
   \STATE $BAL=[0]*L\;\;\;$ // cumulative GPU balancedness
   \FOR{$b=0$ {\bfseries to} $B-1$ {\bfseries and} $\ell=0$ {\bfseries to} $L-1$}
   \STATE $G=[0] * D\;\;\;$ // per-GPU load
   \FOR{$e=0$ {\bfseries to} $E-1$}
   \STATE // expert copies include both base and replicas
   \STATE $r_{el}=P.num\_expert\_copies(e,\ell)\;\;$
   \STATE $g_{el}=P.gpus(e,\ell)\;$ // GPUs with expert copy
   \FOR{$g$ {\bfseries in} $g_{el}$}
   \STATE // per-GPU load
   \STATE $G[g]\,+=floor(W[b][\ell][e]\;/\;r_{el})\;\;$
   \ENDFOR
   \ENDFOR
   \STATE $BAL[\ell]\,+=avg(G)\;/\;max(G)$ // balancedness
   \ENDFOR
   \IF{$r==0$}
   \STATE $base=BAL[\ell]\;/\;B\;\;$ // placement-only
   \ELSE
   \STATE // per-layer average GPU balancedness gain
   \STATE $T[r][\ell] = (BAL[\ell]\;/\;B) - base$
   \ENDIF
   \ENDFOR
    \STATE {\bfseries return} $T$
\end{algorithmic}
\end{algorithm}
    
\end{minipage}

\begin{minipage}{\columnwidth}

\subsection{Solve Replica Allocation via Dynamic Programming}
\label{sec:odalgo2}

The time complexity of the following dynamic programming is $O(L\cdot C\cdot 
|T.keys()|)$, and the space complexity is $O(L\cdot C)$.

\begin{algorithm}[H]
\caption{Replica Allocation Solver Maximizing Replication Benefits}
\label{alg:odalgo2}
\begin{algorithmic}
\STATE {\bfseries Input 1:} map $T$: replica counts to balancedness gains map, acquired with \algoref{alg:odalgo1}
\STATE {\bfseries Input 2:} expert capacity $C$ (base experts \& replicas)
\STATE {\bfseries Output:} selection list $\mathbf{x}$ of length $L$ with $\mathbf{x}[j]\in {0}\cup \mathrm{keys}(T)$
\STATE $R = T.keys()\;\;$ // replica counts
\STATE $L = \mathrm{length}(T[R[0]])\;\;$ // number of MoE layers
\STATE // initialize DP states and selections
\STATE $\mathrm{dp}[0..L][0..C] = -\infty$
\STATE $\mathrm{dp}[0][0] = 0$
\STATE $\mathrm{choice}[0..L][0..C] = None$
\FOR{$\ell=1$ {\bfseries to} $L$}
\STATE $j = \ell-1$
\FOR{$c=0$ {\bfseries to} $C$}
\STATE // option 1: no replication for current layer
\STATE $\mathrm{dp}[\ell][c] = \mathrm{dp}[\ell-1][c]$
\STATE $\mathrm{choice}[\ell][c] = None$
\STATE // option 2: iterate through replica counts
\FOR{$r$ {\bfseries in} $R$}
\STATE // check capacity and reachability
\IF{$c \ge r$ {\bfseries and} $\mathrm{dp}[\ell-1][c-r] > -\infty$}
\STATE // calculate new global balancedness gain
\STATE $cand = \mathrm{dp}[\ell-1][c-r] + T[r][j]$
\STATE // update DP states if a new best is found
\IF{$cand > \mathrm{dp}[\ell][c]$}
\STATE $\mathrm{dp}[\ell][c] = cand$
\STATE $\mathrm{choice}[\ell][c] = r$
\ENDIF
\ENDIF
\ENDFOR
\ENDFOR
\ENDFOR
\STATE $\mathbf{x}[0..L-1] = 0\;\;$ // output initialization
\STATE // acquire per-layer replica count from DP states
\STATE $c = C$
\FOR{$\ell=L$ {\bfseries downto} $1$}
\STATE $r = \mathrm{choice}[\ell][c]$
\IF{$r$ {\bfseries is not} $None$}
\STATE $\mathbf{x}[\ell-1] = r$
\STATE $c = c - r$
\ENDIF
\ENDFOR
\STATE {\bfseries return} $\mathbf{x}$
\end{algorithmic}
\end{algorithm}

\end{minipage}

\begin{minipage}{\columnwidth}
\subsection{Capacity-Aware Interleaved Expert Assignment}
\label{sec:odalgo3}

\begin{algorithm}[H]
\caption{Balanced Intra-Layer Interleaved Replica Assignment}
\label{alg:odalgo3}
\begin{algorithmic}
\STATE {\bfseries Input 1:} number of MoE layers $L$
\STATE {\bfseries Input 2:} number of GPUs $D$
\STATE {\bfseries Input 3:} replica count per-layer $R[0..L-1]$
\STATE {\bfseries Output:} assignment matrix $A$ of shape $L\times D$, where $A[\ell][g]$ is the number of experts on GPU $g$ for layer $\ell$
\STATE {\bfseries Auxiliary Functions:}
\STATE \hspace{1em} $\textsc{MinCutoff}(\mathbf{v}, r)$: returns the $r$-th smallest value in list $\mathbf{v}$ (1-based).
\STATE \hspace{1em} $\textsc{InterleaveSelect}(\mathcal{I}, k)$: given an ordered list of indices $\mathcal{I}$, returns $k$ indices that are as evenly spaced across $\mathcal{I}$ as possible (including endpoints when $k>1$).
\STATE // assign max number of experts uniformly to all GPUs
\STATE // we only need to handle the remainders for each layer
\STATE $A[\ell][g] = \left\lfloor \dfrac{R[\ell]}{D} \right\rfloor\;$ for $\ell\in[0,L)$ and $g\in[0,D)$
\STATE // total \# experts per-GPU across all layers, should keep this consistent across all GPUs
\STATE $G[g]=\sum\limits^{L}_{\ell=0}A[\ell][g]$ for $g\in[0,D)$
\FOR{$\ell = 0$ {\bfseries to} $L-1$}
\STATE // only need to handle remainder replicas at each layer
\STATE $r = R[\ell] \bmod D\;\;$
\IF{$r > 0$}
\STATE $\mathrm{cutoff} = \textsc{MinCutoff}(\mathrm{G}, r)$
\STATE // GPUs with \# experts $<\mathrm{cutoff}$ are always selected
\STATE $M = [g : g\in G \text{ and }g<\mathrm{cutoff}]$
\STATE $q = r - |M|$
\IF{$q > 0$}

\STATE // GPUs with tied \# experts $=\mathrm{cutoff}$
\STATE $K =sort([g : g\in G \text{ and }g=\mathrm{cutoff}])$
\STATE // select among $K$ in evenly-spread manner
\STATE $J = \textsc{InterleaveSelect}(K, q)$ where $J\subseteq K$
\STATE $S = M \cup J$
\ELSE
\STATE $S = M$
\ENDIF
\STATE // assign replicas to GPUs
\FOR{each $g \in S$}
\STATE $A[\ell][g]\,+= 1$
\STATE $\mathrm{G}[g]\, += 1$
\ENDFOR
\ENDIF
\ENDFOR
\STATE {\bfseries return} $A$
\end{algorithmic}
\end{algorithm}

\end{minipage}

\bigskip
\bigskip
\bigskip

\section{Effect of Replication Factor $R$}
\label{sec:ratioscale}

\begin{figure*}[!htb]
    \centering
    \subfigure[\texttt{KE6}]{
        \includegraphics[width=0.329\textwidth]{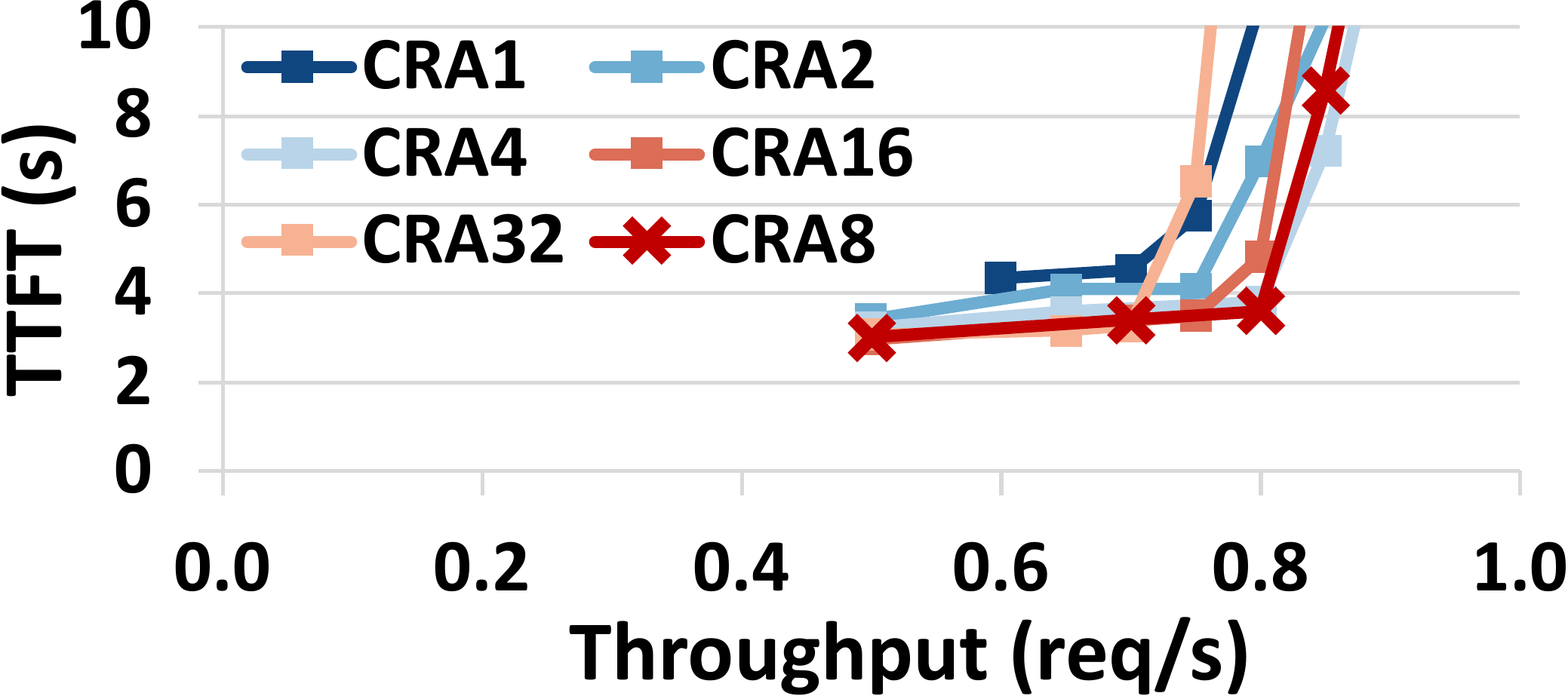}
        \label{fig:ke6a}
    } 
    \subfigure[\texttt{KE8}]{
        \includegraphics[width=0.314\textwidth]{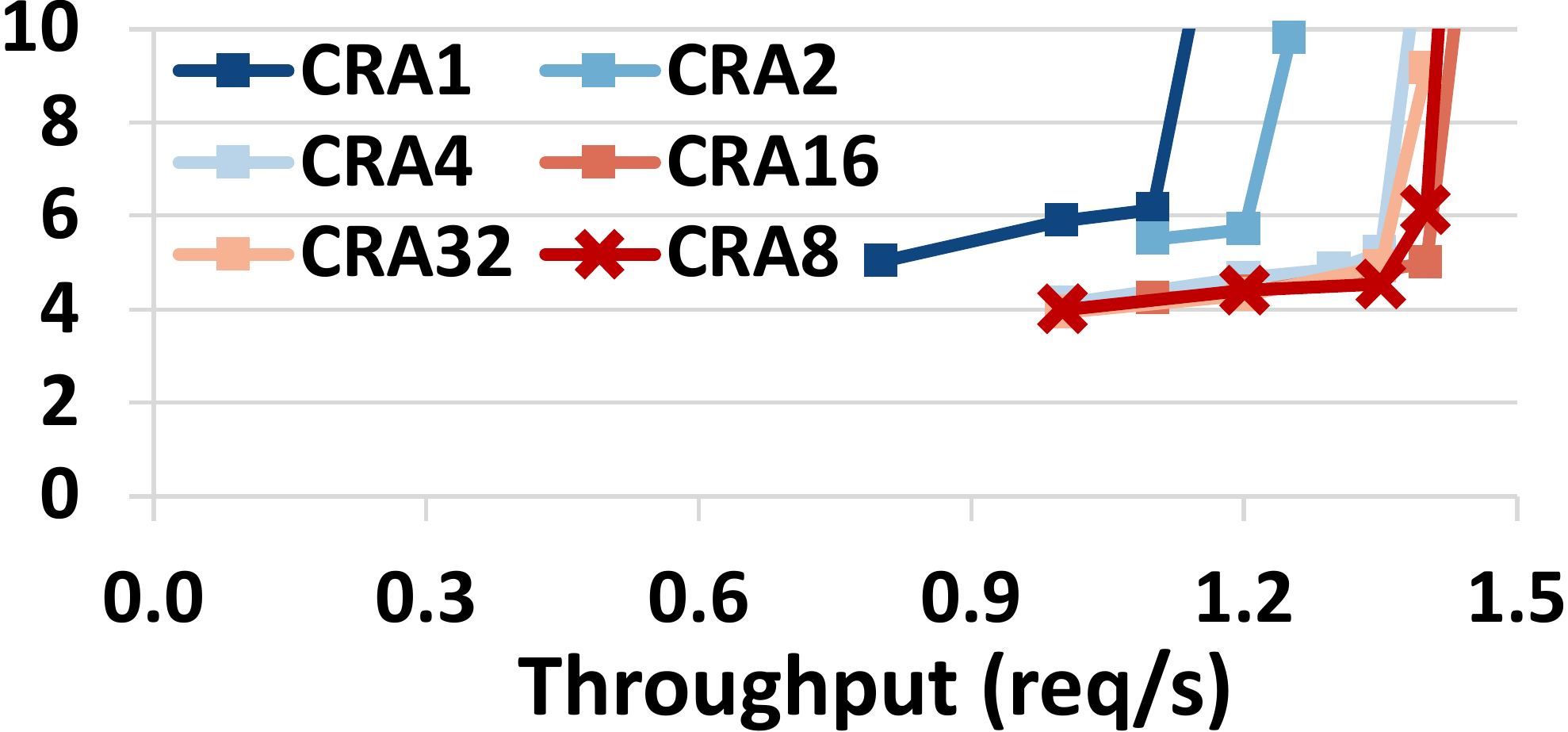}
        \label{fig:ke8a}
    } 
    \subfigure[\texttt{KE12}]{
        \includegraphics[width=0.314\textwidth]{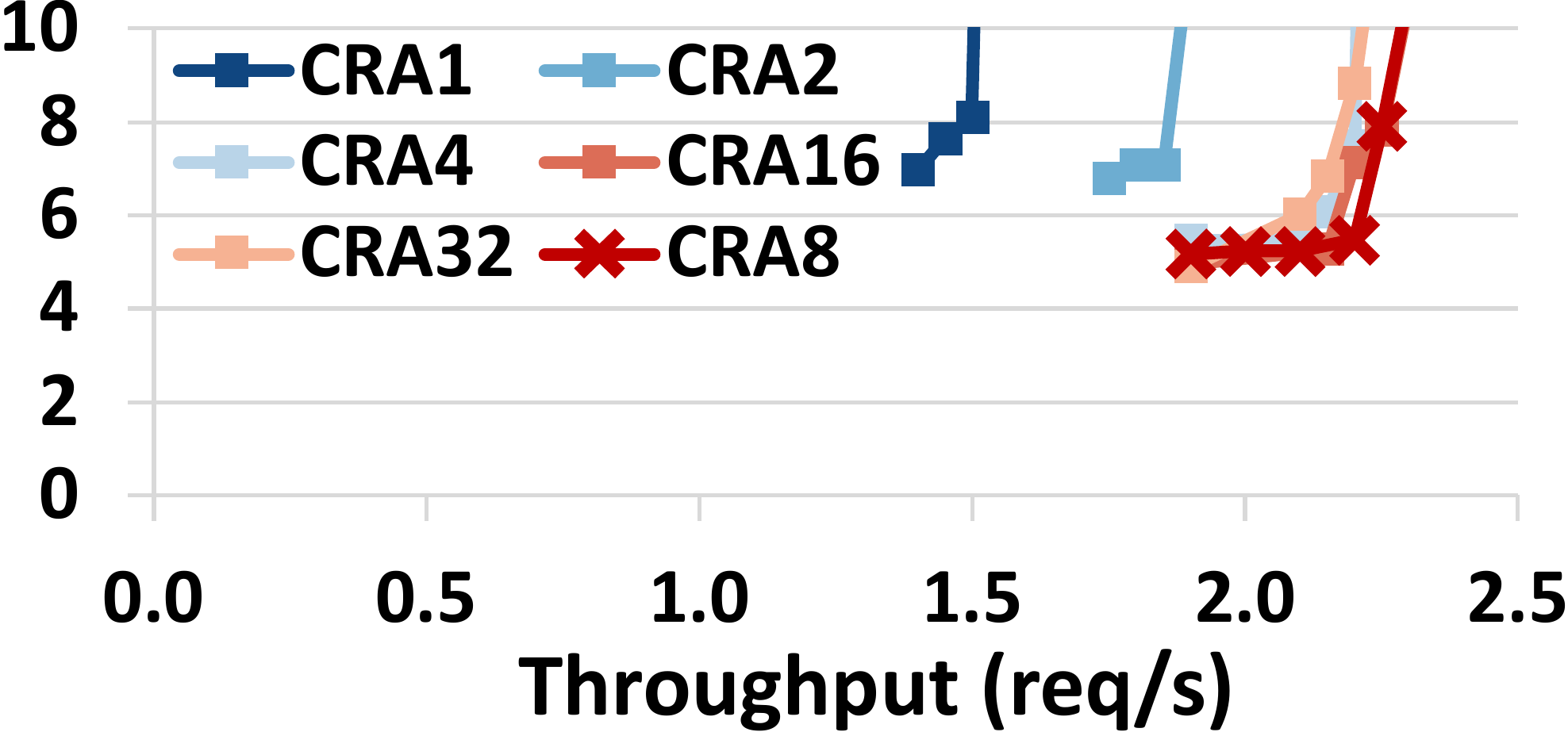}
        \label{fig:ke12a}
    } 
    \subfigure[\texttt{KJ6}]{
        \includegraphics[width=0.329\textwidth]{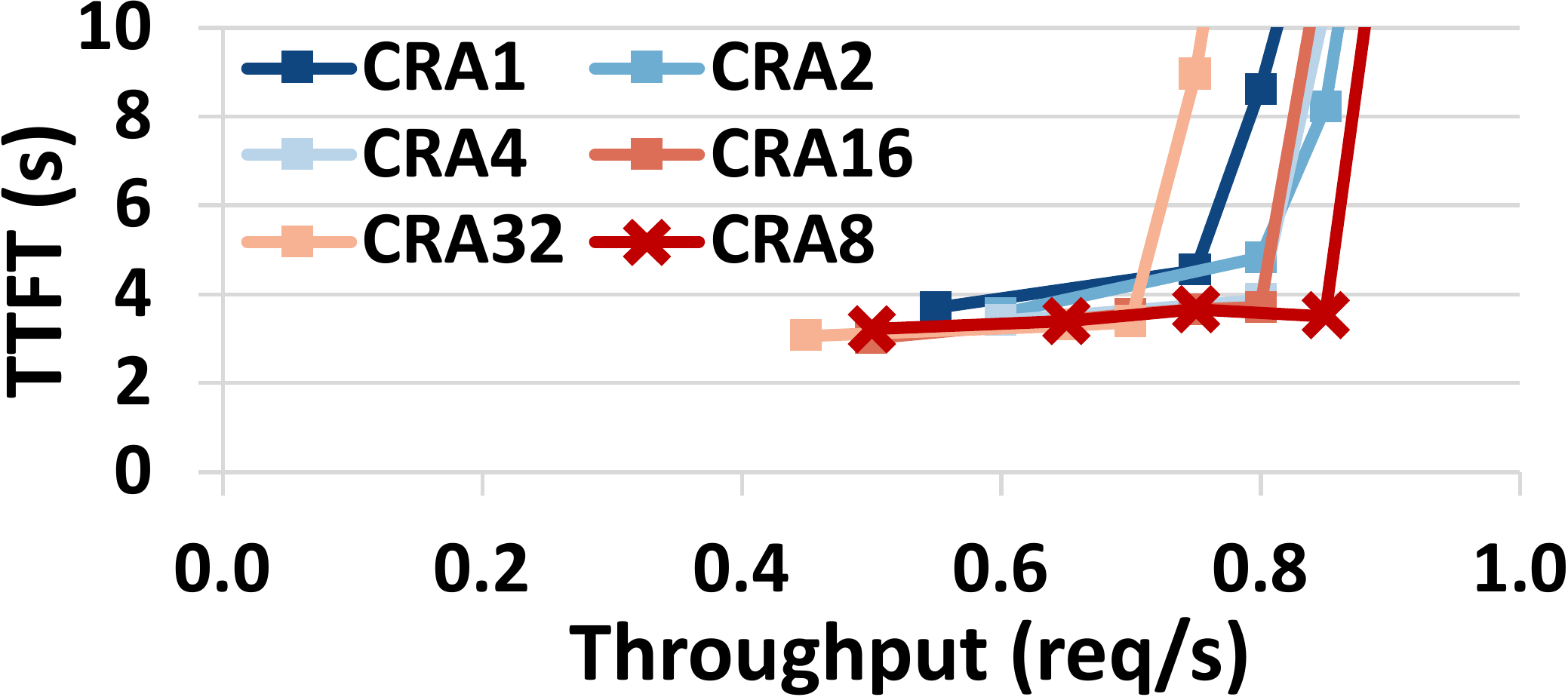}
        \label{fig:kj6a}
    } 
    \subfigure[\texttt{KJ8}]{
        \includegraphics[width=0.314\textwidth]{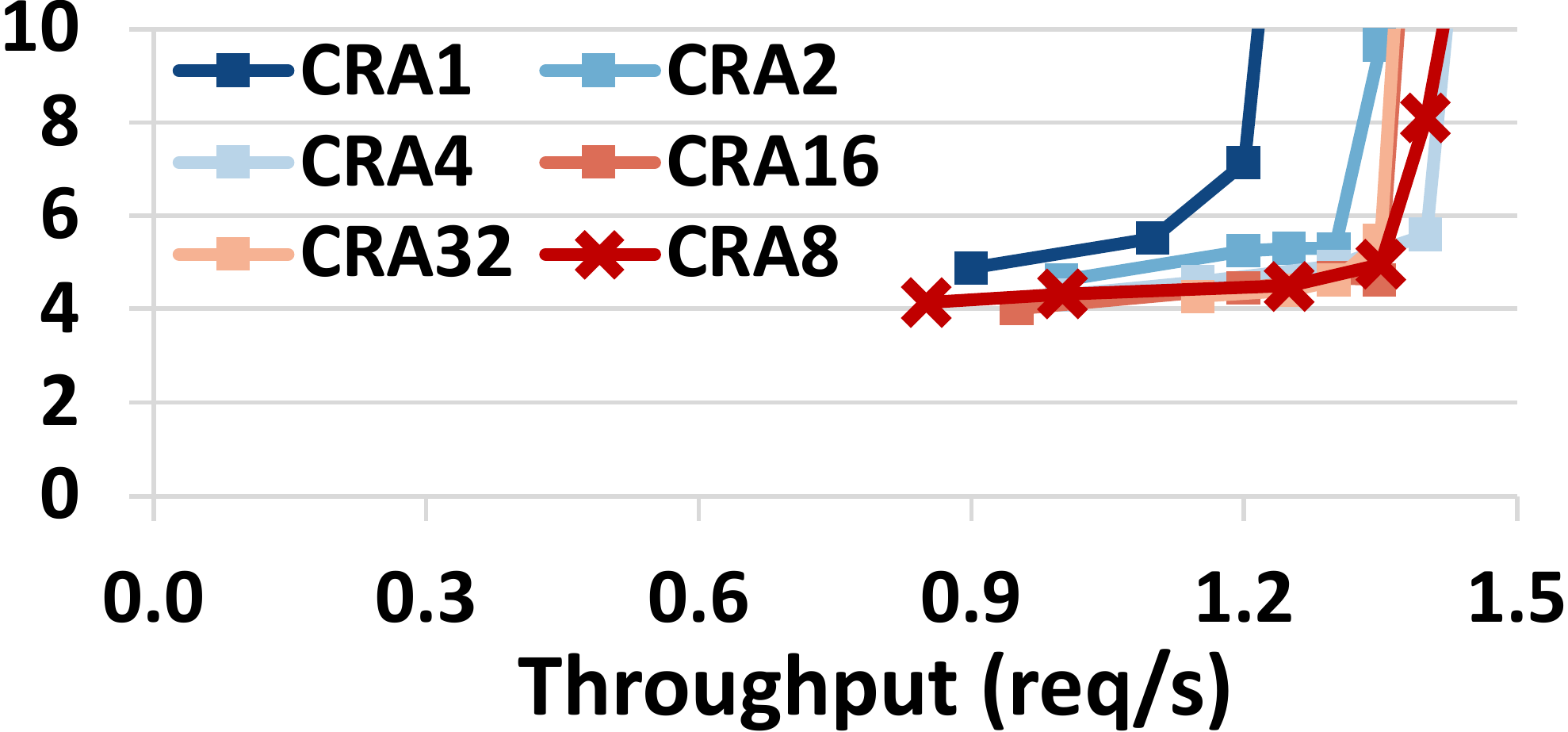}
        \label{fig:kj8a}
    } 
    \subfigure[\texttt{KJ12}]{
        \includegraphics[width=0.314\textwidth]{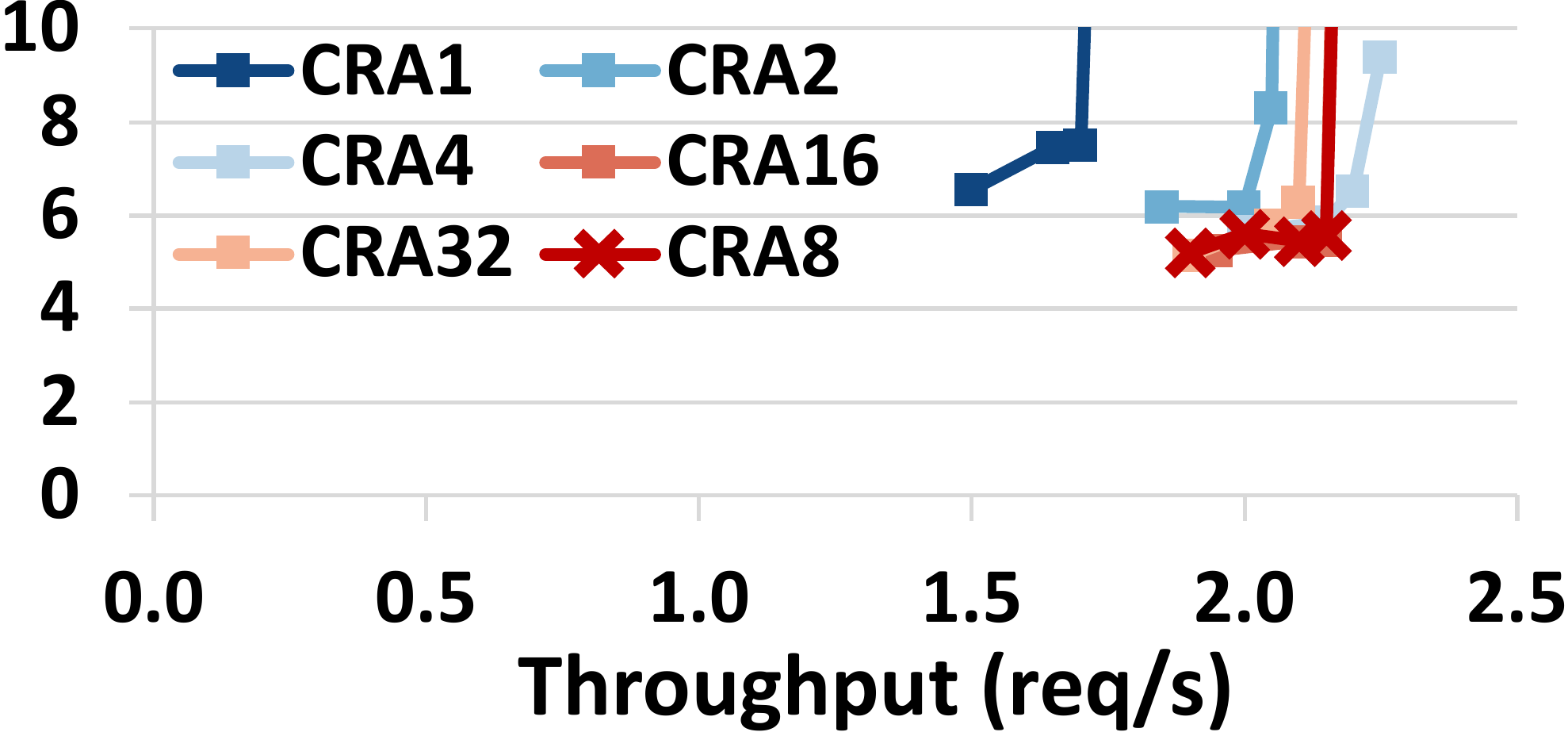}
        \label{fig:kj12a}
    } 
    \subfigure[\texttt{DE8}]{
        \includegraphics[width=0.329\textwidth]{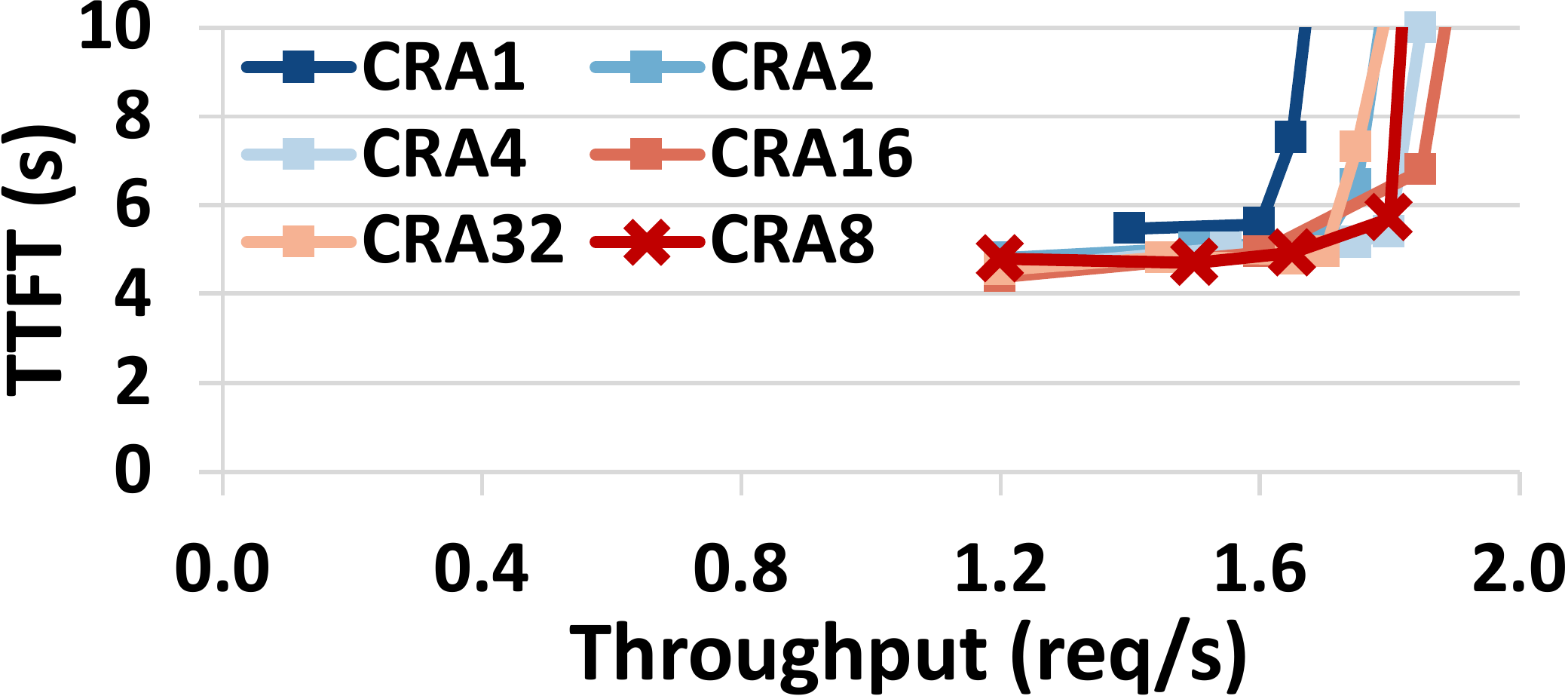}
        \label{fig:de8a}
    } 
    \subfigure[\texttt{DJ8}]{
        \includegraphics[width=0.314\textwidth]{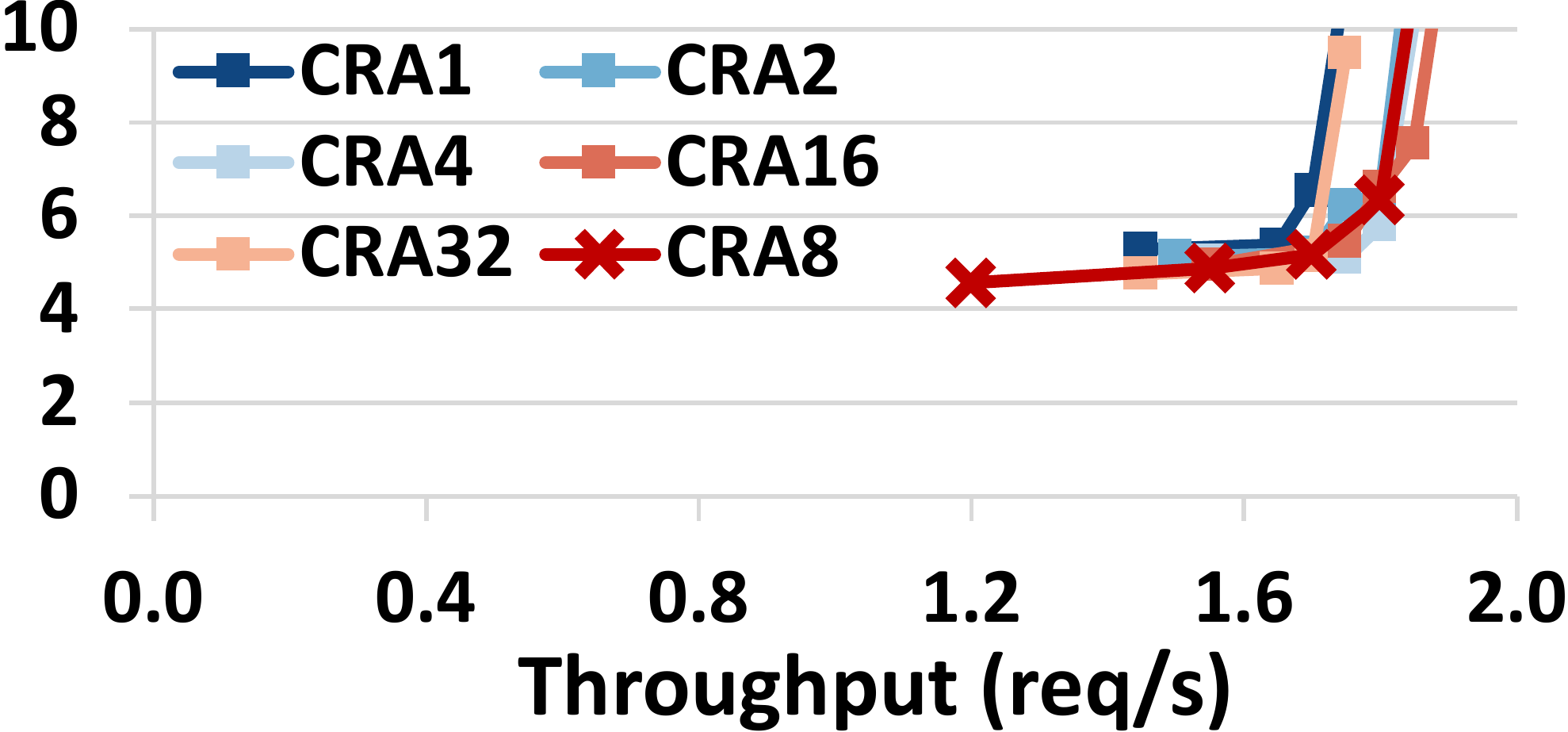}
        \label{fig:dj8a}
    } 
    \subfigure[\texttt{DL8}]{
        \includegraphics[width=0.314\textwidth]{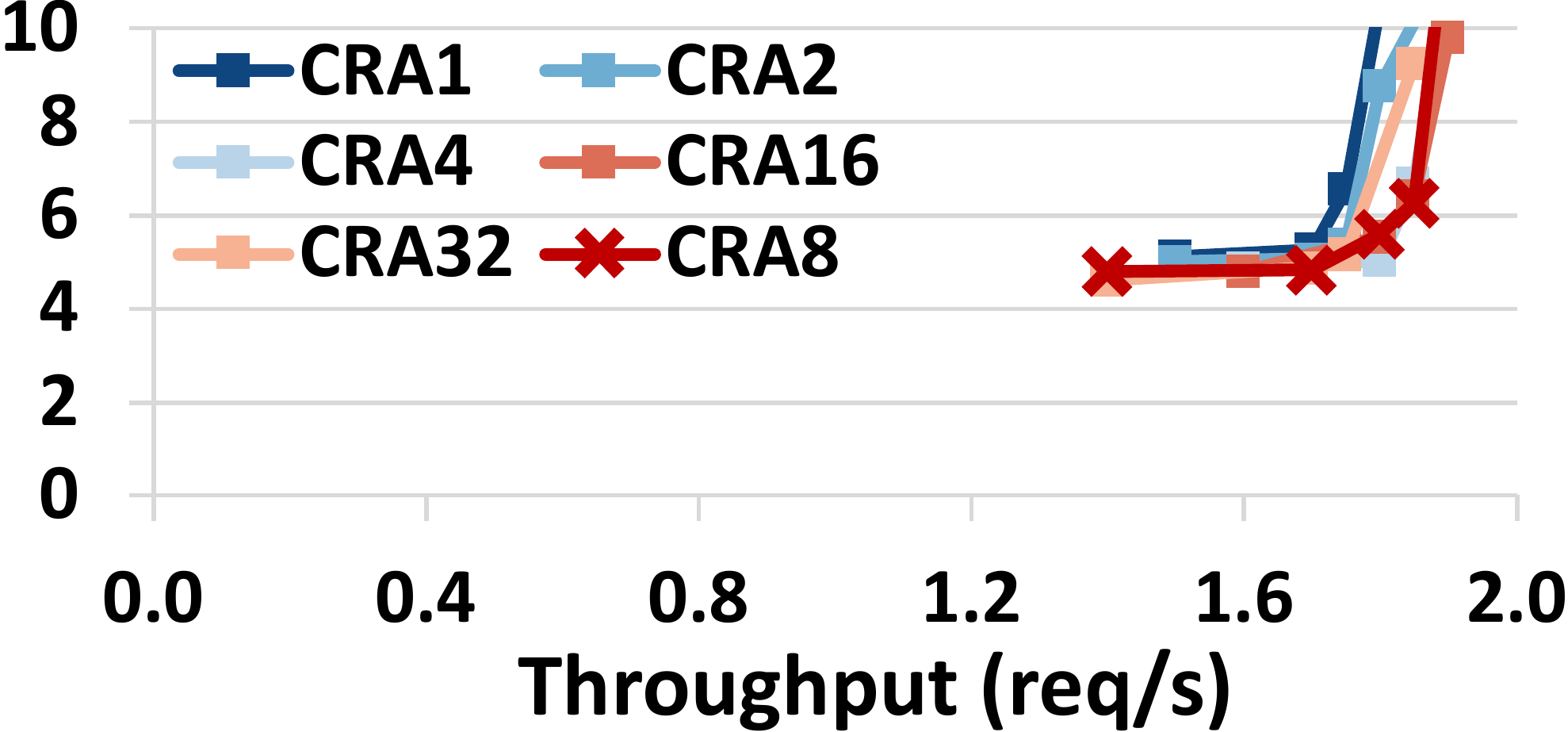}
        \label{fig:dl8a}
    } 
    \caption{End-to-end throughput to TTFT curves on various  replication ratios $R$ over different cluster sizes and configurations.}
    \label{fig:ratioscale}
\end{figure*}

In this section, we evaluate the performance of \proj with different replication factors $R$. \fref{fig:ratioscale} depicts the throughput-TTFT curve for \texttt{\proje} using various $R$ values under different configurations. We observe a trade-off between $R$ and goodput: When $R$ is too small, the number of replicas is insufficient to fully mitigate load imbalance. When $R$ is too large, excessive replication incurs significant memory overhead and causes slowdowns. This aligns with our observations in \sref{sec:excrep}. In our deployment setup, $R=8$ generally achieves the highest goodput across configurations.

\section{Inter-Token Latency Evaluation}
\label{sec:itl}

\begin{figure}[!htb]
    \centering
    \includegraphics[width=1\linewidth]{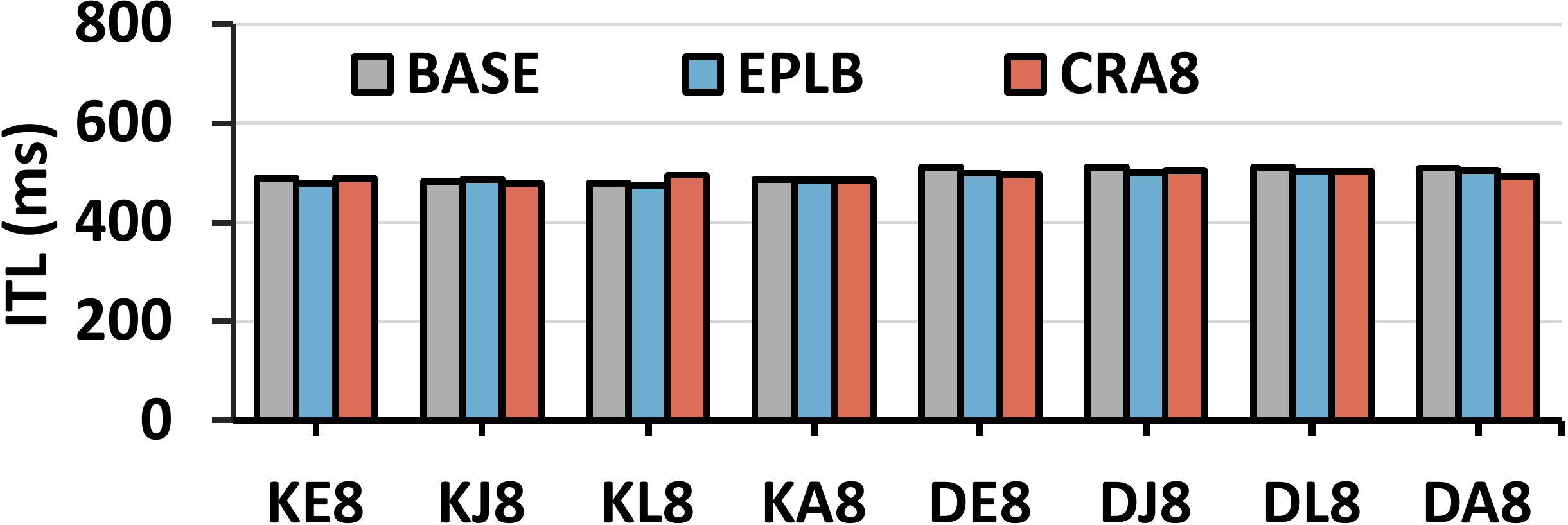}
    \caption{Average inter-token-latency at maximum throughput across different system setups for \texttt{BASE}, \texttt{EPLB} and \texttt{\proje8}.}
    \label{fig:itl}
\end{figure}

\textbf{\proj maintains the same inter-token-latency (ITL). } \fref{fig:itl} depicts the decoding ITL in all configurations at their maximum throughputs. We observe that the decoding ITL is stable across different setups. \cam{During decoding, batch sizes and token counts are significantly lower than during prefill because decoding requests generate only one or a few tokens (with multi-token prediction) while consuming an increasing amount of KV cache. Thus, load imbalance in MoE blocks becomes less significant, and the attention module constitutes a higher fraction of layer time.}

\begin{figure*}[!htb]
    \centering

    \subfigure[\texttt{SYN2} workload mix ($\sigma^2=0.001$)]{
        \includegraphics[width=0.485\textwidth]{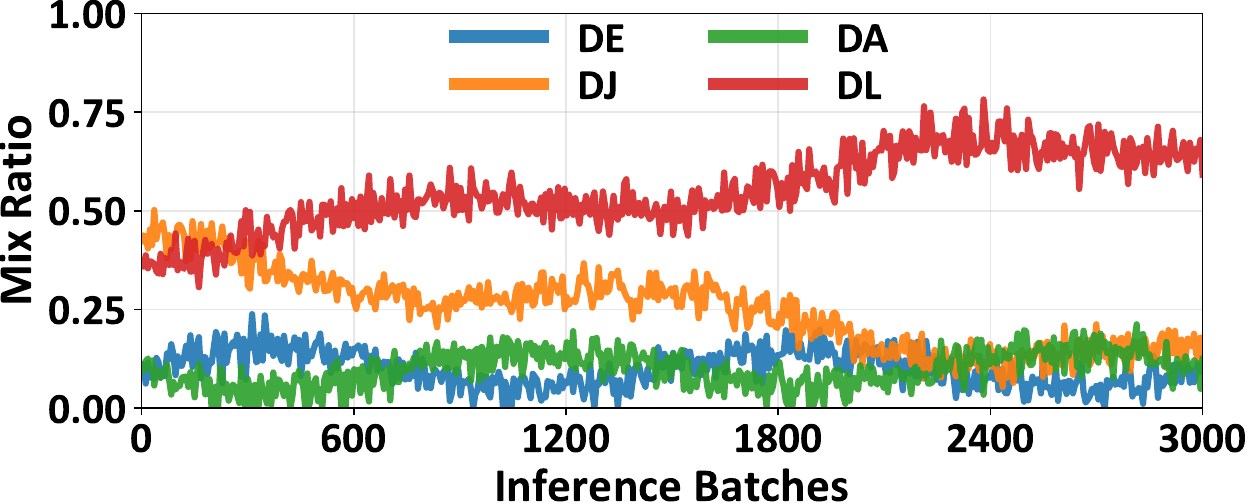}
        \label{fig:shift2mix}
    }
    \subfigure[\texttt{SYN2} balancedness ($\sigma^2=0.001$, no rebalancing)]{
        \includegraphics[width=0.485\textwidth]{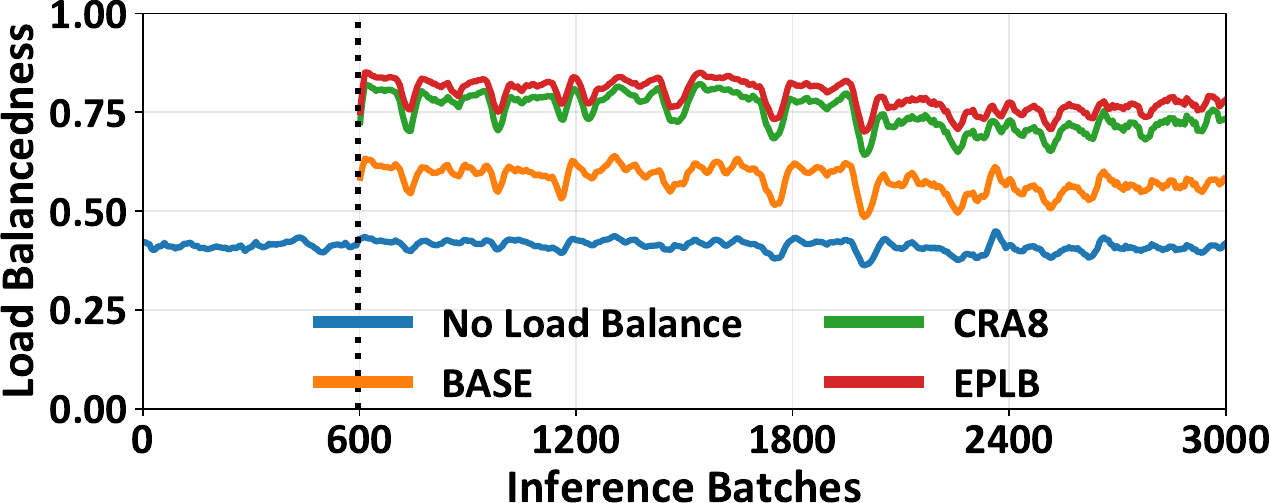}
        \label{fig:shift2bal1}
    } 
    \subfigure[\texttt{SYN2} balancedness ($\sigma^2=0.001$, with rebalancing)]{
        \includegraphics[width=0.485\textwidth]{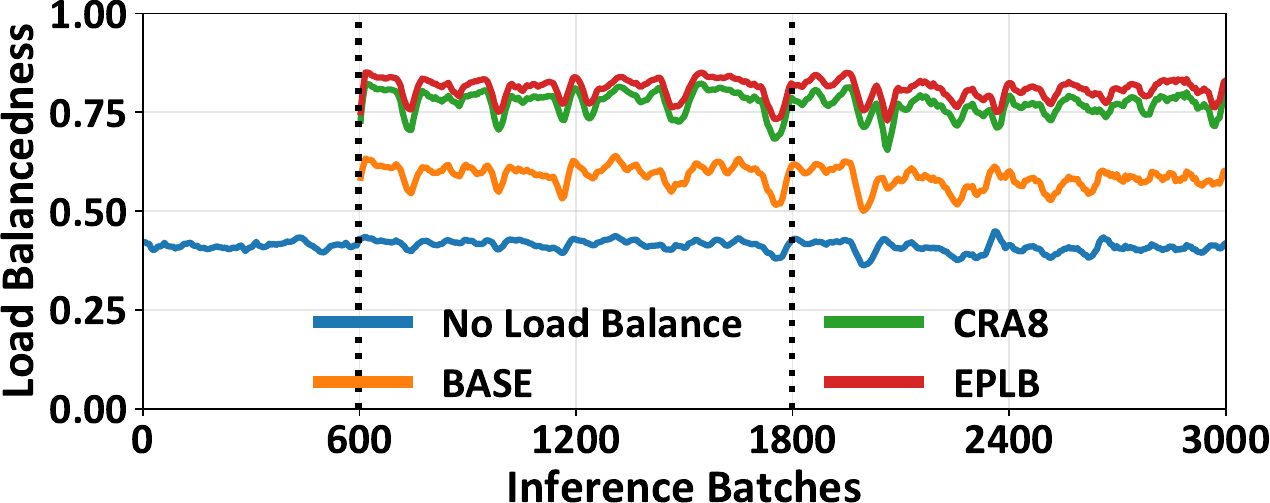}
        \label{fig:shift2bal2}
    } 
    \subfigure[\texttt{SYN2} balancedness ($\sigma^2=0.5$, no rebalancing)]{
        \includegraphics[width=0.485\textwidth]{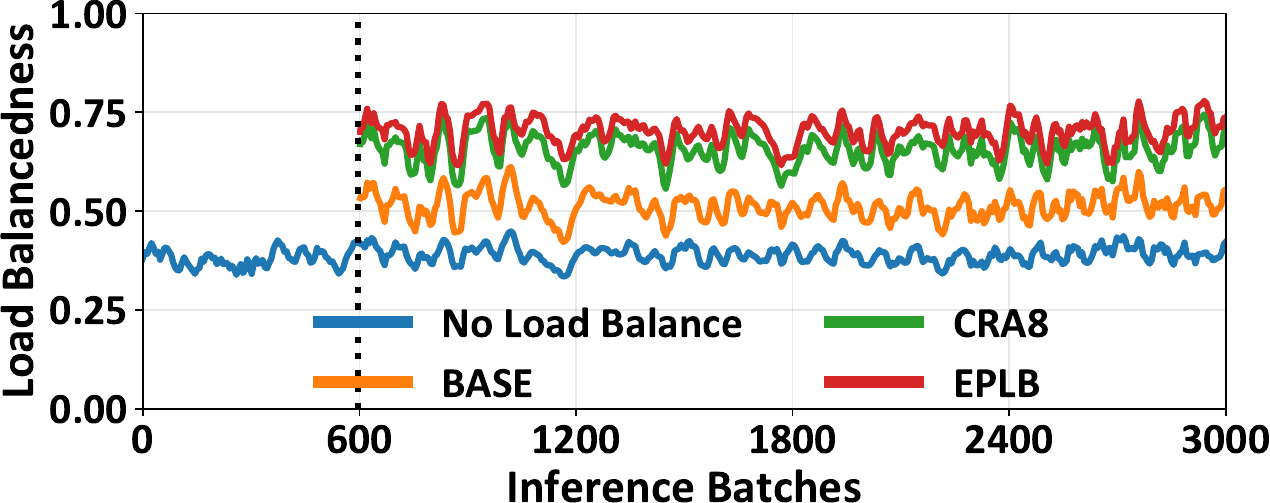}
        \label{fig:shift2bal3}
    } 
    
    \subfigure[\texttt{SYN3} workload mix ($\sigma^2=0.001$)]{
        \includegraphics[width=0.485\textwidth]{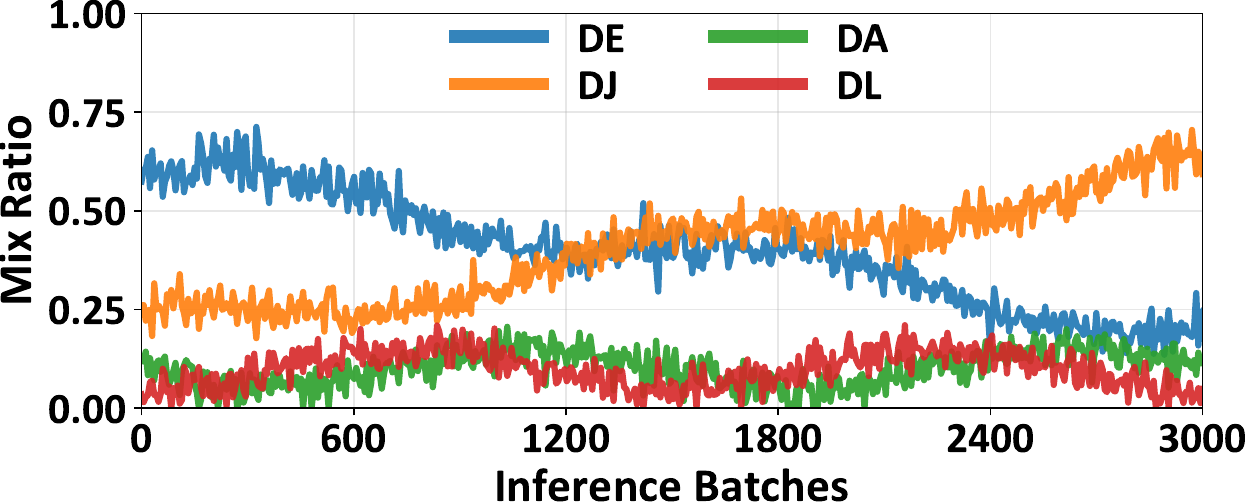}
        \label{fig:shift3mix}
    }
    \subfigure[\texttt{SYN3} balancedness ($\sigma^2=0.001$, no rebalancing)]{
        \includegraphics[width=0.485\textwidth]{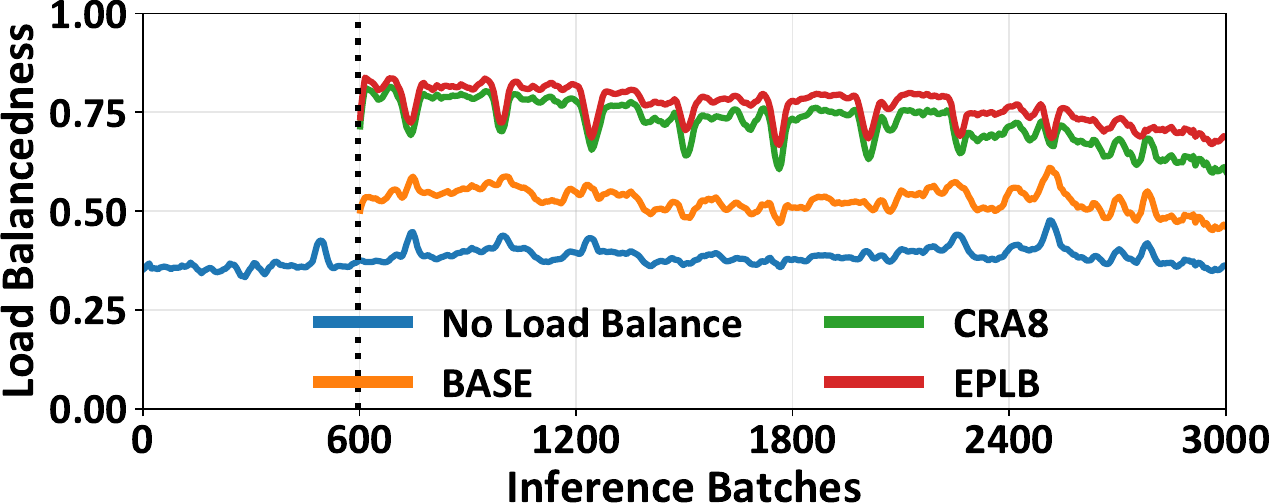}
        \label{fig:shift3bal1}
    } 
    \subfigure[\texttt{SYN3} balancedness ($\sigma^2=0.001$, with rebalancing)]{
        \includegraphics[width=0.485\textwidth]{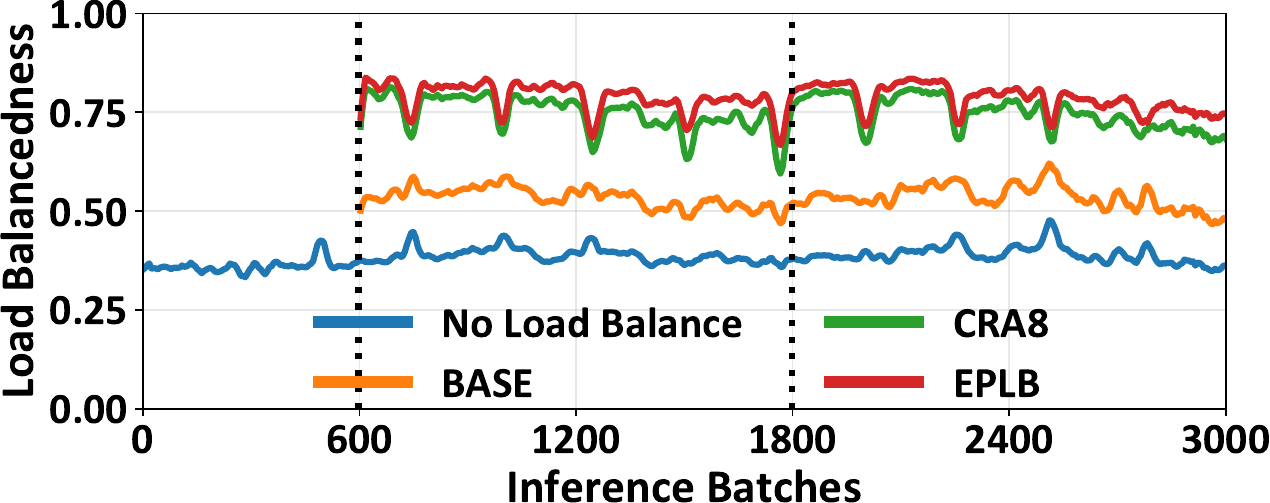}
        \label{fig:shift3bal2}
    } 
    \subfigure[\texttt{SYN3} balancedness ($\sigma^2=0.5$, no rebalancing)]{
        \includegraphics[width=0.485\textwidth]{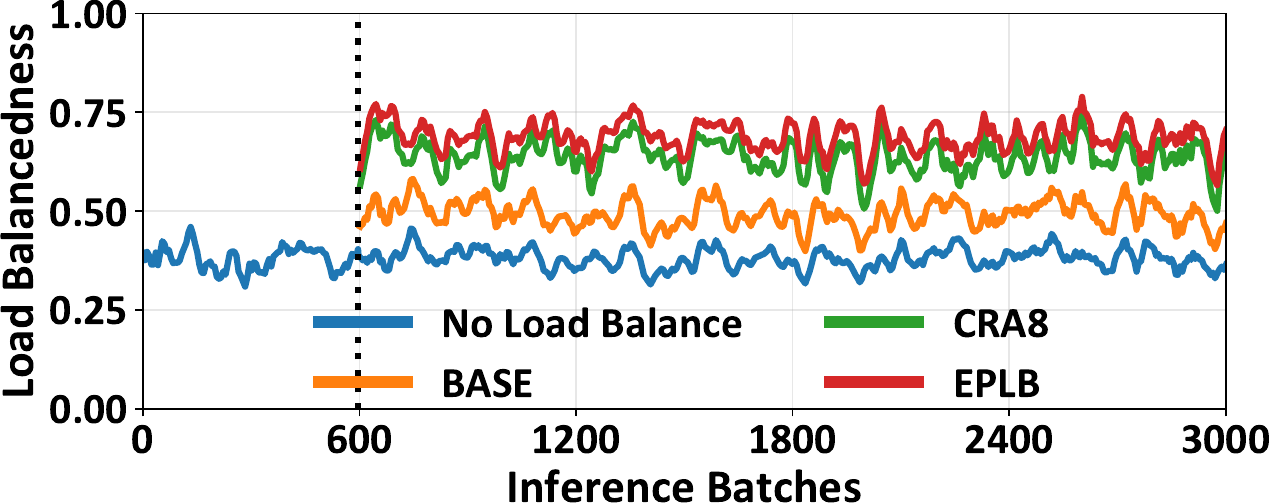}
        \label{fig:shift3bal3}
    } 
    
    \caption{\cam{\texttt{SYN2}, \texttt{SYN3} shifting workload mix ratios and achieved load balancedness over time on DeepSeek-R1-671B (\texttt{D}) across 8 nodes. Figures \ref{fig:shift2mix} and \ref{fig:shift3mix} depict the ratio of each dataset over time with burstiness variance $\sigma^2=0.001$ for \texttt{SYN2} and \texttt{SYN3} respectively. Figures \ref{fig:shift2bal1} - \ref{fig:shift2bal3} and \ref{fig:shift3bal1} - \ref{fig:shift3bal3} depict the achieved load balancedness under various configurations. The dotted lines represent triggered online periodic rebalancing, and data lines are smoothed for clearer presentation. Load-balancing techniques (\texttt{BASE}, \texttt{EPLB}, \texttt{CRA8}) are activated after the initial warmup of 600 iterations. Workloads with higher $\sigma^2$ only exhibit higher degree of burstiness; the shift and diurnal patterns remain identical.}}
    \label{fig:shift2}
\end{figure*}

\begin{figure*}[!htb]
    \centering
    
    \subfigure[\texttt{SYN4} workload mix ($\sigma^2=0.001$)]{
        \includegraphics[width=0.485\textwidth]{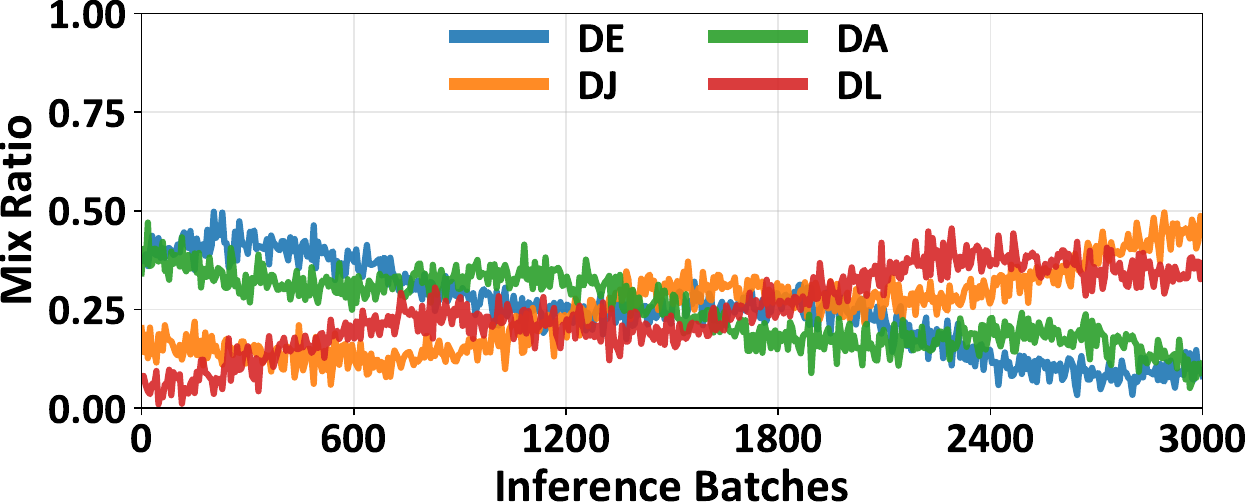}
        \label{fig:shift4mix}
    }
    \subfigure[\texttt{SYN4} balancedness ($\sigma^2=0.001$, no rebalancing)]{
        \includegraphics[width=0.485\textwidth]{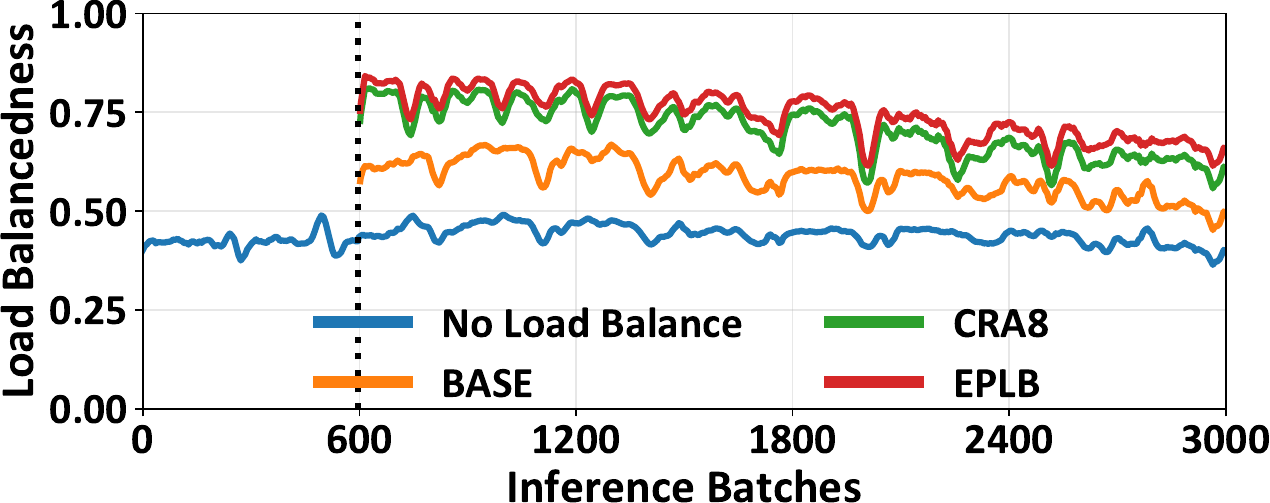}
        \label{fig:shift4bal1}
    } 
    \subfigure[\texttt{SYN4} balancedness ($\sigma^2=0.001$, with rebalancing)]{
        \includegraphics[width=0.485\textwidth]{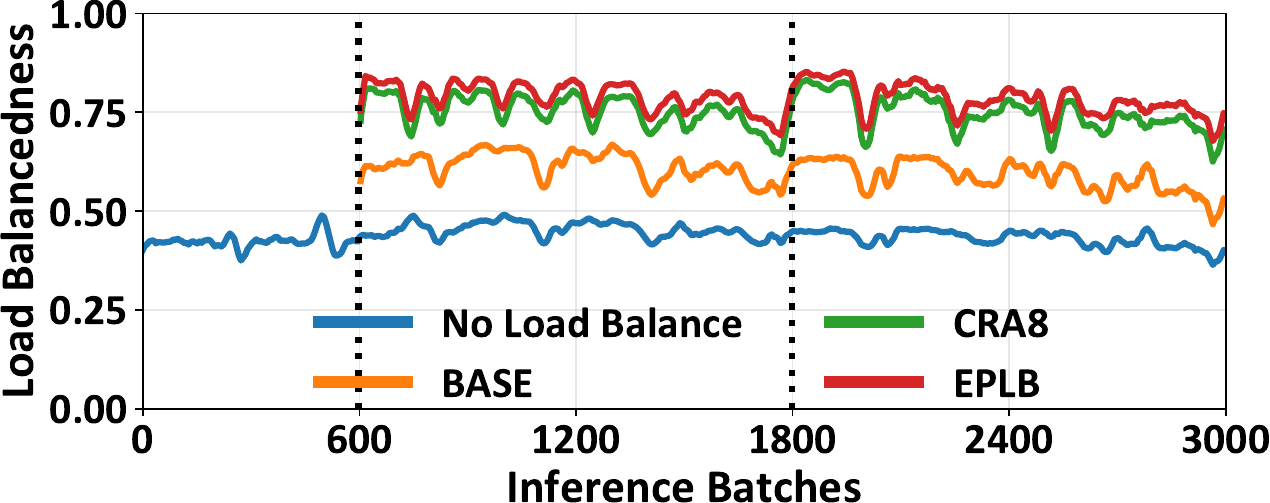}
        \label{fig:shift4bal2}
    } 
    \subfigure[\texttt{SYN4} balancedness ($\sigma^2=0.5$, no rebalancing)]{
        \includegraphics[width=0.485\textwidth]{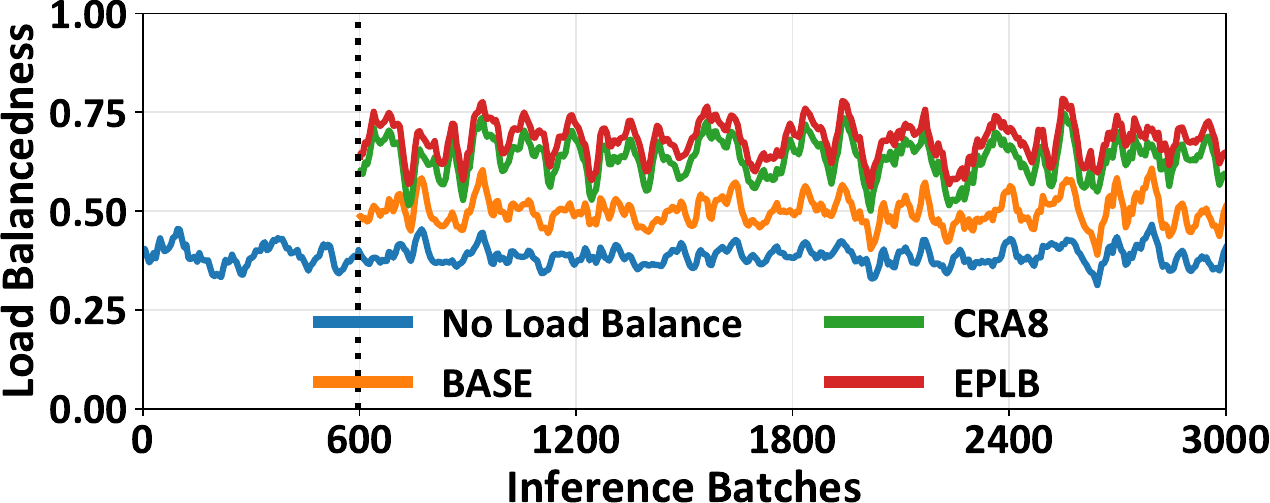}
        \label{fig:shift4bal3}
    } 
    \caption{\cam{\texttt{SYN4} shifting workload mix ratios and achieved load balancedness over time on DeepSeek-R1-671B (\texttt{D}) across 8 nodes. \fref{fig:shift4mix} depicts the ratio of each dataset over time with burstiness variance $\sigma^2=0.001$. Figures \ref{fig:shift4bal1} - \ref{fig:shift4bal3} depict the achieved load balancedness under various configurations. The dotted lines represent triggered online periodic rebalancing, and data lines are smoothed for clearer presentation. Load-balancing techniques (\texttt{BASE}, \texttt{EPLB}, \texttt{CRA8}) are activated after the initial warmup of 600 iterations. Workloads with higher $\sigma^2$ only exhibit higher degree of burstiness; the shift and diurnal patterns remain identical.}}
    \label{fig:shift3}
\end{figure*}

\section{Impact of Workload Shifts - Additional Workloads}
\label{sec:shifteval2}

Figures \ref{fig:shift2} and \ref{fig:shift3} depict the workload mix ratios and the achieved load balancedness on \texttt{SYN2}, \texttt{SYN3} and \texttt{SYN4}. The experiment setup and observations are identical to \fref{fig:shift} in \sref{sec:shifteval}.

\clearpage
\clearpage

\section{Artifact Appendix}
\definecolor{shbg}{RGB}{240, 240, 240}

\lstset{language=sh,
        basicstyle=\linespread{0.9}\ttfamily\footnotesize,
        columns=fullflexible,
        keywordstyle=\ttfamily,
        emph={int, dim3, list},
        emphstyle={\ttfamily},
        xleftmargin=.6cm,
        numbers=left,
        stepnumber=1,
        backgroundcolor=\color{shbg},
          breaklines=true,
          postbreak=\mbox{\textcolor{red}{$\hookrightarrow$}\space},
          breakatwhitespace=true
        }
        
\subsection{Abstract}

This Artifact Appendix describes how to generate an expert placement and replication plan with \proj.
Due to the high hardware demands of our evaluations (up to 12 nodes; see \sref{sec:setup}), we only provide the core implementation of \proj as described in \sref{sec:design} and \aref{sec:odalgo}. Since the MCKP-based replica allocation algorithm (details in \sref{sec:repalloc} and \aref{sec:odalgo2}) relies on explicit user configuration of the replication factor $R$, the core implementation instead employs a heuristic-based algorithm that automatically allocates replicas for each layer when the projected per-replica replication benefit is above a relative threshold. This algorithm makes no prior assumption of the memory budget and can be used in real deployment if automatic replica allocation is desired. In practice, we find that this heuristic-based allocation selects a replication factor $R$ close to 8 across all configurations, which is consistent with our observation in \sref{sec:baseline} and our evaluation setup (\texttt{CRA8}). 
The provided code references EPLB's official implementation \cite{eplb}. It takes a prior expert load distribution as input and outputs: (1) the number of replicas allocated for each MoE layer by \proj, (2) the selected replication ratio $R$ and memory saving relative to EPLB, and (3) the load balancedness of EPLB (placement-only), EPLB, and \proj. We show that \proj achieves comparable load balancedness while consuming significantly less memory than EPLB.

To facilitate the AE process, we provide links to the GitHub repository and the Zenodo archive that contain the \proj source code, expert load distribution traces, and automated workflow scripts to compute and evaluate replication plans. The code runs on Linux, macOS and Windows and is compatible with Intel, Apple Silicon and AMD CPUs. Conda is required. Git-LFS is required when setting up from the GitHub repository.

\subsection{Artifact check-list (meta-information)}

{\small
\begin{itemize}
  \item {\bf Data set: }custom expert load distribution traces collected during the inference of sampled requests from public datasets FinePDFs, RedPajama-Data-1T and Lambada.
  \item {\bf Run-time environment: }Requires Linux, macOS or Windows with Conda installations. Git-LFS is required when setting up from the GitHub repository.
  \item {\bf Hardware: }Intel, Apple Silicon or AMD CPU.
  \item {\bf Metrics: }replication ratio $R$ and load balancedness.
  \item {\bf Output: }plain text files containing raw data of all experiments and PNG figures that illustrate the experiment metrics across configurations.
  \item {\bf Experiments: }\proj replication plan generation. Replay-based balancedness computation.
  \item {\bf How much disk space required (approximately)?: }2 GB.
  \item {\bf How much time is needed to prepare workflow (approximately)?: }5 minutes.
  \item {\bf How much time is needed to complete experiments (approximately)?: }15 minutes.
  \item {\bf Publicly available?: }yes.
  \item {\bf Code licenses (if publicly available)?: }MIT license.
  \item {\bf Data licenses (if publicly available)?: }Apache-2.0 license.
  \item {\bf Archived (provide DOI)?: } \href{https://doi.org/10.5281/zenodo.19022696}{10.5281/zenodo.19022696}
\end{itemize}
}

\subsection{Description}

\subsubsection{How delivered}
Download and extract tarball file \lstinline{CRAFT_core.tar.gz} from \href{https://doi.org/10.5281/zenodo.19022696}{10.5281/zenodo.19022696} or clone the GitHub repository from \href{https://github.com/Accelsnow/CRAFT_core}{https://github.com/Accelsnow/CRAFT\_core}.

\subsubsection{Hardware dependencies}

AE should be run on a host machine running Linux, macOS or Windows.

\subsubsection{Software dependencies}

The verified operating systems and dependencies are listed below:

\begin{itemize}[noitemsep,topsep=0pt,leftmargin=8pt]
\item Ubuntu 24.04.4 LTS, Windows 11, macOS
\item Conda installation
\item Git-LFS (when setting up from the GitHub repository)
\end{itemize}

\subsubsection{Data sets}

The expert load distribution traces included in the GitHub repository are collected by processing sampled requests from the following three publicly available datasets:

\begin{itemize}[noitemsep,topsep=0pt,leftmargin=8pt]
\item FinePDFs
\item Lambada
\item RedPajama-Data-1T
\end{itemize}

\subsection{Installation}
After acquiring the code, execute the following commands:

\begin{lstlisting}
$ cd CRAFT_core
$ conda env create --file environment.yml
$ conda activate craft_core
\end{lstlisting}

If the code is cloned from GitHub without Git-LFS installation, \lstinline{git clone} will issue a warning about missing \lstinline{git-lfs}.
In this case, continue with conda environment creation and activation. After activating the conda environment, run:

\begin{lstlisting}
git reset --hard
\end{lstlisting}

\subsection{Experiment workflow}

To run all experiments and generate the result figure, grant execution permission and run one of the following commands based on the operating system:

Windows:

\begin{lstlisting}
.\run_all.ps1
\end{lstlisting}

Linux and macOS:

\begin{lstlisting}
./run_all.sh
\end{lstlisting}

By default, both workflow scripts are configured with a cluster size of 8 nodes. To run a single experiment with custom configurations, view the full list of configuration options by executing:

\begin{lstlisting}
python craft_core.py --help
\end{lstlisting}

To generate the output figure, place all custom experiment result files under the same directory, then run:

\begin{lstlisting}
python gen_ae_fig.py <custom_directory_path>
\end{lstlisting}

\vfill

\subsection{Evaluation and expected result}
When either \lstinline{run_all.ps1} or \lstinline{run_all.sh} finishes successfully, the output files will be stored under the \lstinline{results} directory in the project root. One \lstinline{.opt} result file is generated for each input \lstinline{.pkl} expert load distribution trace file. Eight input trace files are provided by default under the \lstinline{traces} directory. The naming conventions of the trace and result files follow the configuration syntax described in \sref{sec:setup}. The first letter represents the MoE model:

\begin{itemize}[noitemsep,topsep=0pt,leftmargin=8pt]
\item \texttt{D}: DeepSeek-R1-671B
\item \texttt{K}: Kimi-K2-1000B
\end{itemize}

The second letter represents the workload:

\begin{itemize}[noitemsep,topsep=0pt,leftmargin=8pt]
\item \texttt{E}: FinePDFs dataset deu\_Latn split
\item \texttt{J}: FinePDFs dataset jpn\_Jpan split
\item \texttt{L}: Lambada dataset
\item \texttt{A}: RedPajama-Data-1T dataset arxiv split
\end{itemize}

Each result file contains the following raw data in plain text:

\begin{itemize}[noitemsep,topsep=0pt,leftmargin=8pt]
\item Number of replicas allocated to each MoE layer with CRAFT.
\item The selected CRAFT replication ratio $R$ and the relative memory savings of CRAFT compared to EPLB $\uparrow$.
\item The load balancedness achieved with EPLB (placement-only), EPLB, and CRAFT $\uparrow$.
\end{itemize}

Our experiment workflow also generates a figure that depicts memory saving and load balancedness across configurations at \lstinline{results/ae_fig.png}. For experiments with custom configurations, the figure is saved to \lstinline{<custom_directory_path>/ae_fig.png}.

We expect the results to show that \proj consistently achieves comparable load balancedness to EPLB while consuming significantly less memory.


\end{document}